\documentclass[msc,slp,notimes,romanprepages]{infthesis}

\usepackage[utf8]{inputenc}
\usepackage[T1]{fontenc}
\usepackage{lmodern}
\usepackage[english]{babel}
\usepackage[colorlinks=true, allcolors = black]{hyperref}
\usepackage{fontawesome5}

\usepackage[version=4]{mhchem}
\usepackage{amsmath, amssymb, amsthm}
\usepackage{mathtools}
\usepackage{makecell}
\usepackage{multirow}
\usepackage{booktabs}
\usepackage{tabularx}
\usepackage{array}
\newcolumntype{P}[1]{>{\centering\arraybackslash}p{#1}}
\DeclareMathOperator*{\argmax}{arg\,max}

\usepackage{graphicx}
\usepackage{wrapfig}
\usepackage{subcaption}
\usepackage{verbatim}
\usepackage{siunitx}
\usepackage[dvipsnames]{xcolor}
\usepackage{titlesec}
\usepackage{amsmath} 
\newcommand{\code}[1]{\texttt{\detokenize{#1}}}

\titleclass{\subsubsubsection}{straight}[\subsection]
\newcounter{subsubsubsection}[subsubsection]
\renewcommand\thesubsubsubsection{\thesubsubsection.\arabic{subsubsubsection}}

\titleformat{\subsubsubsection}
  {\normalfont\normalsize\bfseries}{\thesubsubsubsection}{1em}{}
\titlespacing*{\subsubsubsection}{0pt}{3.25ex plus 1ex minus .2ex}{1.5ex plus .2ex}
\makeatletter
\renewcommand\paragraph{\@startsection{paragraph}{5}{\z@}%
  {3.25ex \@plus1ex \@minus.2ex}%
  {-1em}%
  {\normalfont\normalsize\bfseries}}
\renewcommand\subparagraph{\@startsection{subparagraph}{6}{\parindent}%
  {3.25ex \@plus1ex \@minus .2ex}%
  {-1em}%
  {\normalfont\normalsize\bfseries}}
\def\toclevel@subsubsubsection{4}
\def\toclevel@paragraph{5}
\def\toclevel@subparagraph{6}
\def\l@subsubsubsection{\@dottedtocline{4}{7em}{4em}}
\def\l@paragraph{\@dottedtocline{5}{10em}{5em}}
\def\l@subparagraph{\@dottedtocline{6}{14em}{6em}}
\makeatother
\usepackage{tikz}
\usepackage{tikz-3dplot}
\usetikzlibrary{calc,arrows.meta,positioning,decorations.pathmorphing,shapes.geometric}

\definecolor{BurntOrange}{HTML}{E66101}
\definecolor{Cerulean}{HTML}{1F78B4}
\definecolor{BluePastel}{HTML}{15616D}
\definecolor{Orange}{HTML}{FF7D00}

\tikzset{
  nnnode/.style={circle, draw, minimum size=18pt, inner sep=0pt},
  stepbox/.style={rectangle, draw, rounded corners=2pt, inner sep=4pt},
  >={Latex[length=2mm]},
  every label/.style={font=\small\itshape, inner sep=1pt},
  posEdge/.style={line width=0.5pt, draw=BluePastel},
  negEdge/.style={line width=0.5pt, draw=Orange},
}
\pgfmathsetseed{20250826}
\newcommand{\WeightSigma}{0.10}
\newcommand{\OpacityTau}{0.28}
\newcommand{\OpacityGamma}{3.0}
\newcommand{\OpacityFloor}{0.03}
\newcommand{\SparsifyThresh}{0.04}
\newcommand{\edgeWithRandomWeight}[2]{%
  \pgfmathsetmacro{\EPS}{1e-9}%
  \pgfmathsetmacro{\UONEraw}{rnd}%
  \pgfmathsetmacro{\UTWOraw}{rnd}%
  \pgfmathsetmacro{\UONE}{\EPS + (1-2*\EPS)*min(1,max(0,\UONEraw))}%
  \pgfmathsetmacro{\UTWO}{min(1,max(0,\UTWOraw))}%
  \pgfmathsetmacro{\Z}{sqrt(-2*ln(\UONE)) * cos(360*\UTWO)}%
  \pgfmathsetmacro{\W}{\WeightSigma * \Z}%
  \pgfmathsetmacro{\AW}{abs(\W)}%
  \pgfmathsetmacro{\TH}{\SparsifyThresh}%
  \ifdim \AW pt < \TH pt
  \else
    \pgfmathsetmacro{\scaled}{\AW/\OpacityTau}%
    \pgfmathsetmacro{\alphaCore}{1 - exp(-pow(\scaled,\OpacityGamma))}%
    \pgfmathsetmacro{\ALPHA}{\OpacityFloor + (1-\OpacityFloor)*min(1,max(0,\alphaCore))}%
    \ifdim \W pt > 0pt
      \draw[posEdge, opacity=\ALPHA] (#1) -- (#2);
    \else
      \draw[negEdge, opacity=\ALPHA] (#1) -- (#2);
    \fi
  \fi
}

\usepackage[sorting=none,citestyle=numeric-comp]{biblatex}
\usepackage{csquotes}
\usepackage{caption}
\title{%
  \includegraphics[width=80mm]{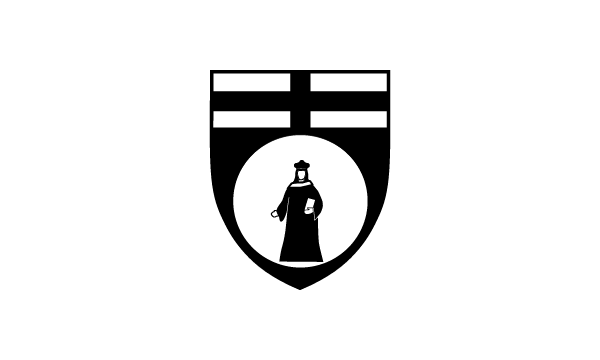}\\[10mm]
  {
  \parbox{0.98\textwidth}{\centering
    Calorimeter shower superresolution with conditional normalizing flows:\\
    Implementation and statistical evaluation%
  }}%
}

\author{Andrea Cosso}
\submityear{November 2025}

\abstract{In High Energy Physics, detailed calorimeter simulations and reconstructions are essential for accurate energy measurements and particle identification, but their high granularity makes them computationally expensive. Developing data-driven techniques capable of recovering fine-grained information from coarser readouts, a task known as calorimeter superresolution, offers a promising way to reduce both computational and hardware costs while preserving detector performance.

This thesis investigates whether a generative model originally designed for fast simulation can be effectively applied to calorimeter superresolution. Specifically, the model proposed in Ref.~\cite{supercalo} is re-implemented independently and trained on the CaloChallenge 2022 dataset based on the Geant4 Par04 calorimeter geometry.

Finally, the model’s performance is assessed through a rigorous statistical evaluation framework, following the methodology introduced in Ref.~\cite{ref_the_ref}, to quantitatively test its ability to reproduce the reference distributions.}

\begin{document}

\begin{preliminary}
\thispagestyle{empty}
\enlargethispage{60mm}
\begin{titlepage}
\begin{center}

\textsc{ \LARGE{Universit\`a\ degli\ Studi\ di\ Genova}}\\
\vspace{1cm}
\includegraphics[scale=0.40]{figures/Unige3.png}\\
\vspace{0.7cm}
\textsc{ \large Scuola di Scienze Matematiche, Fisiche e Naturali} \\
\vspace{1.0cm}

\textsc{ \large Laurea Magistrale in Fisica}\\
\vspace{2.0cm}

\makeatletter
{\LARGE\bfseries\@chapterfont
\setlength{\baselineskip}{0.9\baselineskip}
Calorimeter Shower Superresolution with Conditional normalizing Flows:\\
Implementation and Statistical Evaluation\par}
\makeatother

\vspace*{2.7cm}

\noindent
\begin{minipage}[t]{0.45\textwidth}
\large
Candidato \\
\textbf{Andrea Cosso}
\end{minipage}
\hfill
\begin{minipage}[t]{0.45\textwidth}
\raggedleft \large
Relatori \\
\textbf{Dott.\ Riccardo Torre}\\
\textbf{Dott.\ Marco Letizia}\\
\vspace{1cm}
Correlatore\\
\textbf{Dott.\ Andrea Coccaro}

\end{minipage}

\end{center}

\vspace{1.5cm}
\begin{center}
\textsc{Anno accademico 2024/2025}\\
\end{center}
\end{titlepage}
\newpage
\thispagestyle{empty}
\null
\newpage
\tableofcontents
\newpage
\thispagestyle{empty}
\null
\newpage

\begingroup
\makeatletter
\renewcommand{\abstractname}{Abstract} 
\begin{mainabs}
  \addcontentsline{toc}{section}{Abstract} 
  \@abs 
\end{mainabs}
\makeatother
\endgroup
\cleardoublepage

\newpage
\thispagestyle{empty}
\mbox{}
\newpage

\cleardoublepage
\end{preliminary}

\chapter*{Introduction}
\addcontentsline{toc}{chapter}{Introduction}
\markboth{Introduction}{Introduction}

\begin{figure}%
    \centering
    \subfloat{{\includegraphics[width=0.37\linewidth]{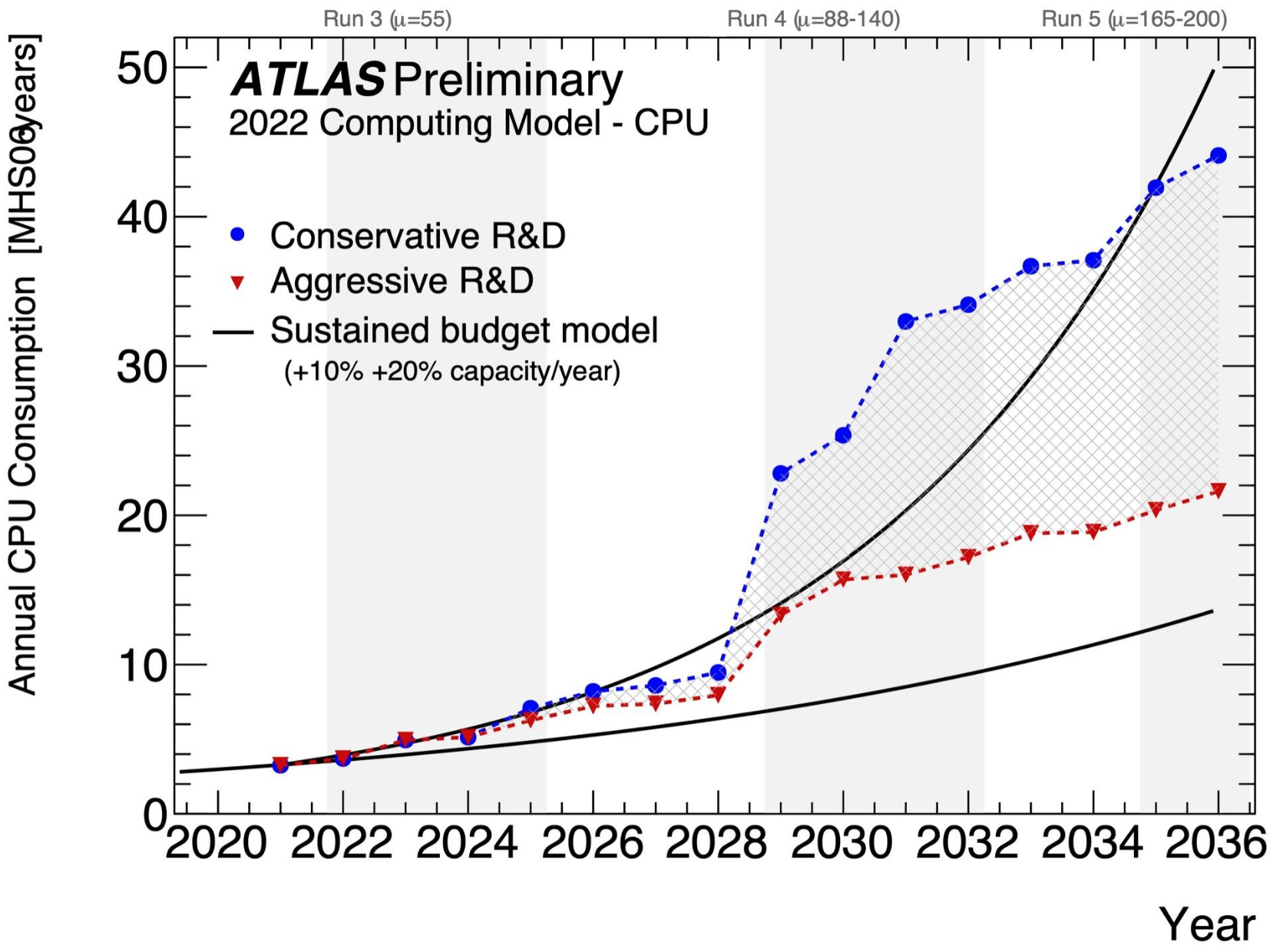} }}%
    \qquad
    \subfloat{{\includegraphics[width=.52\linewidth]{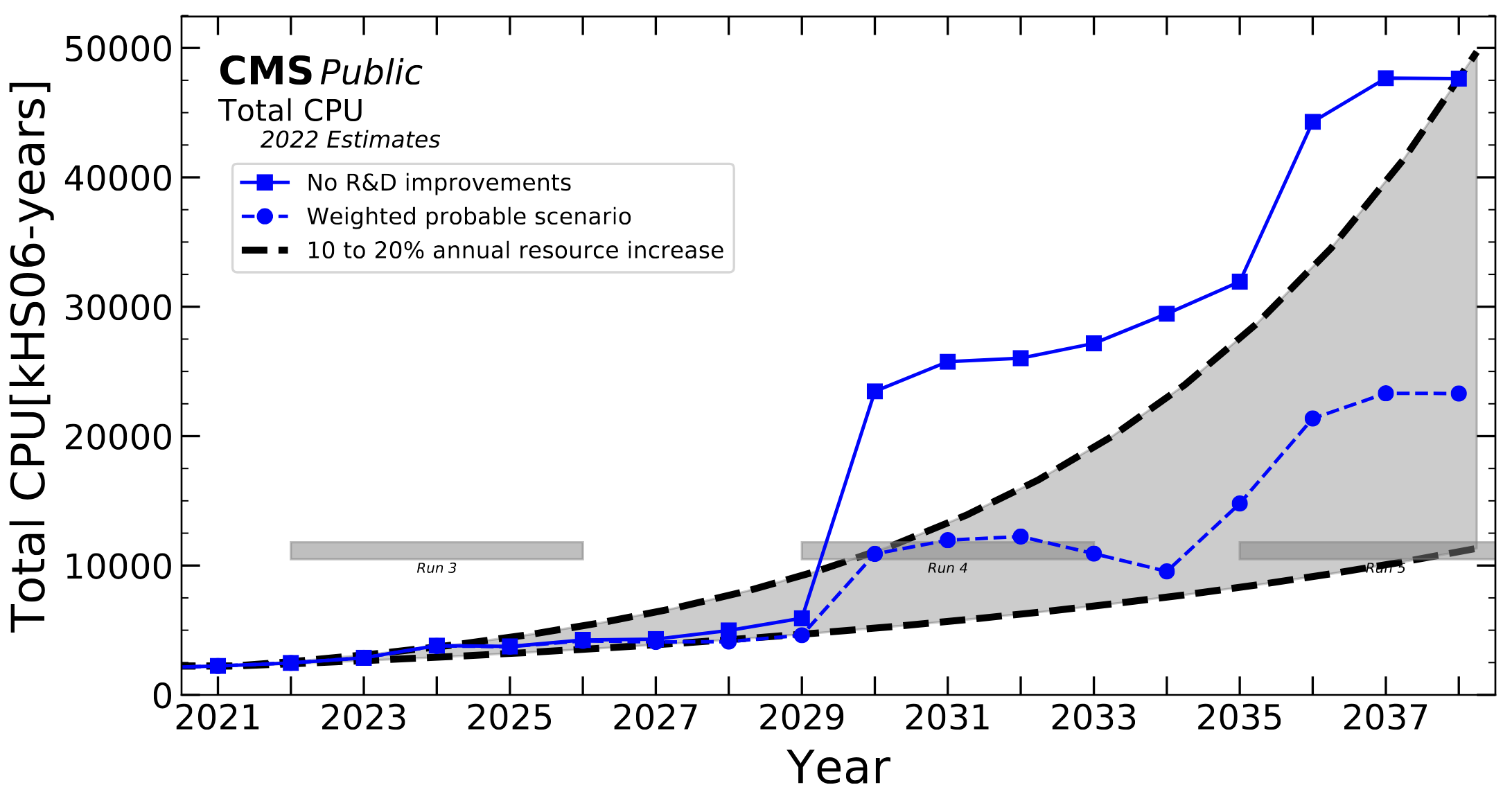} }}%
    \caption{Projected CPU requirements. \textbf{Left:} Atlas \cite{atlas_computing} \textbf{Right:} CMS \cite{CMS_computing}}%
    \label{fig:CPU_consumption}%
\end{figure}

In High Energy Physics (HEP), the Large Hadron Collider (LHC) plays a central role in testing the predictions of the Standard Model and exploring possible signs of new physics. Bridging theoretical predictions and experimental observations requires highly detailed simulations grounded in first-principles physics. In this context, HEP relies extensively on simulations, following a complex pipeline that encompasses event generation, detector simulation, and reconstruction. Detector interactions are typically modeled with high-precision Monte Carlo techniques, most notably implemented in \textsc{Geant4}~\cite{geant4}. Because the accuracy of many LHC measurements depends on such simulations, the increased data volume anticipated in future runs will significantly raise the demand for synthetic events and, consequently, for computational resources. This growing need is expected to become a major computational bottleneck in the near future~\cite{computation1,computation2,computation3}. Indeed, the \emph{High Luminosity LHC} (HL-LHC), which will commence operation in 2030, will increase the luminosity by a factor of 10, from an integrated luminosity of approximately $\simeq 300~\mathrm{fb}^{-1}$ (in Run~3) to $L\simeq 3000~\mathrm{fb}^{-1}$~\cite{hl_lhc}, dramatically increasing the need for computational resources (see Figure \ref{fig:CPU_consumption}). 
A major part of the computational power goes into the detector response simulation, particularly that of the calorimeter. In fact, calorimeters are especially computationally demanding due to the high number of secondary particles that must be tracked and simulated. Therefore, in recent years, with the development of advanced machine learning techniques, there has been a growing interest in developing faster calorimeter simulations. At present, most fast calorimeter simulations are based on parametrized calorimeter responses, thus bypassing the computationally heavy \textsc{Geant4} simulations. The problem with these algorithms is that they lack the fidelity needed to meet the precision requirements of HEP measurements~\cite{parametrize_cal_responce1,parametrize_cal_responce2,parametrize_cal_responce3}. For this reason, there is a strong global effort to develop new generative models capable of addressing the current and future challenges of detector simulation~\cite{calochallenge}.

Efficient super-resolution methods can offer an alternative path: instead of performing full fine-granularity simulations, one can simulate at a coarser discretization and then ``upsample'' to fine resolution using learned models. This reduces simulation cost, memory usage, and data volume while recovering (or approximating) the physics fidelity of fine segmentation.

Preserving fine-grained detector information is essential for the accurate reconstruction and identification of particles from detector signatures, which form the basis of all physics analyses. To illustrate the importance of calorimeter granularity, we can consider an example introduced in Ref.~\cite{photon_superresolution}, involving single high-energy photons at the LHC, of great interest for precision measurements in Quantum Chromodynamics (QCD) and Electroweak theory~\cite{single_photon_importance}. At hadron colliders, the main background source for photons is the electromagnetic decay of high-energy mesons, most frequently $\pi^0 \rightarrow \gamma \gamma$, since neutral pions are commonly produced in hadronic interactions. The signature of such decays at high-energy is often measured as a ``fake single photon'' due to the large Lorentz boost, which results in a small separation between the two photons. Resolving the two distinct signals from the decay of a single meson requires high spatial resolution, achieved by increasing the calorimeter segmentation. This is not always possible, since increasing granularity entails significant technical and financial challenges. Indeed, high-granularity (HG) directly translates into more readout channels, more electronics, higher data rates, increased cooling and power requirements, more material, and greater calibration effort, thus making the costs of HG calorimeters prohibitive~\cite{high_granularity_cal1}. Beyond this example, many physics analyses rely on subtle features of calorimeter showers, such as energy fractions in layers, cluster substructure, the identification of overlapping showers, and similar observables~\cite{photon_superresolution}. Therefore, super-resolution can offer a solution by \emph{virtually} increasing the calorimeter resolution without physically adding channels, leading to better performance in physics tasks such as improved particle identification (ID), more accurate reconstruction of objects (photons, $\pi^0$, jets), enhanced energy and position resolution, and reduced systematic uncertainties.

Another important application of super-resolution arises in the context of aging calorimeters. In real experiments, detectors inevitably age, front-end electronics degrade, calibrations drift, and granularity may effectively degrade. With time and radiation, cells may fail (dead channels) or be disabled due to high levels of noise or false hits that contaminate the measurements~\cite{calorimeter_deterioration,calorimeter_deterioration2,calorimeter_deterioration_CMS}. In this context, a super-resolution model trained to reconstruct fine-grained shower structures from coarse or incomplete data could, in principle, be used to fill in or recover the missing energy pattern in dead or disabled regions. Such an approach would offer significant benefits, including the recovery of otherwise lost information, mitigation of performance degradation in aging detectors, and the continued use of partially degraded calorimeter sections without major recalibration or replacement.

Finally, looking ahead to the High-Luminosity LHC (HL-LHC), pile-up may become a critical issue~\cite{HL_LHC_PU}. In a proton–proton collider like the LHC, protons are grouped in bunches containing roughly $10^{11}$ protons each. These bunches cross each other every 25 ns at the interaction points (ATLAS, CMS, etc.). Each time two bunches cross, more than one proton–proton interaction can occur. All these interactions occur within the same detector readout window, causing their signals to overlap. This phenomenon, known as pile-up, refers to the overlap of signals from multiple interactions occurring in the same or neighboring bunch crossings. During Run 2 at the LHC, the average pile-up was approximately $32$~\cite{CMSPileup,PU_run2_atlas}, while for Run 3 the value was approximately $60$~\cite{PU_run3}. At the HL-LHC, the average number of interactions per bunch crossing is expected to reach 140 to 200~\cite{HL_LHC_PU}, greatly increasing reconstruction complexity. Super-resolution, by \emph{virtually} increasing the calorimeter resolution, can become a powerful tool to improve reconstruction performance. 

In this thesis, we independently replicate the generative model introduced in Ref.~\cite{supercalo} in response to the \emph{Fast Calorimeter Simulation Challenge (CaloChallenge)} initiated in 2022~\cite{calochallenge}, a community challenge for fast calorimetry simulation. Participants were tasked with training their preferred generative models on the provided calorimeter shower datasets, with the goal of accelerating and expanding the development of fast calorimeter simulations, while providing common benchmarks and a shared evaluation pipeline for fair comparison.

The idea explored in Ref.~\cite{supercalo} was to use auxiliary models to generate a coarse representation of a calorimeter shower and implement a super-resolution algorithm to upsample the showers to their fine-grained representation. In this context, our goal is to replicate only the final upsampling model, implemented as a \emph{Normalizing Flow}~\cite{NF,autoregr_and_NF_kingma_2016} using \emph{Rational Quadratic Spline} transformations~\cite{rqs}. The model has been trained on the dataset provided by the \emph{CaloChallenge}, specifically on \emph{Dataset 2}, which was generated with the Par04 example of \texttt{Geant4}. 

The model is then evaluated following the methodology introduced in Ref.~\cite{ref_the_ref}. In the scientific domain, where high levels of precision and accuracy are required, validation presents critical challenges. Many existing validation methods lack a rigorous statistical foundation, making it difficult to provide robust and reliable evaluations. The high precision required in HEP, where accurate modeling of features, correlations, and higher-order moments is essential, demands robust performance assessment of generative models. Two-sample hypothesis testing provides a natural statistical framework for performance evaluation. Reference \cite{ref_the_ref} proposes a robust methodology for evaluating two-sample tests, focusing on non-parametric tests based on univariate integral probability measures (IPMs). The approach extends 1D-based tests, such as the Kolmogorov–Smirnov~\cite{KS1,ks2} or Wasserstein distance~\cite{WD1,WD2}, to higher dimensions by averaging or slicing (i.e.\ projecting data onto random directions on the unit sphere). The recently proposed unbiased Fréchet Gaussian Distance (FGD)~\cite{FGD1,FGD2} and Maximum Mean Discrepancy (MMD)~\cite{MMD1,MMD2} are also included. Reference \cite{ref_the_ref} addressed the challenges posed by state-of-the-art evaluation models, such as their low scalability to higher dimensions and the difficulty of assessing the performance of classifier-based evaluations.

The structure of this thesis follows a progressive path from the general concepts of Machine Learning to their concrete application in calorimeter shower modeling and statistical evaluation. In Chapter~\ref{ch:ml}, the fundamental ideas of Machine Learning are introduced, with a particular focus on their relevance in Physics. The main paradigms of learning are presented together with an overview of neural networks, establishing the basis for the more advanced architectures discussed in the following chapters. Building on these notions, Chapter~\ref{generative_models} explores the principles of generative modeling in HEP, describing the main families of generative models and focusing on \emph{Normalizing Flows} as the central framework adopted in this work. The mathematical formulation of flow-based models is discussed in detail, with emphasis on the \emph{Masked Autoregressive Flow (MAF)} and the \emph{Rational Quadratic Spline (RQS)} transformations used throughout the thesis. The same chapter also introduces the statistical evaluation framework employed to assess model performance. 

The focus then shifts, in Chapter~\ref{chap:calo}, to the physical and experimental context of calorimetry at the Large Hadron Collider (LHC). The working principles of electromagnetic calorimeters are described together with their role in HEP experiments, leading to the definition of the \emph{calorimeter shower super-resolution} problem that motivates this research. Subsequently, Chapter~\ref{chap:implementation} presents the practical realization of the proposed approach, detailing the dataset used (Dataset~2 from the CaloChallenge~2022), the preprocessing and conditional input formulation, and the architecture of the implemented conditional MAF model. The training procedure, hyperparameter configuration, and numerical considerations are also discussed to provide a complete picture of the experimental setup. Finally, Chapter~\ref{chap:conclusion} summarizes the main findings and lessons learned, highlighting the implications of the results for future developments in fast calorimeter simulation and possible directions for extending this work toward more advanced generative modeling frameworks in HEP.

\chapter{Machine Learning Essentials} \label{ch:ml}
\section{Machine Learning in Physics}
Machine learning, a branch of artificial intelligence, focuses on developing algorithms that enable systems to learn patterns directly from data. The central idea is to design models capable of generalizing knowledge gained from previous examples to unseen situations. While the specific objectives depend on the application domain, the task, and the data representation, the main goal is to perform meaningful operations without being explicitly programmed for each case. In recent years, with the increase in computational power, advances in algorithms, and an explosion of available data, machine learning has become an important part of the scientific landscape and beyond. In High Energy Physics, for example, it is employed in a wide range of tasks, including particle identification, event classification, and fast detector simulation.
This section is based on Ref.~\cite{ML_course}.

\subsection{The three paradigms of machine learning}
Learning can be classified in three main categories, or paradigms. In this section we present an introduction to these categories providing some of their applications in physics.

\subsubsection*{Supervised learning} Supervised learning relies on data that is \textit{labeled}, which means that each point in the dataset \(\mathbf{x}\) is associated with a known target \(\mathbf{y}\). The model then tries to learn a function \(\mathbf{y} = f(\mathbf{x})\). Classical examples of supervised learning tasks in physics include:
\begin{itemize}
    \item \textbf{Regression}: Predicting continuous values, such as estimating the energy deposition in a calorimeter based on the particle’s velocity and type.
    \item \textbf{Classification}: Assigning data points to discrete categories, such as classifying particles into types (e.g., electron vs. photon) based on their detector response.
    \item \textbf{Time series forecasting}: Predicting the future behavior of physical systems, such as forecasting the positions of a particle in a magnetic field using historical data.
    \item \textbf{Generation and sampling}: Generating new physical event data that resembles real-world data, such as simulating particle interactions in a detector based on learned distributions.
\end{itemize}

\subsubsection*{Unsupervised learning} Unsupervised learning uses only the data \(\mathbf{x}\) without any target, so an unsupervised algorithm attempts to learn properties from the distribution of \(\mathbf{x}\). Common tasks for unsupervised learning in physics include:
\begin{itemize}
    \item \textbf{Density estimation}: Estimating the underlying probability distribution of particle energies or momenta, for example, learning the distribution of cosmic ray intensities in different regions of space.
    \item \textbf{Anomaly detection}: Identifying unusual events in detector data, such as detecting anomalous particle interactions that do not conform to expected physics models.
    \item \textbf{Generation and sampling}: Generating new samples of physical events, such as producing synthetic events for Monte Carlo simulations or generating data simulating the behavior of new particles.
    \item \textbf{Imputation of missing values}: Filling in missing data caused by sensor failures or incomplete measurements, such as predicting the missing hit data within a detector grid based on neighboring readings.
    \item \textbf{Denoising}: Removing noise from experimental data, such as cleaning up noisy signals in a particle’s trajectory reconstructed from detector hits.
    \item \textbf{Clustering}: Grouping similar events or particles based on their characteristics, such as clustering events in a detector based on energy deposits and spatial arrangement (e.g., hadron vs. electron showers).
\end{itemize}

\subsubsection*{Reinforcement learning} 
Reinforcement learning mimics the trial-and-error process humans use to learn tasks, by training software agents to make decisions that optimize an objective. Every action that works toward the goal is reinforced, while actions opposing the goal are ignored. \\
\\
While the first two paradigms are well explored in physics, reinforcement learning is less common and still under exploration. 

\subsection{Data discussion}
One of the cornerstones of the success of machine learning algorithms is the availability and quality of data. In this section, we briefly discuss these aspects within the scientific domain. A fundamental distinction can be made between \emph{observational} and \emph{experimental} sciences.

\paragraph{Observational sciences} 
In these fields, the amount of available data is often limited by external constraints. For example, in medicine, statistical limitations and privacy regulations restrict data accessibility, while in astronomy or climatology, the quantity and quality of data depend on the duration, frequency, and precision of observations, as well as on the nature of the systems being studied.

\paragraph{Experimental sciences} 
In contrast, experimental sciences are primarily limited by our ability to produce and collect data. Examples include the luminosity achievable at a particle collider or the performance of the detector technology used for data acquisition. Since the experimental apparatus is designed and controlled by the experimenter, it can, to some extent, be optimized to maximize the amount and quality of collected data.

\subsection{Statistical learning}
The term \emph{statistical learning} does not have a single, universally accepted definition. In this thesis, it refers to the branch of machine learning grounded in statistical theory and inference. Historically, machine learning has been primarily concerned with prediction, whereas statistics has emphasized inference and uncertainty quantification. Statistical learning thus represents the intersection of these two perspectives, combining principles from statistics, computer science, and data science. In this section, we introduce the main concepts, methods, and challenges of statistical learning, together with the fundamental components of learning algorithms.

\subsubsection{The Loss function} \label{sec:loss}
One of the basic assumptions of statistical learning is that every learning process can be formulated as an optimization problem, usually in terms of the minimization of one or more \emph{objective functions}, called \emph{Loss} or \emph{Cost} functions. The objective functions usually depend on the data $\{(\mathbf{x}_i, \mathbf{y}_i)\}$, model $f(\, \cdot \, ; \mathbf{\theta})$ and parameters $\mathbf{\theta}$. The following discussion focuses on the supervised learning case for simplicity, although the same principles can be extended to unsupervised settings. 

The loss function (or simply \emph{loss}) should give a measure of the distance between the true output and the observed one in case of supervised learning, a measure of how well the model is describing the underlying data distribution in unsupervised learning, or it can be something more complex such as a cumulative score function in case of reinforcement learning. The loss function should satisfy two key requirements: 
\begin{itemize}
    \item Should be as simple as possible to evaluate and minimize, since computational efficiency is important.
    \item Should be as close as possible to the \emph{figure of merit} that one aims to optimize for the given problem.
\end{itemize}
The choice of the loss function is a crucial aspect in the result of the learning task, since different numerical optimization algorithms and different functions can lead to different results. 

The figure of merit is the function used for the evaluation of the model. In practice, it is not always possible to use the exact figure of merit as the loss function, since it may be difficult to evaluate or to optimize directly. The loss function should therefore approximate or be closely related to a figure of merit that is meaningful for the problem under study, ensuring—at least in principle—that minimizing the loss corresponds to minimizing the true figure of merit, at least at a theoretical level, the minimum of the loss function must coincide with the minimum of the figure of merit. A brief overview of two widely used loss functions for supervised and unsupervised tasks is presented below, while additional examples are provided in Appendix~\ref{app:loss}. 

It is important to note that the model output depends on the parameters $\mathbf{\theta}$, while $\hat{\mathbf{\theta}}$ denotes the optimal parameters that minimize the loss function through the optimization process. 
\paragraph{Mean Squared Error (MSE)} is the average of the squared differences between the predicted and the true output. It is one of the simplest yet effective losses used for the supervised tasks. Usually employed in regression tasks due to its relation with \textbf{Maximum Likelihood Estimation (MLE)}, it is sensitive to the tails and penalizes outliers. Its functional form can be written in a simple way, for an output vector $\mathbf{y}$ (i.e. the target vector) in $n$ dimensions and $N$ samples: 
\begin{displaymath}
    \mathcal{L}_{\text{MSE}} = \frac{1}{N}\sum_{i = 1}^{N} \left\lVert \mathbf{y}_i - f(\mathbf{x}_i; \mathbf{\theta}) \right\rVert^2_2
\end{displaymath}
where $\left\lVert \cdot \right\rVert_2$ is the Euclidean 2-norm, defined by $\left\lVert \mathbf{x} \right\rVert_2 \vcentcolon = \sqrt{\displaystyle \sum_{i=1}^{n} x_i^2}$

\paragraph{Negative Log-Likelihood (NLL)} 
In probabilistic modeling, the output of a model is treated as a random variable whose distribution is parametrized by the model parameters. In other words, instead of predicting a single deterministic value, the model specifies a Probability Density Function (PDF) $p(\mathbf{X}; \mathbf{\theta})$ in the unsupervised case, or a conditional PDF $p(\mathbf{Y} \mid \mathbf{X}; \mathbf{\theta})$ in the supervised case. 

The learning objective is then defined as the \emph{negative log-likelihood} of the observed data given the model parameters, that is, the negative logarithm of the probability assigned by the model to the observed samples. For a supervised task, the loss can be written as:
\begin{displaymath}
    \mathcal{L}_{\text{NLL}}^{\text{(sup)}} = - \sum_{i = 1}^{N} \log p(\mathbf{y}_i \mid \mathbf{x}_i; \mathbf{\theta}),
\end{displaymath}
while for the unsupervised case it becomes:
\begin{equation}
    \mathcal{L}_{\text{NLL}}^{\text{(unsup)}} = - \sum_{i = 1}^{N} \log p(\mathbf{x}_i; \mathbf{\theta}).
\end{equation}

Under the assumption of Gaussian outputs, it can be shown that the NLL loss reduces to the Mean Squared Error (MSE) loss plus a constant term. This explains why the MSE loss is widely used in practice: when the data distribution is approximately Gaussian, one can bypass the explicit probabilistic formulation and obtain the maximum-likelihood estimate of the parameters simply by minimizing the MSE.

\subsubsection{An introduction to the statistical model}
A statistical model, defined by a set of mathematical operations acting on the input data, can be seen as a sophisticated if-then rule that, by using a set of parameters, can be trained to solve the task it was built for. The parameters are simply called \emph{model parameters} or \emph{model weights} (those that multiply features) and \emph{bias} (the additive term). To clarify the difference between weights and biases, take as an example a simple polynomial model, defined by  
\begin{displaymath}
    y = \sum_{i = 1}^{d} \omega_i x_i + b,
\end{displaymath}
where $y$ is the \emph{output} of the model, $\{\omega_i\}$ the set of weights and $b$ the bias. We refer to a \emph{hypothesis} as a specific instance of the model obtained by fixing the parameter values $\boldsymbol{\theta} = (\omega_1, \dots, \omega_d, b)$. The collection of all such hypotheses, obtained by varying these parameters, defines the \emph{hypothesis space} of the model. The parameters that are not optimized during training, but instead fixed before the training process begins, are called \emph{hyperparameters}. They introduce the need of \emph{model selection} and the \emph{validation set} that will be discussed later in the text. An example of hyperparameter can be found in the simple polynomial model as the maximum degree of the polynomial $d$: it is fixed in the training procedure but will need to be chosen in the model selection instance, since a priori one might not know what its optimal value is. Other examples of hyperparameters will be discussed later in the text. It is important to note that hyperparameters are not bound to the hypothesis space but can be parameters of the optimization algorithm, such as the number of \emph{training steps} and the \emph{learning rate} which will be further discussed in a dedicated section.

\subsubsubsection*{Training, Validation and Testing}
The training procedure of a machine learning model begins with the definition of distinct datasets, each serving a specific purpose in the learning process. Although this is a simplified description, the structure can change when techniques such as \emph{cross-validation}\footnote{Cross validation goes beyond the scope of this thesis; a detailed introduction can be found in Ref.~\cite{ML_course}} are employed.  
Typically, the available data are divided into three independent and non-overlapping sets:
\begin{itemize}
    \item \textbf{Training set:} used to fit the model parameters by minimizing the loss function.
    \item \textbf{Validation set:} used to monitor the model's performance during training and guide model selection.
    \item \textbf{Test set:} used to evaluate the final performance of the model after training is complete.
\end{itemize}

To understand the need for this partitioning, it is useful to briefly outline the training process.  
During training, the model receives one or more input samples (organized in \emph{batches} or \emph{mini-batches}, as discussed later in the optimization section) from the training set. The model’s predictions are compared to the true targets through the loss function, and the resulting error is used to update the weights via the \emph{backpropagation} algorithm\footnote{Geoffrey Hinton won the Nobel Prize in 2024 for the introduction of backpropagation~\cite{backprop}}. This iterative process continues until convergence.  

After each update, the model’s performance is assessed on the validation set by computing the validation loss. Unlike the training loss, this quantity is not used to update the parameters; instead, it provides a measure of how well the model generalizes to unseen data and serves as a criterion for \emph{model selection}, i.e., for choosing the model that best balances fit and generalization. The concept of model selection will be further discussed in a dedicated section.  

Finally, the test set is employed only once the training and model selection are completed. It provides an unbiased estimate of the model’s performance on new, unseen data, typically evaluated through metrics that depend on the specific task.

\subsubsubsection*{Capacity} Capacity is defined as the ability of the model, or a particular hypothesis, to accurately describe a dataset. We can distinguish between \emph{representational capacity} and \emph{effective capacity}:

\paragraph{Representational capacity} is defined as the capacity of the model to accurately describe a large variety of true data models. It does not depend on the data and is related to the number of parameters, number of features, complexity of the model's functional form, and so on. In the example of the polynomial model, a high-degree polynomial can accurately describe multiple true data models with different degrees. For this reason, we say that a high degree polynomial has a higher representational capacity than a low degree one.

\paragraph{Effective capacity} n the other hand, takes into account the data, regularization techniques, optimization methods, and other secondary factors. It is a more empirical definition of capacity and can be defined as the practical ability of the model to capture the important features in the training data, given additional effects, such as the finite training dataset size, noise, regularization techniques and optimization algorithms.

\paragraph{Optimal capacity} although its definition heavily depends on the specific problem and on the figure of merit, the optimal capacity can be determined by optimizing the trade-off between learning in detail the training (\emph{overfitting}) data and the capabilities of generalizing to new, unseen data.

\subsubsection*{Generalization, overfitting and underfitting} 
In machine learning, \emph{generalization} can be defined as the ability of a model to describe previously unseen data. Given a function that measures the error, such as the losses described earlier, its value computed on the training set is called \emph{training error}, while the values computed on the validation and test set are called \emph{validation error} and \emph{test error} respectively. The \emph{generalization error} is the error the model makes on unseen data, which quantifies how well it generalizes beyond the training samples. Since the true generalization error cannot be measured directly, it is commonly approximated by the \emph{test error}. The validation error cannot be a robust measure of the generalization error, in fact, the validation set is used for model selection and thus is seen by the model during optimization. Even with a perfect fit of the model to the data, the generalization error always has a non-zero lower bound. This \emph{irreducible} generalization error is commonly referred to as \emph{Bayes error} and is related to the fact that the noise in the training data prevents the model from learning the true underlying model. 

To train an ML algorithm, there are two crucial objectives: 
\begin{itemize}
    \item The model should be able to describe well the training data it used to estimate its parameters. This translates into the smallest possible training error.
    \item The model should be able to generalize well to new, unseen data. So the generalization error must also be as small as possible and, possibly, minimizing the \emph{generalization gap}, defined as the difference between the training and generalization errors.
\end{itemize}
When the first objective cannot be satisfied, we say that the model is \emph{underfitting}, while the challenge associated with the second objective is called \emph{overfitting}, which means that the model has learned "too well" the training dataset and has a big generalization gap.

\paragraph{Underfitting:} Occurs when a model does not have enough capacity to even describe the training data, or the optimization task hasn't converged. Note that, even though the theoretical optimal value for the loss is known a priori, a proper ``scale'' for the actual problem as well as the associated Bayes error is not known. For this reason, generally, it is not a trivial task to understand whether we are underfitting or not. Underfitting is typically recognized \emph{a posteriori}: when the model capacity or training configuration is adjusted, the training error decreases, indicating that the previous model was too simple to capture the underlying structure of the data. Underfitting is related to the effective capacity, not to the representational capacity; indeed, even if a theoretical hypothesis is general enough to be, in principle, capable of modeling the data, the actual optimization task may become too difficult, and the model could not converge to the right parameters, leading to underfitting.

\begin{figure}
    \centering
    \includegraphics[width=0.8\linewidth]{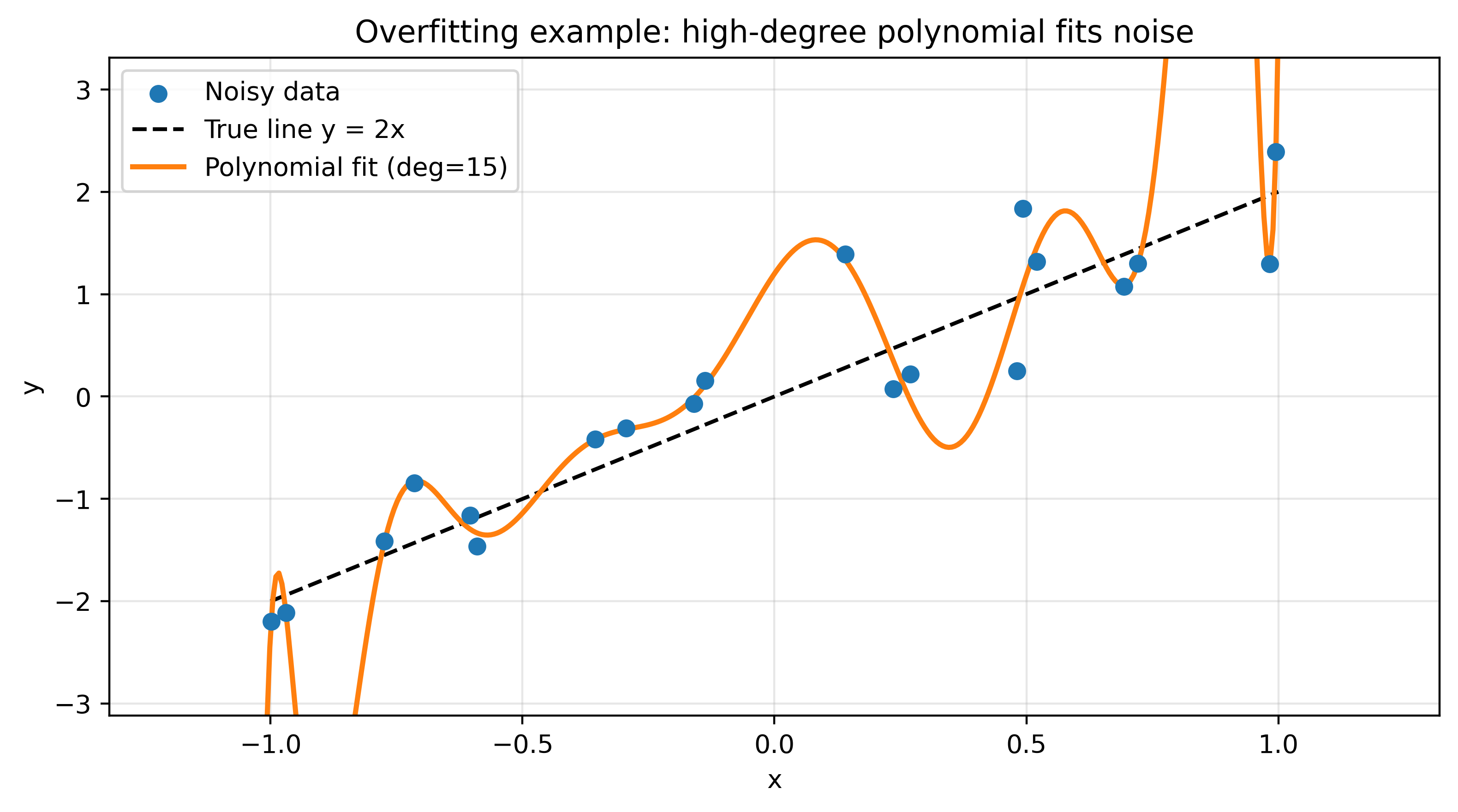}
    \caption{Overfitting example. By fitting noisy data with sufficiently high degree polynomials, the curve is able to parametrize noise.}
    \label{fig:overfitting}
\end{figure}

\paragraph{Overfitting:} Capacity can increase by changing the number of parameters or by changing the hyperparameters. When the number of parameters of the model approaches the number of data points in the training set, the model can start describing the data perfectly, almost independently of the specific model. At this stage, the train error could become arbitrarily small due to the fact that the model can learn even the noise in the data. The result is a very good fit for the training data, with a very poor generalization, which translates to a large generalization gap. This is what we call \textbf{overfitting}. In contrast to underfitting, the overfitting can be identified during training. To do this, we look at the \textbf{learning curve}, i.e., a plot of the training and validation losses versus the training steps. A typical indicator of overfitting is that the training error continues to decrease while the validation error remains constant or increases. In this instance, the model can be too sensitive to the noise or the capacity too large. A simple example of overfitting can be found in Figure \ref{fig:overfitting}, where it is clear that a sufficiently high degree polynomial can fit noise in the data. 

The essence of training a machine learning model is to find the perfect balance between overfitting and underfitting, either by training strategies or by choosing the right model (this is what we defined as effective capacity). This concept is commonly referred as the \emph{Bias-Variance trade-off}. 

\subsubsection*{Regularization} 
We should briefly discuss some of the most commonly used techniques to find the balance between overfitting and underfitting. Acquiring the balance by merely engineering the model is usually too difficult, if not impossible, in most cases. One way to find such balance is to start with a really simple model with``average'' results and then slowly increase the capacity until the generalization gap starts to increase, while the training error is still decreasing. At this point, the model is starting to overfit. We can push a bit further in the overfit direction by adding a \emph{regularizer}. Regularization is any technique that aims at lowering the generalization gap without affecting (at all or in a marginal part) the train error. In practice, this is almost never possible, so the train error is affected, and it increases. For this reason, the model capacity is usually increased along with regularization to decrease the generalization gap, while maintaining the train error constant. 

Regularization can be seen as prior knowledge for the model or as a penalty to the loss and can be applied in different ways. 
A general way to apply regularization is to add a penalty term to the objective function for training. Note that this is not always true; indeed, some form of regulations cannot be written this way, for example \emph{early stopping} and \emph{dropout} that will be discussed later in the text. In mathematical form we can write a penalty term as:
\begin{displaymath}
    \tilde{\mathcal{L}}(\mathbf{\theta}) = \mathcal{L}(\mathbf{\theta}) + \lambda \Omega(\mathbf{\theta}) \text{\,  with \, } \lambda \in [0,+\infty ) ,
\end{displaymath}
where typically, $\Omega$ is chosen to affect the parameters but not the biases, and $\lambda$ may depend on the stage of the algorithm to address different problems depending on the stage of the calculation. In order to clarify the concept of regularization, some illustrative examples are presented. This discussion is intended as an introduction to the most widely used regularization methods, rather than an exhaustive overview.

\paragraph{L1 and L2 regularizers}
L2 and L1 regularizers are the most known penalty terms in deep learning. The solution to linear models using L2 regularizer is called \emph{Ridge regression}, while the solution using L1 is called \emph{Lasso regression}. L2 penalizes large weights by the L2 norm of the weight tensor (for this reason, L2 is also known as \emph{weight decay}). L2 is expressed in formulae as
\begin{displaymath}
    \mathcal{L}_{\text{L2}} = \lambda \left \lVert \mathbf{\omega} \right \rVert_2 = \lambda \sum_{i=0}^{d}\omega_i^2 \text{\,  with \, } \lambda \in [0,+\infty \rparen.
\end{displaymath}
L1 regularization, on the other hand, promotes sparsity (i.e.\ encourages many weights to be exactly zero) by penalizing the L1 norm:
\begin{displaymath}
    \mathcal{L}_{\text{L1}} = \lambda \left \lVert \mathbf{\omega} \right \rVert_1 = \lambda \sum_{i=0}^{d}|\omega_i| \text{\,  with \, } \lambda \in [0,+\infty \rparen
\end{displaymath}

We now consider regularization methods that are not formulated as explicit penalty terms in the loss function.
\paragraph{Data Augmentation}
Data augmentation refers to a set of techniques that aim at increasing the diversity and amount of data available for the training process without collecting new data. It is based on the creation of modified samples of data, so that the model can improve robustness and expressivity, without entering the overfitting regime. Being particularly useful when the dataset dimension is small or collecting data is expensive, data augmentation is heavily dependent on the problem at hand. A typical example can be given by geometric transformation for images, such as rotation, flip, zoom, and color space transformation such as contrast enhancement, brightness scale adjustment, or noise injection. Another possible way is to generate new synthetic data using Generative AI models or non-parametric algorithms to increase the diversity and amount of training data. It may be stressed that data augmentation may not be considerate adequate for some possible machine learning applications.

\paragraph{Early stopping}
Early stopping is one of the most simple, yet effective regularization techniques; it does not require any additional computational cost compared to the previously discussed techniques. The key idea is to stop the training process when the model's performances on the validation set start to statistically deteriorate. This is done by monitoring the validation loss during training and stopping the process when it starts to increase, or it remains constant. It can be shown how early stopping can be effectively considered as a regularization technique by making explicit its connection with L2 regularization but is beyond the scope of this thesis. 

\paragraph{Dropout}
The last regularization method described is different from the others, since, during training, it temporarily modifies the structure (architecture) of the model by randomly deactivating some parts of it. Typical values of dropout rate (the percentage of the parameter to disconnect) range between $0.2$ and $0.5$ and, being it a hyperparameter, the actual best value is found empirically with model selection. Dropout prevents overfitting by avoiding certain``areas'' of the model to specialize excessively on certain features of the data, forcing the model to develop a more resilient representation of the acquired knowledge. 

Dropout can also be seen as a way of training an ensemble of models altogether, with every training step focusing on a different incarnation of the ensemble. 

It is important to note that Dropout is only implemented during the training process, during inference, the entire model is active, and the output must be scaled by the dropout rate to compensate for the larger effective model size.

\subsubsection*{Numerical optimization}
Numerical optimization is quintessential in machine learning, since all the training process is based on the optimization of the objective function. In the training process, the optimization consists in a recursive operation where, after each iteration, the parameters are adjusted to minimize the loss function. For this reason, optimization is at the heart of machine learning, and bridges the gap between theoretical models and practical, effective learning algorithms. 
A numerical approach is usually the only way to optimize complex loss functions; in fact, it allows finding the optimal parameters in high-dimensional spaces where analytical solutions are intractable. Take for example the GPT-4, with an estimated number of parameters that ranges between $1.7$ and $1.8$ \emph{trillion} parameters. An analytical solution to a minimization of a function of this many variables is practically impossible, hence the numerical optimization is the only way to proceed.
Optimization problems can be divided into two categories:
\subparagraph{Convex optimization problems} are those where the objective function forms a convex set, meaning that any segment between two points in the graph of the functions does not cut the graph. An important property of convex functions is that any local minimum is also a global minimum, simplifying the optimization process.

\subparagraph{Non-Convex optimization problems} on the other hand do not exhibit this property, leading to local minima or saddle points. This is the most frequent problem one has to address when training machine learning model, especially with neural networks that are highly non-convex. in this case, the choice of optimization algorithm becomes crucial as the problem shifts from finding the global minimum to finding a minimum which is "good enough" for expected performances. \\
\\
We now introduce the most widely used optimization algorithms, focusing on \emph{Adam} but with a brief overview on the intermediate algorithms such as momentum-based methods, stochastic or adaptive gradient techniques.

\subsubsubsection*{Adam optimizer} \label{sec:adam}
Building upon the principles of \emph{gradient descent} (GD), \emph{Adam} is one of the most widely used optimization algorithms in deep learning. To understand its role, it is useful to first introduce the basic concept of gradient descent.
GD is the prototype of a first-order algorithm (it only uses the first-order derivative at each step) and due to its simplicity and effectiveness, is particularly well suited for large-scale optimization tasks. Being able to navigate through convex and non-convex landscapes, GD iteratively updates the parameters by moving in the direction opposed to the steepest ascent without information on the curvature. The step size is controlled by a hyperparameter called \emph{Learning Rate} and the n+1-th step is given by:  
\begin{displaymath}
    \hat{\mathbf{\omega}}_{n+1} = \hat{\mathbf{\omega}}_{n} - \alpha \nabla f(\hat{\mathbf{\omega}}_{n})
\end{displaymath}
where $\alpha$, the learning rate, is crucial in determining the convergence and stability of the algorithm. A proper value of $\alpha$ is crucial: if it is too small, the algorithm takes too many iterations to converge, while if it is too large, the algorithm might overshoot the minimum, leading to divergence. The choice of $\alpha$ is a trade-off between training speed and stability, and it is often tuned using the validation data. Other hyperparameters are the initial guess $\hat{\mathbf{\omega}}_{0}$ and the total number of iterations.

A natural extension of gradient descent is \emph{Stochastic Gradient Descent (SGD)}, that modifies the way gradient is computed from the data and the way model parameters are updated. Instead of computing the gradient and updating the weights for all training samples as in standard GD, SGD computes the gradient for each training sample individually or a small subset of samples, called \emph{mini-batch}. The reason to introduce this generalization is twofold: first, it can significantly increase the computational efficiency, especially for large datasets, since the gradient is computed on a smaller subset of data. Second, it introduces a certain level of noise in the gradient, which can help the algorithm escape local minima and saddle points, potentially leading to better generalization. The full batch version of GD is often called \emph{Batch Gradient Descent}. 

From a practical point of view, the model weights are usually updated once per \emph{epoch}, where an epoch refers to a complete pass through the entire training dataset. So in the update the resulting gradient is averaged over all the single samples or over all the mini-batches. Notice that this generalization of the gradient descent algorithm (i.e., extending the update to a batch of samples) can be done also for the more advanced algorithms introduced later in the text, so usually they are used both.

The stochastic and mini-batch modification of GD have some cons, such as the fact that the convergence can be less stable due to the noise introduced and the fact that the choice of the mini-batch size and learning rate can significantly affect the performance of the algorithm.

From the GD update rule, we can interpret the second term as a velocity vector, in that case proportional to the gradient of the loss function. In the presence of high curvature or noisy gradients, this can lead to oscillation and slow convergence. \emph{Momentum-based methods} aim at addressing this issue by introducing a velocity term that accumulates the gradient over time, allowing the algorithm to build up speed in directions of consistent descent and dampen oscillations in directions of high curvature. The added term acts as a low-pass filter on the gradients, smoothing out rapid changes and allowing the algorithm to maintain a consistent direction of descent. The additional term is parametrized by a hyperparameter usually called $\beta$, where a higher value of $\beta$ gives more weight to past gradients, leading to smoother updates, while a lower value makes the algorithm more responsive to recent gradients.  

To increase responsiveness to the loss curvature has been introduced \emph{Nesterov Accelerated Gradient (NAG)}, that instead of calculating the gradient at the current position, computes it at a future position of the parameters as anticipated by the current momentum. In the presence of sharp bends in the loss landscape, this subtle shift allows for a better choice of the parameter update, leading to faster convergence and reduced oscillations.

The next step towards the Adam algorithm is the introduction of algorithms known as \emph{adaptive learning rate algorithms}  that instead of modifying the gradient function, adapts the learning rate. The first example of these algorithms is \emph{Adagrad}, that adapts the learning rate for each parameter individually by scaling it inversely proportional to the square root of the sum of all past gradients. Parameters that have been updated frequently receive smaller learning rates, while those that have been updated infrequently receive larger learning rates. This is particularly useful in the case of sparse dataset since it allows for an automatic feature scaling. It is also useful in the case of a large number of features, since the learning rate can be scaled according to the  varying importance of different features. 

To overcome the limitations of Adagrad, which is not discussed here as it lies beyond the scope of this thesis, \emph{RMSprop (Root Mean Square Propagation)} has been introduced. RMSprop modifies the accumulation mechanism by replacing it with a moving average.

Combined with mini-batch update, Adam is probably the most widely used optimizer algorithm in deep learning, since it merges the benefits of RMSprop with the ones of momentum-based GD. It accumulates both the first and the second order gradient moments to update the parameters using both the adaptive learning rate and a gradient function with momentum as in momentum-based GD. By using the component notation, the update rules for Adam are given by: 
\begin{equation}
    (\hat{\omega}_{i})_{n+1} = (\hat{\omega}_i)_{n} - \frac{\alpha}{\sqrt{(\tilde{v}_{ii})_n} + \epsilon}(\tilde{m}_i)_n \text{,} \quad (\tilde{m}_i)_n = \frac{(\tilde{m}_i)_n}{1- \beta_1^n} \text{,} \quad (\tilde{v}_{ii})_n = \frac{(\tilde{m}_{ii})_n}{1- \beta_2^n}
\end{equation}
with
\begin{equation}
    (m_i)_n = \beta_1 (m_i)_{n-1} + (1-\beta_1) \partial_i \mathcal{L}(\hat{\mathbf{\omega}}_{n}) \text{,} \quad (v_{ii})_n = \beta_2 (v_{ii})_{n-1} + (1-\beta_2) (\partial_i \mathcal{L}(\hat{\mathbf{\omega}}_{n}))^2
\end{equation}
and
\begin{equation}
    (m_i)_0 = 0 \text{,} \quad (v_{ii})_0 = 0
\end{equation}
where $(\tilde{m}_i)_n$ and $(\tilde{v}_{ii})_n$ are bias-corrected estimates of the first and second moments ($(m_i)_n$ and $(v_{ii})_n$) respectively, $\beta_1$ and $\beta_2$ are the exponential decay rate for these moment estimates which are usually set close to one, like 0.9 and 0.999 respectively and $\epsilon$ is the usual constant added for numerical stability, usually set around $10^{-8}$. The bias correction ensures accurate initial estimates when $(m_i)_n$ and $(v_{ii})_n$ might be biased towards zero, especially when $\beta_1$ and $\beta_2$ are set close to one.
The effectiveness of Adam comes from the fact that it is capable not only of adjusting the trajectory direction with the memory of past gradients, but also to adjust the step size according to the geometry of the data. 

An example of second order optimizer is reported in appendix. 

\section{Introduction to neural networks}
Neural Networks are a class of machine learning models originally inspired by how the biological systems process information. The first concept of neural network arose in mid-$\text{20}^{\text{th}}$ century but only in recent decades the field saw concrete advancements in performance and architectures. As will be further shown later in the text, neural networks are made up of interconnected nodes or neurons that, via the learning process, are capable of performing complex tasks.

Only in the last few years have we witnessed breakthrough in computer vision, natural language processing and speech recognition that have revolutionized the way we interact with technology and the integration into society continues, with the rise of ethical considerations. The scope of this section is to introduce the basic concepts of neural network before moving on to more advanced models. 
\subsection{Perceptron}
A Perceptron~\cite{rosenblatt1958perceptron} is the typical building block of a neural network (NN) architecture. Introduced in 1958 by F. Rosenblatt to model a human neuron, the perceptron is single artificial neuron, capable of manipulating multiple inputs (i.e. real numbers) to produce an output. The manipulation consists in a weighted sum of the inputs plus a bias term, where the weights are to be considered parameters, and the result is passed through an \emph{activation function} that produces the output. In the original formulation, the activation function is the Heaviside step function and, interestingly enough, the perceptron was originally intended to not be a program, but an actual, physical machine and was subsequently implemented in a custom-built hardware designed for image recognition, and known as the Mark I perceptron \cite{perc_as_physical_machine}. 

\begin{figure}[h]
  \centering
  \begin{minipage}[b]{0.47\textwidth}
    \vspace{2pt}
    \noindent\textbf{Forward pass}
    \begin{itemize}
      \item Take inputs $x_1,\dots,x_n$
      \item Compute $z=\sum_i \omega_i x_i + b$, where b is the bias term.
      \item Output $y=h(z)$
    \end{itemize}
  \end{minipage}\hfill
  \begin{minipage}[b]{0.47\textwidth}
    \centering
    \begin{tikzpicture}[node distance=1.8cm]
      \node (x1) at (-3,  1.0) {$x_1$};
      \node (x2) at (-3,  0.0) {$x_2$};
      \node at (-3,-0.35) {$\vdots$};
      \node (xn) at (-3, -1.0) {$x_n$};

      \node[nnnode] (sum) at (-0.8,0) {$\sum$};
      \node[nnnode] (act) at ( 0.6,0) {$h$};
      \node (y) at (2.0,0) {$y$};

      \draw[->] (x1) -- node[above,sloped,inner sep=2pt] {\scriptsize $\omega_1$} (sum);
      \draw[->] (x2) -- node[above,sloped,inner sep=2pt] {\scriptsize $\omega_2$} (sum);
      \draw[->] (xn) -- node[below,sloped,inner sep=2pt] {\scriptsize $\omega_n$} (sum);
      \draw[->] (sum) -- node[above] {\scriptsize $+\,b$} (act);
      \draw[->] (act) -- (y);
    \end{tikzpicture}
  \end{minipage}
  \caption{Description of the forward pass (left) and the perceptron schematic (right).}
 \label{fig:perceptron}
\end{figure}
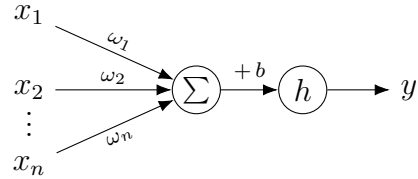

The perceptron can be formulated in mathematical form by writing:
\begin{displaymath}
    y = h\left(\sum_{i=1}^{n} \omega_i x_i + b\right)
\end{displaymath}
where $h(x)$ is the \emph{activation function} and $b$ the bias term.
A schematic, accompanied by the forward pass illustration is reported in Figure \ref{fig:perceptron}.

\subsubsection*{Activation functions}
Activation functions are responsible for one of the key strengths of Neural Networks: \emph{non-linearity}. In fact, without any activation function the output would be a linear combination of the input, limiting the expressiveness of the network.
Each activation function has its own use, and it should be chosen based on the problem at hand.
\begin{itemize}
    \item \textbf{Sigmoid or logistic activation function:} defined as
            \begin{displaymath}
                f(x) = \frac{1}{1+e^{-x}},
            \end{displaymath}
            by construction we have $f(x) \in \lparen 0,1 ) \text{ } \forall x$, which is important for tasks in which the output must be interpreted as a probability, such as classification tasks in which, usually, the output nodes represent the probability of the input to be in the associated class. 
    \item \textbf{Hyperbolic tangent activation function:} similar to the sigmoid activation function but with output in $\lparen -1,1 \rparen$. It has the advantage of mitigating the problem of vanishing gradients (i.e., the gradients used to update the parameters become exponentially small due to small derivatives of the activation functions multiplied many times for the weights update) and it is defined by:
    \begin{displaymath}
        f(x) = \frac{e^x - e^{-x}}{e^x + e^{-x}}
    \end{displaymath}
    \item \textbf{Rectified Linear Unit (ReLU) activation function}~\cite{relu}: it is defined by: 
    \begin{displaymath}
        f(x) = \text{max}(0,x)
    \end{displaymath}
    so it outputs the input if positive or zero otherwise. Given its simplicity and its computational efficiency it is a very popular choice. 
\end{itemize}

\subsubsection*{Multi-output perceptron}
\begin{figure}[h]
  \centering
  \begin{tikzpicture}[>=Latex, scale=0.95, transform shape] 
    \node (x1) at (-3,  1.2) {$x_1$};
    \node (x2) at (-3,  0.0) {$x_2$};
    \node (x3) at (-3, -1.2) {$x_3$};

    \node[nnnode] (s1) at (-0.9,  0.6) {$\sum$};
    \node[nnnode] (a1) at ( 0.9,  0.6) {$h$};
    \node          (y1) at ( 2.5,  0.6) {$y_1$};

    \node[nnnode] (s2) at (-0.9, -0.6) {$\sum$};
    \node[nnnode] (a2) at ( 0.9, -0.6) {$h$};
    \node          (y2) at ( 2.5, -0.6) {$y_2$};

    \draw[->] (x1) -- (s1);
    \draw[->] (x2) -- (s1);
    \draw[->] (x3) -- (s1);

    \draw[->] (x1) -- (s2);
    \draw[->] (x2) -- (s2);
    \draw[->] (x3) -- (s2);

    \draw[->] (s1) -- node[above] {\scriptsize $+\,b_1$} (a1);
    \draw[->] (s2) -- node[above] {\scriptsize $+\,b_2$} (a2);

    \draw[->] (a1) -- (y1);
    \draw[->] (a2) -- (y2);
  \end{tikzpicture}
  \caption{Multi-output perceptron: three inputs feed two output units; each output performs a weighted sum, adds a bias, then applies activation $h(\cdot)$.}
  \label{fig:perceptron-multiout}
\end{figure}
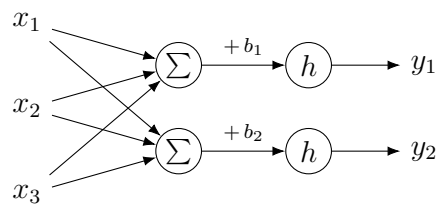

As the name suggests, it is a trivial generalization of the perceptron capable of generating multiple outputs, allowing to address problems with multiple target variables or output labels. This is also useful for tasks in which the output variables are correlated, as the perceptron can learn the relationship between them. A mathematical formulation of a multi-output perceptron can be written like this: 
\begin{displaymath}
    y_i = h(\sum_{j}\omega_{ij} X_j)
\end{displaymath}
where $i = 1, \dots, m$ with m the number of output nodes and $h(\cdot)$ is the activation function. Notice how in this formulation there is no explicit presence of the bias term. That is because the bias term is incorporated in the vector $\mathbf{X}$ and will be multiplied by a weight (non-trainable) fixed to one: $\omega_b = 1$. To be more explicit the input vector will be represented as $ \mathbf{X} = \lparen x_1, \dots, x_n , b_1, \dots, b_m \rparen$ and the weight matrix will have $\omega_{i,n+1} = 1 \text{ fixed } \forall i$ such that in the product the bias term will always be multiplied by one. 
A schematic representation is shown in Figure \ref{fig:perceptron-multiout} where we made explicit the bias summation.

\subsection{Multi-layer perceptron}
The natural extension of a multi-output perceptron is to add multiple layers of nodes, or neurons in between the input and the output. A collection of nodes operating at the same depth is called a \emph{layer} and the layers in between the input and the output are called \emph{hidden layers}. The hidden layers are meant to extract features from the \emph{input layer} and send them to the \emph{output layer}. So naturally increasing the number of hidden layers or the number of nodes in each of them will increase the complexity. This architecture is called MLP (Multi-Layer Perceptron) but different names could be found in literature, such as DNN (Deep Neural Network) or feed-forward neural networks. One can then experiment with different connection patterns between the nodes of consecutive layers; however, the simplest and most common configuration is the \emph{fully connected} neural network, in which every node in a layer is connected to every node in the next layer. As we will see in the section dedicated to the \emph{normalizing Flows}, this is not the only choice for an MLP architecture. 

\begin{figure}[h]
  \centering
  \begin{tikzpicture}[
    scale=0.92, transform shape,
    nnnode/.append style={minimum size=6.5pt, inner sep=0pt}
  ]
    \def\vstepIn{0.34}   
    \def\vstepH{0.36}    
    \def\vstepOut{0.36}  

    \def\xIn{0}
    \def\xHone{2.6}
    \def\xHtwo{5.2}
    \def\xOut{7.8}

    \foreach \i [evaluate=\i as \y using {(\i - (20+1)/2)*\vstepIn}] in {1,...,20}{
      \node[nnnode] (I-\i) at (\xIn,\y) {};
    }
    \foreach \j [evaluate=\j as \y using {(\j - (15+1)/2)*\vstepH}] in {1,...,15}{
      \node[nnnode] (HOne-\j) at (\xHone,\y) {};
    }
    \foreach \k [evaluate=\k as \y using {(\k - (15+1)/2)*\vstepH}] in {1,...,15}{
      \node[nnnode] (HTwo-\k) at (\xHtwo,\y) {};
    }
    \foreach \m [evaluate=\m as \y using {(\m - (10+1)/2)*\vstepOut}] in {1,...,10}{
      \node[nnnode] (O-\m) at (\xOut,\y) {};
    }

    \foreach \i in {1,...,20}{
      \foreach \j in {1,...,15}{
        \edgeWithRandomWeight{I-\i}{HOne-\j}
      }
    }
    \foreach \j in {1,...,15}{
      \foreach \k in {1,...,15}{
        \edgeWithRandomWeight{HOne-\j}{HTwo-\k}
      }
    }
    \foreach \k in {1,...,15}{
      \foreach \m in {1,...,10}{
        \edgeWithRandomWeight{HTwo-\k}{O-\m}
      }
    }
  \end{tikzpicture}
  \caption{The scheme of an MLP, the input layer has 20 nodes, 2 hidden layers of 15 nodes each and 10 nodes in the output layer. Edge color shows sign (blue = positive, orange = negative) and opacity scales with the weight magnitude. Image inspired by Ref.~\cite{3b1b_neuralnetworks}}
  \label{fig:mlp}
\end{figure}
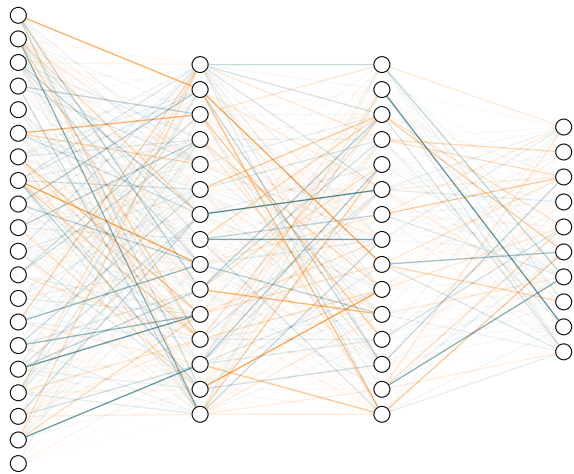

The hidden layers transform the multi-output perceptron into a universal function approximator, since it approximate any function $f$ takes an input variable $\mathbf{x}$ into the output variable $\mathbf{y}$. The MLP approximates the function $f$ by defining a mapping $\mathbf{y} = g(\mathbf{x}; \mathbf{\theta})$ and finding the optimal parameters $\mathbf{\theta}$ that result in the best approximation of the function $f$ by $g$. 
A schematic representation of a fully connected MLP is shown in Figure~\ref{fig:mlp}, where the opacity of the connections (sometimes called \emph{edges}) represents their magnitude, the colors represent the sign of the weight associated with the edge.

In the next chapter, more advanced architectures will be introduced, many of which are based on the foundations of the MLP.

\chapter{From Generation to Validation: Principles and Evaluation Metrics} \label{generative_models}

\section{Generative models in physics} 

Generative models have become an important tool in many areas of physics because they can learn complex, high-dimensional probability distributions directly from data. In experimental and theoretical physics, many problems involve sampling from or approximating such distributions, which are often too expensive to compute with traditional methods. For example, Monte Carlo simulations are widely used to generate events, propagate particles through detectors, or simulate radiation showers. While these simulations are accurate, they are extremely time-consuming and computationally expensive. Generative models can act as \emph{fast simulators}, reproducing realistic samples with a fraction of the computational cost \cite{ML_at_lhc}.

Beyond fast simulation, generative models can also be used for a variety of other physics tasks. In data analysis, they can help to perform \emph{likelihood-free inference} by learning the mapping between theory parameters and observable data, allowing one to estimate or constrain physical parameters even when the exact likelihood function is not available. In anomaly detection, they can identify unusual or rare events that deviate from the learned data distribution, potentially pointing to new physics signals that differ from the Standard Model. In theoretical modeling, they can learn complicated probability densities that describe, for example, parton distribution functions or energy flow in jets. 

In detector physics, and especially in calorimetry, the use of generative models is motivated by the large amount of data and the fine spatial resolution of modern detectors. Accurate simulation of electromagnetic or hadronic showers requires modeling complex correlations between thousands of detector cells. Traditional simulation tools such as \textsc{Geant4} provide high accuracy but are computationally heavy. Generative models, in contrast, can reproduce similar distributions much faster once trained, enabling large-scale simulation and fast event generation for studies at the High-Luminosity LHC and future experiments.

For all these reasons, the development of reliable and interpretable generative models has become a growing area of research in high-energy physics. They provide an opportunity to reduce simulation costs, accelerate data analysis, and improve the understanding of complex systems by learning directly from data, while maintaining the physical consistency required in scientific applications \cite{ML_at_lhc}.

The classical foundations are \emph{generative adversarial networks} (GANs) \cite{GAN}, \emph{variational autoencoders} (VAEs) \cite{VAE, VAE_image}, and \emph{normalizing flows} (NFs) \cite{NF}. In recent years, \emph{diffusion and score-based models} \cite{diffusion1,diffusion2} together with \emph{conditional flow-matching} models (CFMs), based on neural ODE dynamics \cite{CFM1,CFM2}, have become leading approaches for high-fidelity and fast calorimeter shower generation. This trend is documented by the CaloChallenge review \cite{calochallenge}, in which the best performances were obtained with continuous flow matching and diffusion based models.

\subsection*{An overview of the main architectures}

In this subsection, we give a short introduction to the most common generative model architectures used in modern machine learning. The goal is to present their main ideas and training principles, without going into full technical detail. Less emphasis will be placed on \emph{normalizing flows}, as a complete theoretical and practical discussion is provided in the next dedicated section, due to their central role in this thesis.

\subsubsection*{Variational Autoencoders (VAEs).}

Variational Autoencoders~\cite{VAE,VAE_image} are probabilistic models that describe data generation as a two-step process: first, a latent variable \( z \) is sampled from a simple prior distribution \( p(z) \), usually a standard normal; second, the observed data \( x \) is generated from this latent variable through a decoder distribution \( p_\theta(x|z) \), where \(\theta\) are the model parameters. The challenge is that the true posterior distribution \( p_\theta(z|x) \) is generally intractable. VAEs introduce an encoder \( q_\phi(z|x) \), parametrized by \(\phi\), that approximates the posterior. The model is trained by maximizing the so-called \emph{Evidence Lower Bound} (ELBO):
\[
\mathcal{L}_{\text{VAE}} = 
\mathbb{E}_{q_\phi(z|x)} [\log p_\theta(x|z)] 
- 
\mathrm{KL}\!\left[q_\phi(z|x) \,\|\, p(z)\right].
\]
The first term encourages the decoder to reconstruct the input data correctly from the latent representation, while the second term (Kullback-Leibler divergence~\cite{KL_divergence}) regularizes the latent space, pushing the approximate posterior \( q_\phi(z|x) \) close to the prior \( p(z) \). This prevents the encoder from overfitting and ensures that meaningful samples can be generated by drawing \( z \) directly from \( p(z) \). In practice, the model is trained end-to-end by using the \emph{reparameterization trick}~\cite{VAE}, which allows gradients to pass through the stochastic latent variable during optimization.

\begin{figure}[h]
    \centering
    \includegraphics[width=0.55\linewidth]{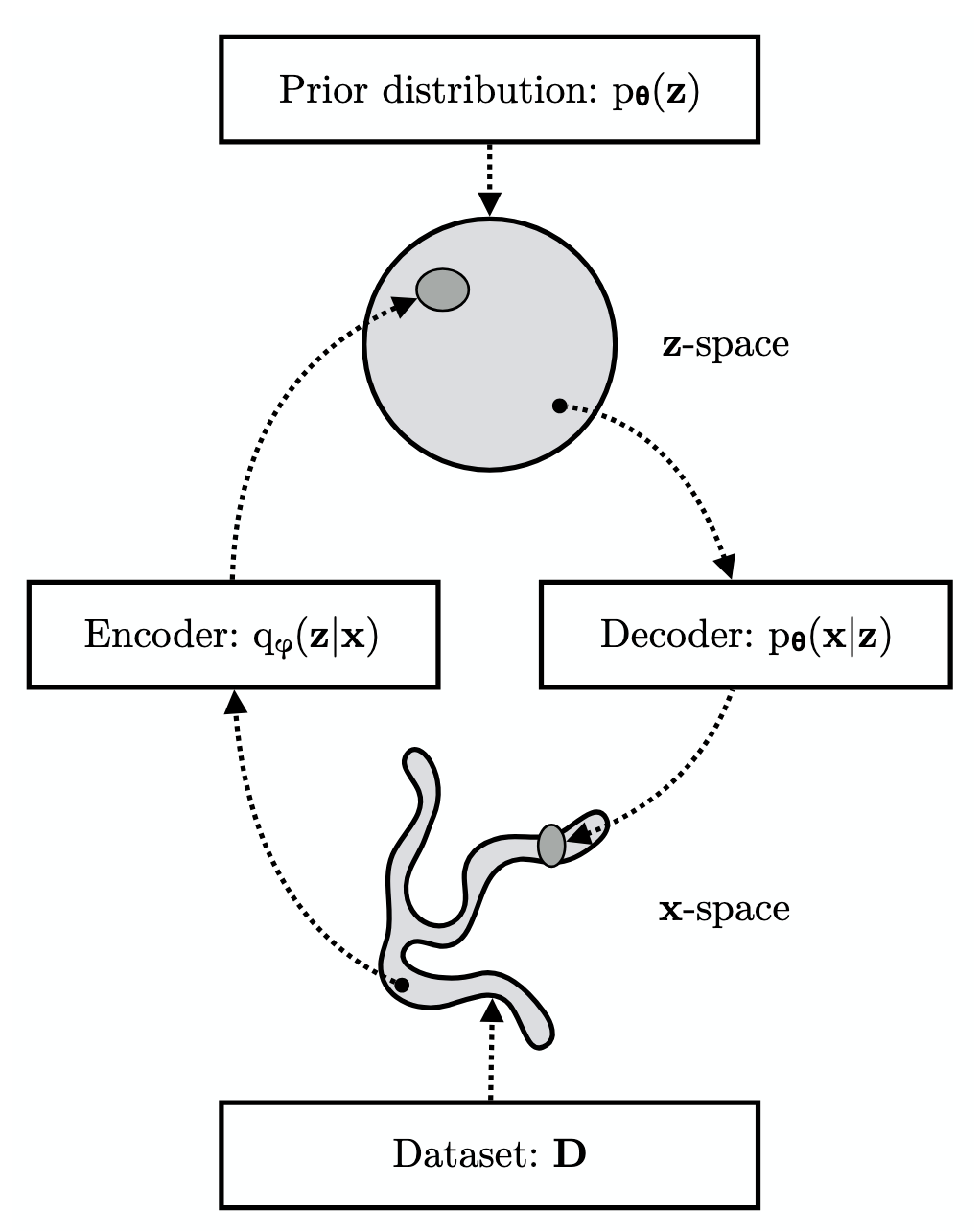}
    \caption{Overview of a variational autoencoder (VAE)~\cite{VAE_image}. The encoder \( q_{\phi}(\mathbf{z}|\mathbf{x}) \) maps data \(\mathbf{x}\) to latent variables \(\mathbf{z}\), while the decoder \( p_{\theta}(\mathbf{x}|\mathbf{z}) \) reconstructs the data from samples drawn from the latent space.}
    \label{fig:vae}
\end{figure}

VAEs are widely used when both data generation and uncertainty quantification are needed. However, the Gaussian assumptions and the variational approximation may lead to blurry samples or oversimplified distributions. Despite these limitations, they remain a cornerstone in generative modeling due to their stability and probabilistic formulation.

\subsubsection*{Generative Adversarial Networks (GANs).}
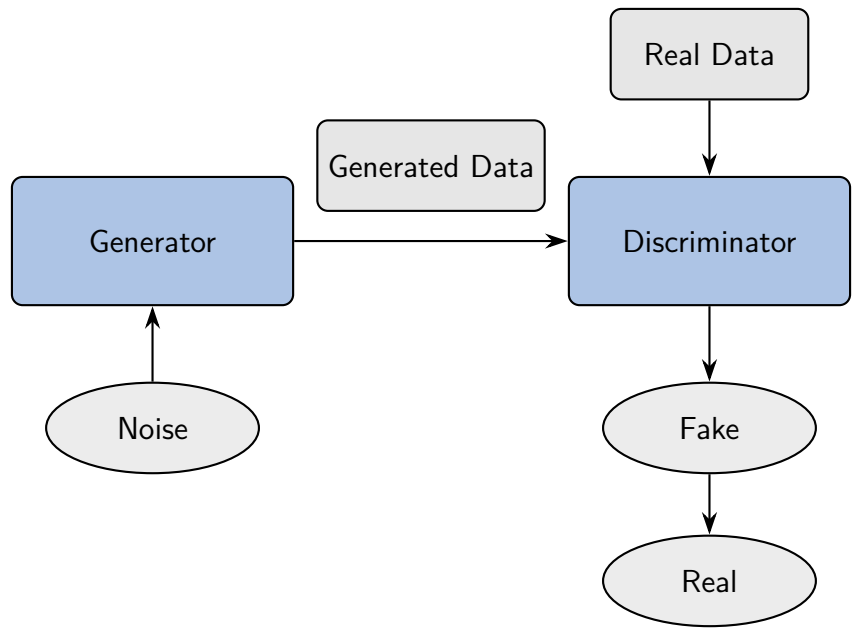
\begin{figure}[h]
\centering
\begin{tikzpicture}[
  font=\sffamily,
  node distance=2.2cm and 3.6cm,              
  >={Stealth[length=3mm,width=2mm]},          
  block/.style={rounded corners, draw=black, line width=0.8pt, fill=RoyalBlue!30,
                minimum width=3.7cm, minimum height=1.7cm, align=center},                  
  small/.style={rounded corners, draw=black, line width=0.8pt, fill=gray!20,
                minimum width=2.6cm, minimum height=1.2cm, align=center},
  oval/.style={ellipse, draw=black, line width=0.8pt, fill=gray!15,
               minimum width=2.8cm, minimum height=1.2cm, align=center}
]

\node[block] (G) {Generator};
\node[block, right=of G] (D) {Discriminator};

\node[small, above=1cm of D] (RealData) {Real Data};
\node[small] at ($ (G)!0.5!(D) + (0,1cm) $) (GenData) {Generated Data};

\node[oval, below=1cm of G] (Noise) {Noise};
\node[oval, below=1cm of D] (Fake) {Fake};
\node[oval, below=0.8cm of Fake] (Real) {Real};

\draw[->, thick] (Noise.north) -- (G.south);
\draw[->, thick] (G.east) -- (D.west);
\draw[->, thick] (RealData.south) -- (D.north);
\draw[->, thick] (D.south) -- (Fake.north);
\draw[->, thick] (Fake.south) -- (Real.north);

\end{tikzpicture}
\caption{Schematic of a Generative Adversarial Network (GAN): the generator maps noise to data samples, which are evaluated by the discriminator alongside real data to predict \emph{fake} or \emph{real}. The diagram is strongly inspired by Ref.~\cite{gan_diagram}.}
\label{fig:gan-architecture}
\end{figure}

Generative Adversarial Networks~\cite{GAN} represent a different philosophy. Instead of explicitly modeling a probability distribution, they define an \emph{implicit} generative process through a neural network \( G_\theta(z) \), which maps latent variables \( z \sim p(z) \) to synthetic samples \( x' = G_\theta(z) \). The quality of the generated samples is judged by a second network, called the discriminator \( D_\psi(x) \), which tries to distinguish between real data and generated samples. The two networks are trained in an adversarial game with the objective:
\[
\min_G \max_D 
\mathcal{L}_{\text{GAN}} =
\mathbb{E}_{x \sim p_{\text{data}}(x)}[\log D_\psi(x)] +
\mathbb{E}_{z \sim p(z)}[\log (1 - D_\psi(G_\theta(z)))].
\]
The discriminator learns to assign high scores to real samples and low scores to fake ones, while the generator learns to produce samples that the discriminator cannot distinguish from real data. At equilibrium, the generator reproduces the data distribution \( p_{\text{data}}(x) \) as closely as possible. A diagram explaining the working principles of GANs is reported in Figure \ref{fig:gan-architecture}.

GANs are powerful because they can produce very sharp and realistic samples, but their training can be unstable. The minimax optimization often suffers from non-convergence and mode collapse, where the generator only reproduces a subset of the data. Many extensions have been proposed, such as the Wasserstein GAN (WGAN), which replaces the standard cross-entropy objective with the Wasserstein distance between real and generated distributions to stabilize training.

In physics applications, GANs have been used for fast detector simulation, jet generation, and calorimeter shower modeling. Their ability to learn complex, high-dimensional correlations makes them suitable for these tasks, but the lack of a tractable likelihood and the sensitivity to hyperparameters make quantitative validation more challenging compared to likelihood-based models such as VAEs and normalizing flows.

\subsubsection*{Normalizing Flows (NFs).}

Normalizing flows learn \emph{invertible transformations} that map a simple base distribution to the data distribution. Because the mappings are bijective, they admit an explicit probability density via the change-of-variables formula, enabling \emph{exact} log-likelihoods and exact sampling. These properties make flows attractive for physics, where tractable densities aid statistical validation and anomaly detection, and fast sampling accelerates simulation. A detailed treatment of architectures and training is provided in Section~\ref{sec:NF}.

\subsubsection*{Diffusion Models}

\begin{figure}[h]
    \centering
    \includegraphics[width=0.7\linewidth]{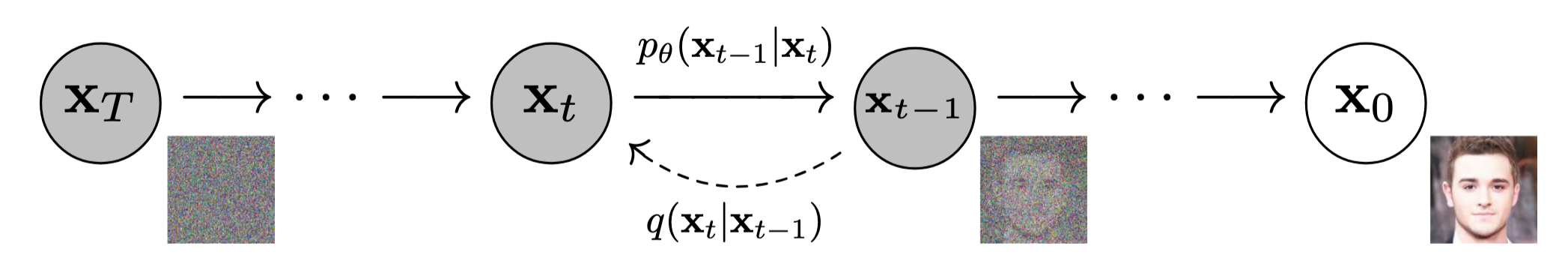}
    \caption{Diagram of the \emph{diffusion} process. Image from Ref.~\cite{diffusion2}}
    \label{fig:diffusion_model}
\end{figure}

Diffusion models represent a more recent and powerful approach to generative modeling. Their main idea is to model the data distribution as the result of a gradual denoising process. Training is based on learning how to reverse a \emph{diffusion} that progressively adds Gaussian noise to the data.  
Let \( x_0 \) denote a data sample and \( x_t \) the same sample after \( t \) diffusion steps. The forward process adds noise according to
\[
q(x_t|x_{t-1}) = 
\mathcal{N}\!\left(x_t; \sqrt{1 - \beta_t}\,x_{t-1}, \beta_t I\right),
\]
where \(\beta_t\) controls the noise level. After many steps, the data become nearly Gaussian. The model then learns the reverse process \( p_\theta(x_{t-1}|x_t) \), which gradually removes noise to reconstruct a clean sample. In the continuous-time limit, this process can be described by a stochastic differential equation (SDE) whose drift is parametrized by a neural network trained to predict the added noise \cite{diffusion1,diffusion2,diffusion3}. In Figure \ref{fig:diffusion_model} a diagram illustrates the diffusion process. 

Diffusion models have recently shown outstanding performance in generating realistic and diverse samples across many domains, including high-energy physics. Their training is stable, they provide good mode coverage, and they can capture highly non-linear correlations in calorimeter showers. However, the sampling process is relatively slow because it requires solving many denoising steps. This limitation has motivated the development of faster alternatives such as \emph{Conditional flow-matching models}, which combine the stability of diffusion training with the efficiency of deterministic flows.

\subsubsection*{Conditional Flow Matching (CFM) Models}

\begin{figure}[h]
    \centering
    \includegraphics[width=0.5\linewidth]{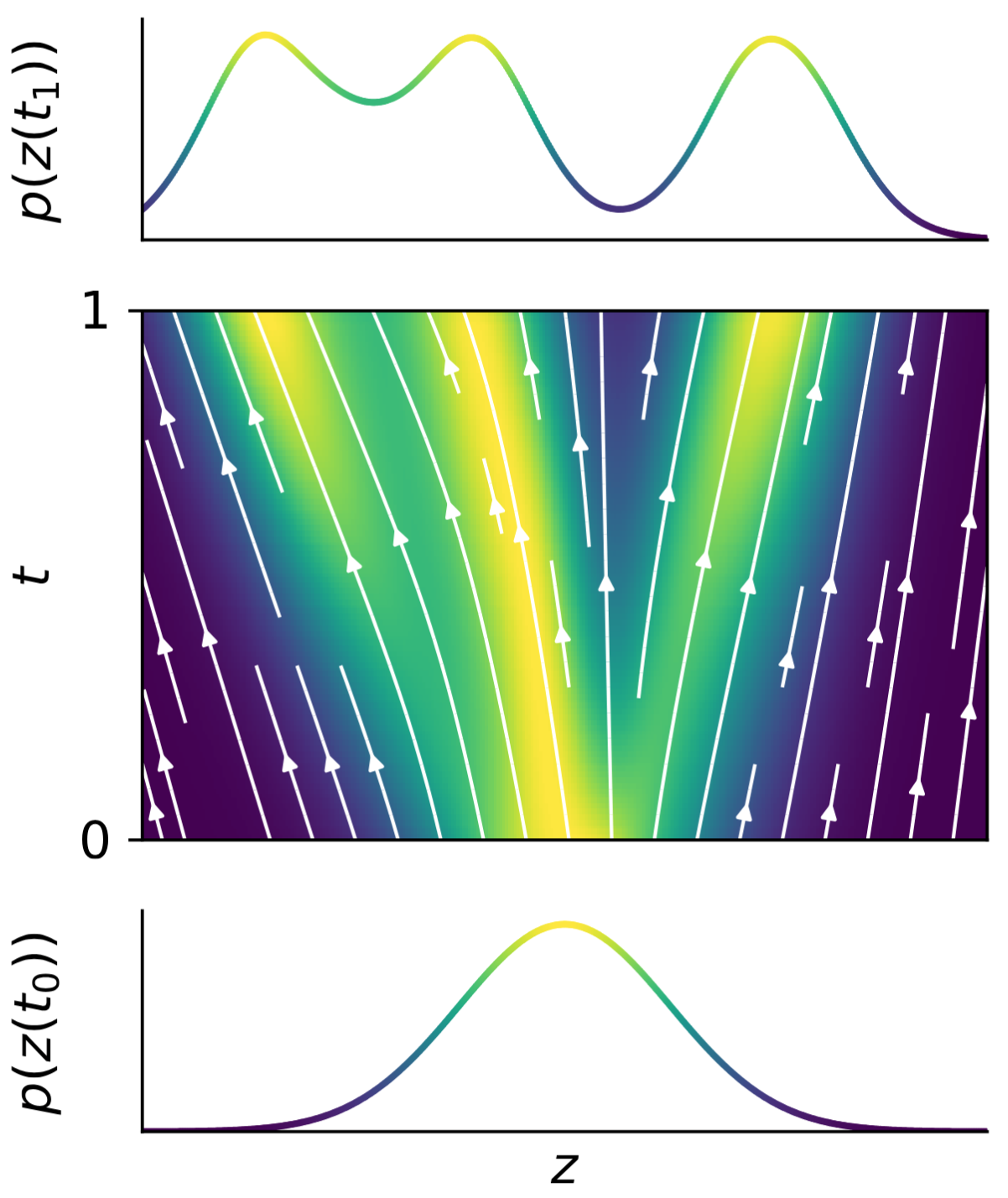}
    \caption{
        Illustration of density flow in a conditional flow-matching framework, adapted from~\cite{CFM2}. The figure shows the continuous evolution of the probability density $p(z(t))$ governed by an ODE solver performing optimal transport between a simple Gaussian base distribution $p(z(t_0))$ and the complex target distribution $p(z(t_1))$. The central panel depicts the vector field that drives the transformation, while the top and bottom panels show the marginal densities at the start and end of the flow.
    }
    \label{fig:cmf}
\end{figure}

Conditional Flow Matching (CFM) models are a recent class of generative models that unify ideas from normalizing flows and diffusion models. Instead of learning a sequence of discrete transformations (as in standard NFs) or a stochastic denoising process (as in diffusion models), CFMs learn a continuous-time deterministic transformation that transports samples from a simple base distribution to the data distribution. This transformation is defined by an ordinary differential equation (ODE) in time:
\[
\frac{dx_t}{dt} = v_\theta(x_t, t),
\]
where \(v_\theta(x_t, t)\) is a neural network that predicts the instantaneous velocity of each point along the flow. The model is trained so that this velocity field correctly transforms the base distribution into the data distribution. A simple example is shown in Figure \ref{fig:cmf}. 

The key idea of \emph{flow matching} \cite{Flow_matching} is to train the network to match the true optimal transport field between two distributions, avoiding the need to estimate log-determinants or to solve a stochastic process during training. The \emph{conditional} version (CFM) extends this framework by conditioning the flow on auxiliary information, such as the particle type or incident energy in calorimeter simulations \cite{conditional_FM}. This conditioning allows the model to generate showers consistent with specific physical parameters, which is crucial for detector modeling.

CFM models combine the main advantages of diffusion models (stable training and good mode coverage) with those of normalizing flows (fast sampling and deterministic inference). For this reason, they currently represent one of the most promising approaches for high-fidelity and efficient generation in HEP, as shown by the latest CaloChallenge results \cite{calochallenge}.

\section{Normalizing Flows: Formalism and Overview} \label{sec:NF} 
Normalizing Flows are a class of neural density estimators. They emerged as a powerful branch of generative models as they can approximate complex distributions from which to sample. They also provide, by construction, density estimation. 
\subsection{The core idea}
As introduced in the previous section, the basic principle is to learn a \emph{target} 
distribution by applying a chain of invertible transformations to a (known) \emph{base} distribution. The purpose
of an NF is to estimate the unknown underlying distribution of some data. 
Since the parameters of both the base
distribution and the transformation are fully known, one can sample from the target distribution by generating 
some samples from the base distribution and then applying the transformation. This is known as the \emph{generative direction} of the flow. Furthermore, since the transformations are invertible, one can obtain the probability of a true sample by inverting the transformations. This is called the \emph{normalizing direction}. 
\subsection{The formalism of normalizing Flows}
To better understand the formalism behind Normalizing Flows, we can define a normalizing flow as a parametric \emph{diffeomorphism} $f_{\theta}$ (also called a \emph{bijector}) between a latent space with known distribution $\pi_{\phi}(z)$ and a data space of interest with unknown distribution $p(x)$. The foundation of a NF is the change-of-variables formula for a PDF: let us define $Z,X \in \mathbb{R}^D$ and $\pi_{\phi}, p : \mathbb{R}^D \to \mathbb{R}$ such that $Z \sim \pi_{\phi}(z)$ and $X \sim p(x)$. We assume the distribution $\pi$ to be characterized by some parameters $\phi$ (typically $\pi$ is chosen to be a multivariate Gaussian, so $\phi$ typically contains the means and the correlation matrix). Let $f_{\theta}$ be the parametric diffeomorphism (bijective map) such that $f_{\theta} : Z \to X$ with inverse $g_{\theta}$ and $\theta = \{\theta_i\}$ with $i=0, \dots, N$, where $N$ is the number of parameters. Then the two PDFs are related by: 

\begin{equation} \label{eq:change_of_var}
    p(x) = \pi_{\phi}(f_{\theta}^{-1}(x))|\det \mathbf{J}_{f_{\theta}}(x)|^{-1} = \pi_{\phi}(g_{\theta}(x))|\det \mathbf{J}_{g_{\theta}}(x)| 
\end{equation}
where $\mathbf{J}_{f_{\theta}}(z) = \frac{\partial \mathbf{f}_{\theta}}{\partial z}$ and 
$\mathbf{J}_{g_{\theta}}(x) = \frac{\partial \mathbf{g}_{\theta}}{\partial x}$ are the Jacobians of $f_{\theta}(z)$ and 
$g_{\theta}(x)$ respectively. 
To keep the flow computationally efficient, the determinant of the Jacobian must be simple and easy to compute. Therefore, transformations with triangular Jacobian matrices are preferable so that the determinant can be written as the product of the elements on the main diagonal. This keeps the computation of the Jacobian determinant efficient. We can leverage the relation in Eq.~\ref{eq:change_of_var} to extract samples from the unknown, complex distribution $p$ by drawing samples from the simple distribution $\pi_{\phi}$ and applying the function $g_{\theta}$, provided that the function $f_{\theta}$ is expressive enough. 
Constructing arbitrarily complicated non-linear invertible bijectors can be difficult, but one approach is to note that the composition of invertible functions is itself invertible, and the determinant of the Jacobian of the composition is the product of the determinants of the Jacobians of the individual functions. Then, for the generative direction, we can choose 
\begin{displaymath}
    \mathbf{f} = \mathbf{f}_1 \circ \dots \circ \mathbf{f}_N
\end{displaymath}
and the determinant of the Jacobian matrix: 
\begin{displaymath}
    \det \mathbf{J_f} = \displaystyle\prod_{i} \det \mathbf{J_{f_i}}(x).
\end{displaymath}
Also note that the inverse function can be easily written as
\begin{displaymath}
    \mathbf{g} = \mathbf{g}_N \circ \dots \circ \mathbf{g}_1.
\end{displaymath}

One can then perform a maximum likelihood estimation of the parameters $\Phi = \{\phi, \theta\}$:
the log-likelihood of the observed data $D = \{x^I\}_{I=1}^N$ is given by
\begin{equation}
    \log p(D|\Phi) = \sum_{I = 1}^{N} \log p(x^I|\Phi) = \sum_{I = 1}^{N} \left( \log \pi_{\phi} (g_{\theta}(x^I)) + \log |\det \mathbf{J}_{g_{\theta}}(x^I)| \right),
\end{equation}
and the best estimate is given by:
\begin{equation}
    \hat{\Phi} = \argmax_{\Phi} \log p(D|\Phi)
\end{equation}

The diffeomorphism $f_{\theta}$ should also satisfy some other properties:
\begin{itemize}
    \item It should be computationally efficient, both in the normalizing direction and in the generative one.
    \item The Jacobians should be easy to compute. 
    \item It should be sufficiently expressive to model the target distribution.
\end{itemize}
Typically, a NF is implemented using NNs to determine the parameters of the bijectors. 

\begin{figure}[h]
    \centering
    \includegraphics[width=0.8\linewidth]{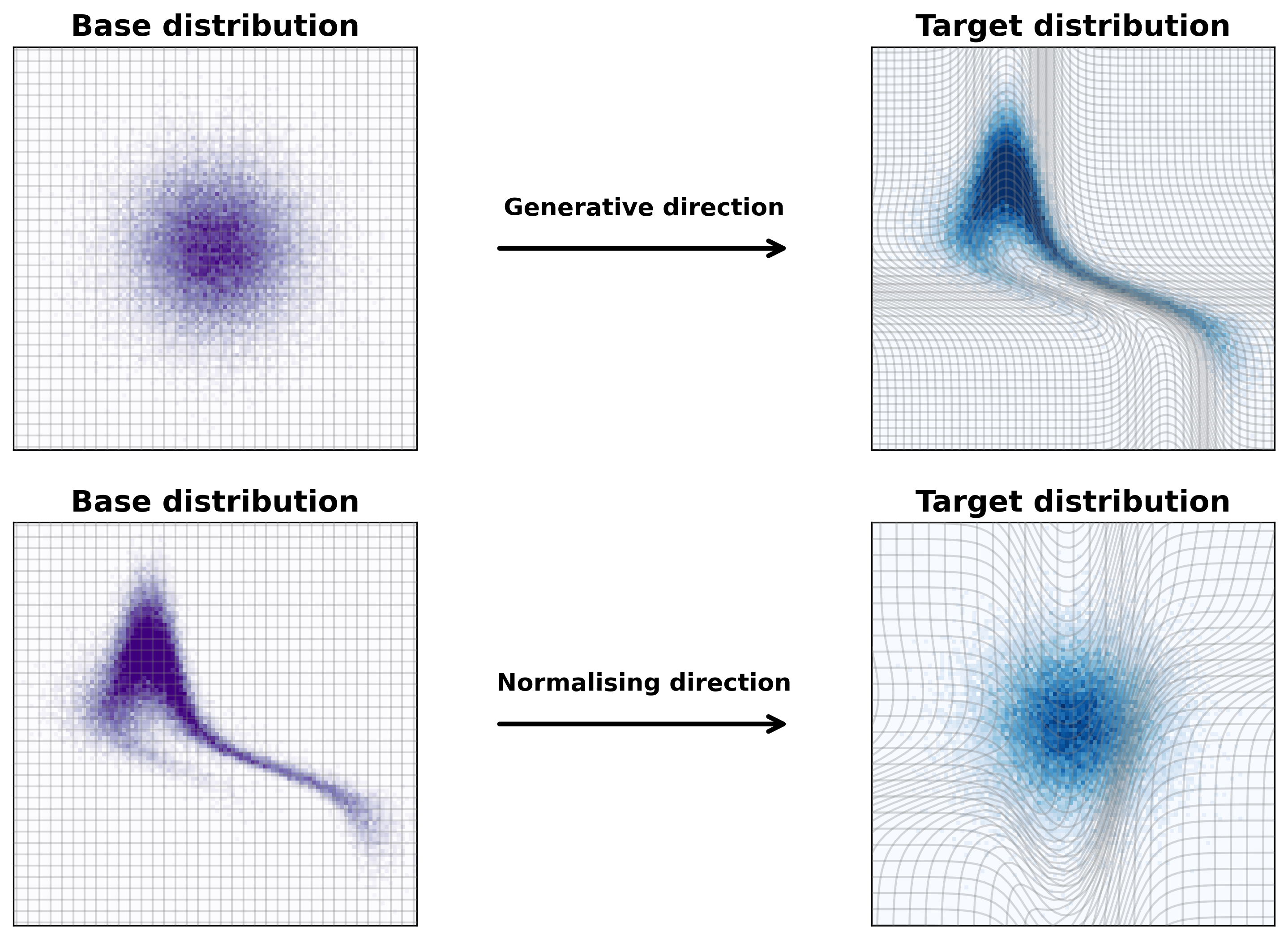}
    \caption{Illustration of the bidirectional mapping in Normalizing Flows
In the \emph{generative direction} (top), a simple latent variable sampled from a base distribution (typically a standard Gaussian) is transformed through a sequence of invertible mappings into a complex \emph{target distribution} in data space. Conversely, in the \emph{normalizing direction} (bottom), data samples are mapped back to the latent space, allowing for exact likelihood evaluation via the change-of-variables formula. The deformation of the background grid visually represents the smooth and invertible transformations that characterize flow-based models.}

    \label{fig:nf}
\end{figure}

An illustrative example of the bidirectional mapping performed by Normalizing Flows is shown in Figure \ref{fig:nf}. In the \emph{generative direction}, a simple latent variable drawn from a base distribution, a standard Gaussian in the specific example, is transformed through a sequence of invertible mappings into a complex target distribution representing the data space. Conversely, the \emph{normalizing direction} corresponds to the inverse transformation, where observed data are mapped back to the latent space, enabling exact likelihood evaluation via the change-of-variables formula. The deformation of the background grid highlights how these transformations smoothly warp the space while preserving invertibility, providing an intuitive geometric interpretation of the flow mechanism.

\subsection{Coupling and Autoregressive flows}
normalizing flows can be divided into two main architectural structures: \textit{Coupling-layer flows} and \textit{Autoregressive flows}. In the former, the input vector is separated into two or more pieces and then transforms some of them with a function of the others; in the latter, the input dimensions are ordered and each of them is transformed according to the previous ones. This distinction will become more clear after the discussion of different examples. It is important to note that, in normalizing flows, the parameters of the transformation are typically determined by neural networks, which are generally not invertible. In the two distinct structures this problem is addressed in different ways and will be thoroughly discussed below. 
Although coupling and autoregressive flows may appear different in structure, they are closely related. In fact, autoregressive flows can be seen as a limiting case of coupling flows in which the partition of the input is performed at every single dimension. In coupling layers, a subset of variables remains fixed while the other subset is transformed conditionally. In autoregressive flows, this conditioning is extended to all previous variables, providing maximal flexibility at the cost of slower computation. Conversely, coupling flows trade a small loss in expressiveness for significantly faster parallel computation. This connection was first discussed in~\cite{maf,autoregr_and_NF_kingma_2016}.

\subsubsection*{Coupling-layer examples}

\paragraph{RealNVP} 
The name comes from the fact that it uses Real-valued Non-Volume Preserving transformations~\cite{RealNVP}. 
A general principle that will be thoroughly discussed in the more technical chapters is that the determinant of a triangular matrix is given by the product of the elements on its main diagonal. This will be very important from a numerical point of view since we will have to calculate the determinant of the Jacobian matrix of the transformations. 

The RealNVP implements an invertible transformation (chosen in the original paper to be an affine transformation) based on a simple but powerful idea. The input vector is split into two parts:
\[
x = (x_{1},\dots,x_{d},\,x_{d+1},\dots,x_{D})
\equiv (x_{A},\,x_{B}),
\]
with
\[
A = \{1,\dots,d\},\; B = \{d+1,\dots,D\}.
\]

The first split is used to compute the transformation parameters, while the second split is transformed according to these parameters. The forward (generative) transformation can be written as:
\begin{align*}
y_{A} &= x_{A}, \\[4pt]
y_{B} &= x_{B} \,\odot\, \exp\bigl(s(x_{A})\bigr) + t(x_{A}),
\end{align*}
where
\[
s:\mathbb{R}^{d}\to\mathbb{R}^{D-d}, 
\quad
t:\mathbb{R}^{d}\to\mathbb{R}^{D-d}.
\]
\noindent
In components, for each \(i \in B\):
\[
y_{i} = x_{i}\,e^{s_{i-d}(x_{1:d})} + t_{\,i-d}(x_{1:d}).
\]
\noindent
The inverse transformation is equally simple and can be written as: 
\begin{align*}
x_{A} &= y_{A}, \\[4pt]
x_{B} &= \bigl(y_{B} - t(y_{A})\bigr)\,\odot\,\exp\bigl(-\,s(y_{A})\bigr).
\end{align*}

Even though the functions \(s\) and \(t\) are implemented by neural networks that are not themselves invertible, the overall transformation remains invertible. This is guaranteed because the parameters of the transformation depend only on the untransformed subset \(x_A\). The Jacobian of the transformation is triangular, and its log-determinant can be computed efficiently as:
\[
\log |\det \textbf{J}| = \sum_i s_i(x_A).
\]
This property makes RealNVP numerically stable and computationally efficient, forming the foundation for many later flow-based models such as GLOW, briefly discussed below, and NICE~\cite{nice}. 

\paragraph{GLOW} 
The GLOW architecture~\cite{kingma2018glowgenerativeflowinvertible} builds upon the RealNVP model and extends it with improved expressiveness and training stability. It is composed of a sequence of flow steps, each consisting of three transformations applied in order: an \emph{activation normalization} (actnorm), an \emph{invertible $1{\times}1$ convolution}, and an \emph{affine coupling layer}. The overall transformation remains invertible, and the log-determinant of the Jacobian can be computed efficiently, allowing exact likelihood estimation.

\subparagraph{Actnorm.} 
In place of traditional batch normalization, GLOW introduces an \emph{actnorm} layer that performs a channel-wise affine transformation of the activations:
\[
\mathbf{y} = \mathbf{s} \odot \mathbf{x} + \mathbf{b},
\]
where $\mathbf{s}$ and $\mathbf{b}$ are learnable scale and bias parameters. They are initialized using a single minibatch so that each output channel has zero mean and unit variance, ensuring numerically stable initialization. Afterward, these parameters become trainable and data-independent. Because the transformation is affine, the inverse and the log-determinant of the Jacobian are easy to compute:
\[
\log|\det \textbf{J}| = \sum_i \log|s_i|.
\]

\subparagraph{Invertible $1{\times}1$ convolution.} 
The invertible $1{\times}1$ convolution replaces the fixed channel permutations used in RealNVP, allowing the model to learn more flexible dependencies across channels. It can be seen as a learnable generalization of a permutation, where the weight matrix $\mathbf{W} \in \mathbb{R}^{c \times c}$ is initialized as a random rotation matrix to ensure invertibility. The log-determinant of this transformation for a tensor of shape $(h, w, c)$ is given by:
\begin{equation}
\log \left| \det \left( \frac{d \: \mathrm{conv2D}(\mathbf{h};\mathbf{W})}{d \: \mathbf{h}} \right) \right| 
= h \cdot w \cdot \log |\det(\mathbf{W})|.
\end{equation}
This operation efficiently mixes information across feature channels while preserving invertibility.

\subparagraph{Coupling layer.} 
The final component in each GLOW block is the \emph{affine coupling layer}, which follows the same principle as in RealNVP. The input is split into two parts: one remains unchanged, while the other is transformed using scale and translation parameters predicted by a neural network conditioned on the unchanged part. This ensures that the transformation remains invertible and that the Jacobian determinant is easy to compute. The coupling layers, combined with learned channel mixing through $1{\times}1$ convolutions, allow GLOW to capture complex dependencies between input dimensions.

Overall, GLOW provides a stable and efficient framework for flow-based generative modeling, improving over RealNVP in terms of expressiveness and convergence. It remains one of the key references for invertible neural networks and density-based generative modeling.

\subsubsection*{Autoregressive networks}
This section provides an introduction to \textit{Autoregressive networks}~\cite{autoregr_and_NF_kingma_2016,Coccaro_2024}. Further details are left to the next section. We can introduce autoregressive models as a generalization of coupling flows in which the transformation is implemented by a DNN. Each output $i$ is modeled by the DNN according to the previously transformed dimensions. Let $h(\cdot;\theta) : \mathbb{R} \rightarrow \mathbb{R}$ be a bijector parametrized by $\theta$. Then we can define the autoregressive model function $\mathbf{g}:  \mathbb{R}^D \rightarrow \mathbb{R}^D$ such that $\mathbf{y} = \mathbf{g}(\mathbf{x})$, where each entry of $\mathbf{y}$ is conditioned on the previous output: 

\begin{equation}
    y_i = h(x_i; \Theta(y_{1:i-1}))
\end{equation}
where $y_{1:i-1}$ is a short notation for $(y_1, ...,y_{i-1} )$ and $i = 2, ..., D$, with $D$ the number of dimensions. The $\Theta$ function is called a \textit{conditioner}. The inverse transformation is then given by: 
\begin{equation}
    x_i = h^{-1}(y_i; \Theta_i(y_{1:i-1}))
\end{equation}
We could have chosen a conditioner that depends only on the untransformed dimensions of the input: 
\begin{equation}
    y_i = h(x_i, \Theta(x_{1:i-1}))
\end{equation}
The Jacobian matrix of an autoregressive transformation is triangular, giving a big advantage in the calculation of the determinant, which now becomes the product of the elements in the principal diagonal: 
\begin{equation}
    \det(\mathbf{J}_{\mathbf{g}})= \prod_{i=1}^D \frac{\partial y_i}{\partial x_i}
\end{equation}

\section{Masked Autoregressive Flow (MAF)}
In this section, we introduce a specific approach to autoregressive networks, built on the realization (pointed out by Kingma et al. (2016) \cite{autoregr_and_NF_kingma_2016}) that autoregressive models, when used to generate data, correspond to a deterministic transformation of an external source of randomness (typically obtained by random number generation). This transformation, due to the autoregressive property, has a tractable Jacobian by design and, for certain autoregressive transformations, is also invertible, precisely corresponding to a \emph{normalizing flow} introduced earlier in the text (Section \ref{sec:NF}). 

The specific implementation introduced in this section is the \emph{Masked Autoregressive Flow (MAF)}~\cite{maf} using the \emph{Masked Autoencoder for Distribution Estimation (MADE)}~\cite{made} as the building block. It corresponds to a generalization of the \emph{RealNVP}, and it is closely related to the \emph{Inverse Autoregressive Flow (IAF)}~\cite{kingma2016iaf,Coccaro_2024}.

The key idea of a MAF is to improve model fit by stacking multiple instances of the model into a deeper flow. Given autoregressive models $M_1, M_2, \dots, M_n$, an estimate of the objective PDF is found by transforming the output of the first block $M_1$ with the subsequent block $M_2$; the output of $M_2$ is then transformed by $M_3$ and so on until the last block. The autoregressive blocks are typically chosen to be MADE blocks that will be further discussed later in the section. In other words, we call \emph{MAF} an implementation of stacking MADE blocks into a flow.  

In the original implementation, each MADE block was responsible for outputting the parameters of an affine transformation $\vec{\alpha}$ and $\vec{\mu}$. In the \emph{generative direction}, the transformations are written as: 

\begin{displaymath}
    x_i = u_i \cdot e^{\alpha_i} + \mu_i
\end{displaymath}
where $\mu_i = f_{\mu_i}(\mathbf{x}_{1:i-1})$, $\alpha_i = f_{\alpha_i}(\mathbf{x}_{1:i-1})$ and $u_i \sim \mathcal{N}(0,1)$\footnote{As a parenthesis this is precisely how to generate samples with normalizing flows. Random values of $\mathbf{u}$ are drawn from a random generator and passed through the normalizing flow to get the sample.}. This is not the only possible choice and, in the next paragraphs, one of the most powerful alternatives, the \emph{Rational Quadratic Spline (RQS)}, will be further discussed since it will be one of the fundamental aspects of the implementation. The following is an extract of Ref.~\cite{maf}. 

An important point to note is that \emph{MADE} removes the need to compute activations sequentially \emph{within} a layer: thanks to its masking scheme (detailed in the next section), all units can be evaluated in parallel while still respecting the autoregressive dependencies. However, a \emph{MAF} remains autoregressive at the level of the transformation: each output component $z_i$ depends only on the prefix $\mathbf{x}_{<i}$. This structure yields a triangular Jacobian, allowing the \emph{forward} mapping $\mathbf{x}\!\to\!\mathbf{z}$ and its log-determinant to be computed efficiently and in parallel. A single MADE pass predicts all per-dimension transformation parameters (affine $(\mu_i, \alpha_i)$ or spline $\vec{\mathbf{k}}_i$), and the total log-determinant is simply the sum $\sum_i \log\lvert \partial T_i / \partial x_i \rvert$ used in the likelihood evaluation.

In contrast, \emph{sampling} proceeds in the inverse direction, $\mathbf{z}\!\to\!\mathbf{x}$, and is inherently sequential:
\[
x_i = T_i^{-1}\!\big(z_i;\, \mathbf{x}_{<i}, \mathbf{c}\big), \qquad i = 1, \dots, D,
\]
because the parameters for step $i$ depend on the previously generated components $\mathbf{x}_{<i}$. For affine transformations this reduces to $(z_i - \mu_i)/e^{\alpha_i}$, whereas with RQS transformations each step requires inverting a monotone one-dimensional spline, computationally inexpensive, but still ordered.

This reflects the well-known trade-off between MAF and IAF: a \emph{MAF} allows parallel and efficient likelihood evaluation but sequential sampling, whereas an \emph{IAF} reverses this behavior, providing fast sampling but requiring sequential likelihood computation.

\subsection{MADE: Masked Autoencoder for Distribution Estimation}
Building its foundation on the autoencoder structure, the MADE~\cite{made} network introduces the autoregressive constraints by masking the autoencoder parameters. Before the introduction of MADE, given its iterative nature, evaluating $p(\mathbf{x})$ with a deep autoregressive model would cost $O(D)$ times more than a simple neural network, while MADE preserves the efficiency of a single pass through a regular autoencoder. 

\subsubsection*{Autoencoders} To better understand what follows, a brief discussion of how an \emph{autoencoder}~\cite{ae} works is necessary. For simplicity, we will concentrate on the case of binary observations, where for every $D$-dimensional input $\mathbf{x}$, each input dimension $x_d$ belongs to $\{ 0,1\}$. 

An autoencoder attempts to learn a hidden representation $\mathbf{h}(\mathbf{x})$ of its input $\mathbf{x}$ such that, from it, we can obtain a reconstruction $\hat{\mathbf{x}}$ which is as close as possible to $\mathbf{x}$. We can write
\begin{displaymath}
    \mathbf{h}(\mathbf{x}) = \mathbf{g}(\mathbf{b}+ \mathbf{Wx}) \text{ ,} \quad \hat{\mathbf{x}} = \text{sigm}(\mathbf{c+Vh(x)})
\end{displaymath}
where $\mathbf{V}$ and $\mathbf{W}$ are weight matrices, $b$ and $c$ are bias vectors, $g$ is a non-linear activation function and sigm is a shorthand for the sigmoid activation function described in Chapter~\ref{ch:ml}. 

Before getting more in depth on how the MADE block is built, we need to understand how to use autoregression as distribution estimation.

\subsubsection*{Autoregression as Distribution Estimation}
The question we need to answer is how we can write $p(\mathbf{x})$ in such a way that it can be computed based on the output of a properly corrected autoencoder. 
The key idea is to use the decomposition of a PDF in product of its nested conditionals:
\begin{displaymath}
    p(\mathbf{x}) = \prod_{d=1}^D p(x_d \vert \mathbf{x}_{<d})
\end{displaymath}
where $\mathbf{x}_{<d} = [x_1, \dots, x_{d-1}]^T$. 

By defining $p(x_d \vert \mathbf{x}_{<d}) = \hat{x}_d$ we can now get a hint on how to generalize the autoencoder to perform a density estimation task: each output must be a function taking as input $\mathbf{x}_{<d}$ only and outputting the probability of observing a value $x_d$ at the $d$-th dimension. Notice that the autoencoder forms a proper distribution if each output unit $\hat{x}_d$ only depends on the previous units $\mathbf{x}_{<d}$. This property is called \textbf{autoregressive property}, since computing the negative log-likelihood is equivalent to sequentially predicting (regressing) each dimension of the input $\mathbf{x}$.

\subsubsection*{Introducing masks} \label{sec:masks}
The question now becomes how to implement the autoregressive property by modifying the autoencoder. Since output $\hat{x}_d$ must depend only on the preceding inputs $\mathbf{x}_{<d}$, it means that there must be no computational path between output $\hat{x}_d$ and any of the inputs $x_d, \dots, x_D$. In other words, for each of these paths, at least one connection (in the matrix \textbf{W} or \textbf{V}) must be 0. As the name suggests, to zero the connections, we introduce some \textbf{binary masks} to be elementwise-multiplied to each weight matrix. To remove a connection, we simply set the corresponding entry in the mask to 0. Then the masked autoencoder can be described by 
\begin{displaymath}
    \mathbf{h}(\mathbf{x}) = \mathbf{g}(\mathbf{b}+ (\mathbf{W}_l \odot \mathbf{M}^{\mathbf{W}^l})\mathbf{x}) \text{ ,} \quad \hat{\mathbf{x}} = \text{sigm}(\mathbf{c}+ (\mathbf{V} \odot \mathbf{M^V}) \mathbf{h}(\mathbf{x}))
\end{displaymath}
where the subscript $l = 0, \dots, L$ is the hidden layer index and $L$ is the number of hidden layers. To better understand how the process works, let us start from a simplified version of a standard autoencoder, which has a single hidden layer, and then generalize the concept to a deep autoencoder. 

To impose the autoregressive property, the process starts from assigning to each unit in each hidden layer an integer $m$ between 1 and $D-1$ inclusively. The $k^{\text{th}}$ hidden unit's number $m(k)$, gives the maximum number of input units to which it can be connected. Notice that $m(k) = D$ (with $k \in \{0, \dots, K\}$ and $K$ is the number of units in the hidden layer) is disallowed since this hidden unit would be connected to all the inputs and therefore could not contribute to the modeling of any of the conditionals $p(x_d \vert \mathbf{x}_{<d})$. Similarly, $m(k) = 0$ is excluded since it would create constant hidden units. As the number of hidden units can be greater than the number of inputs $D$, the values are drawn from a uniform discrete distribution defined on integers from 1 to $D-1$.
The constraints on the maximum number of input units connected to every hidden node are encoded in the matrix masking the connections between the nodes. The rules are different for the hidden layer and the output layer; for the hidden layer we write: 
\begin{displaymath}
    M_{k,d}^{\mathbf{W}} = 1_{m(k)\geq d} = 
    \begin{cases}
      1  \quad \text{if} \;m(k) \geq d\\
      0 \quad  \text{otherwise}
    \end{cases}
\end{displaymath}
and for the output mask: 
\begin{displaymath}
    M_{d,k}^{\mathbf{V}} = 1_{d > m(k)} = 
    \begin{cases}
      1  \quad \text{if} \;d > m(k)\\
      0 \quad  \text{otherwise}
    \end{cases}
\end{displaymath}
with $d \in \{1, \dots, D \}$ and $k \in \{1, \dots, K \}$. The difference in the two expressions comes from the fact that we need to encode the constraint that the $d^{\text{th}}$ output unit is only connected to $\mathbf{x}_{<d}$. Therefore, the output weights can only connect the $d^{\text{th}}$ output to hidden units with $m(k) < d$. Notice that the second rule (the one for the output layer), implies that, as desired, the first output $\hat{x}_1$ will not be connected to any of the hidden units.

The autoregressive property, with this construction, can be demonstrated by first noticing that since the masks $\textbf{M}^\textbf{V}$ and $\textbf{M}^\textbf{W}$ represent the network connectivity, their product $\textbf{M}^\textbf{V,W} = \textbf{M}^\textbf{V} \textbf{M}^\textbf{W} $ represents the connectivity between the input and output layer. Specifically $\textbf{M}^\textbf{V,W}_{d',d}$ represents the network path between $d^\text{th}$ input and ${d'}^\text{th}$. Thus, to demonstrate the autoregressive property, we need to show that $\textbf{M}^\textbf{V,W}$ is strictly lower diagonal. By definition, we have: 
\begin{displaymath}
    \textbf{M}^\textbf{V,W}_{d',d} = \sum_{k = 1}^K \textbf{M}^\textbf{V}_{d',k} \textbf{M}^\textbf{W}_{k,d} = \sum_{k = 1}^K 1_{d' > m(k)} 1_{m(k) \geq d}
\end{displaymath}
If $d' \leq d$ there are no values for $m(k)$ such that it is both strictly less than $d'$ and greater or equal to $d$, thus $\textbf{M}^\textbf{V,W}_{d',d}$ is 0. This demonstrates the autoregressive property of the masked autoencoder if masks are built with the illustrated rules. 

The question now becomes how to generalize this logic to a deep autoencoder. This can be simply generalized by assigning a maximum number of connected inputs to each of the hidden units and constructing similar masks to satisfy the autoregressive property. 
For networks with $L>1$ hidden layers, superscripts are introduced to index layers. The first hidden layer matrix is now $\textbf{W}^1$, the second layer matrix is $\textbf{W}^2$, and so on. This notation will be extended also to the integers assigned to the hidden units; the maximum number of connected inputs to the $k^\text{th}$ unit in the $l^\text{th}$ layer is denoted with $m^l(k)$, with $k \in \{1, \dots, K^l\}$ and $K^l$ being the number of units in the $l^{\text{th}}$ hidden layer. 

Referring to the first layer mask already discussed, the generalization to an $L = 2$ autoencoder can be done by making sure that each unit $k'$ (in the second hidden layer) is only connected to first layer units connected to at most $m^2(k')$ inputs, i.e. the first layer units such that $m^1(k) \leq m^2(k')$. By following this argument, the rules are generalized to a deep autoencoder: 

\begin{displaymath}
    M_{k',k}^{\mathbf{W}^l} = 1_{m^l(k')\geq m ^{l-1}(k)} = 
    \begin{cases}
      1  \quad \text{if} \;m^l(k') \geq m^{l-1}(k)\\
      0 \quad  \text{otherwise}
    \end{cases}
\end{displaymath}
and
\begin{displaymath}
    M_{d,k}^{\mathbf{V}} = 1_{d > m ^{L}(k)} = 
    \begin{cases}
      1  \quad \text{if} \;d > m^{L}(k)\\
      0 \quad  \text{otherwise}
    \end{cases}
\end{displaymath}

In the discussion the ordering in the first and last layer was left to the natural ordering introduced by the input vector $\mathbf{x}$. This is not mandatory and will be very important for the \emph{order-agnostic training}. Indeed, have shown that training an autoregressive model on \emph{all} orderings can be beneficial. It can be achieved by assigning integer values to the input layer and using them to build the masks. Note that the integer values must be the same (in the same order) for the input and output vector.

\begin{figure}
    \centering
    \includegraphics[width=0.7\linewidth]{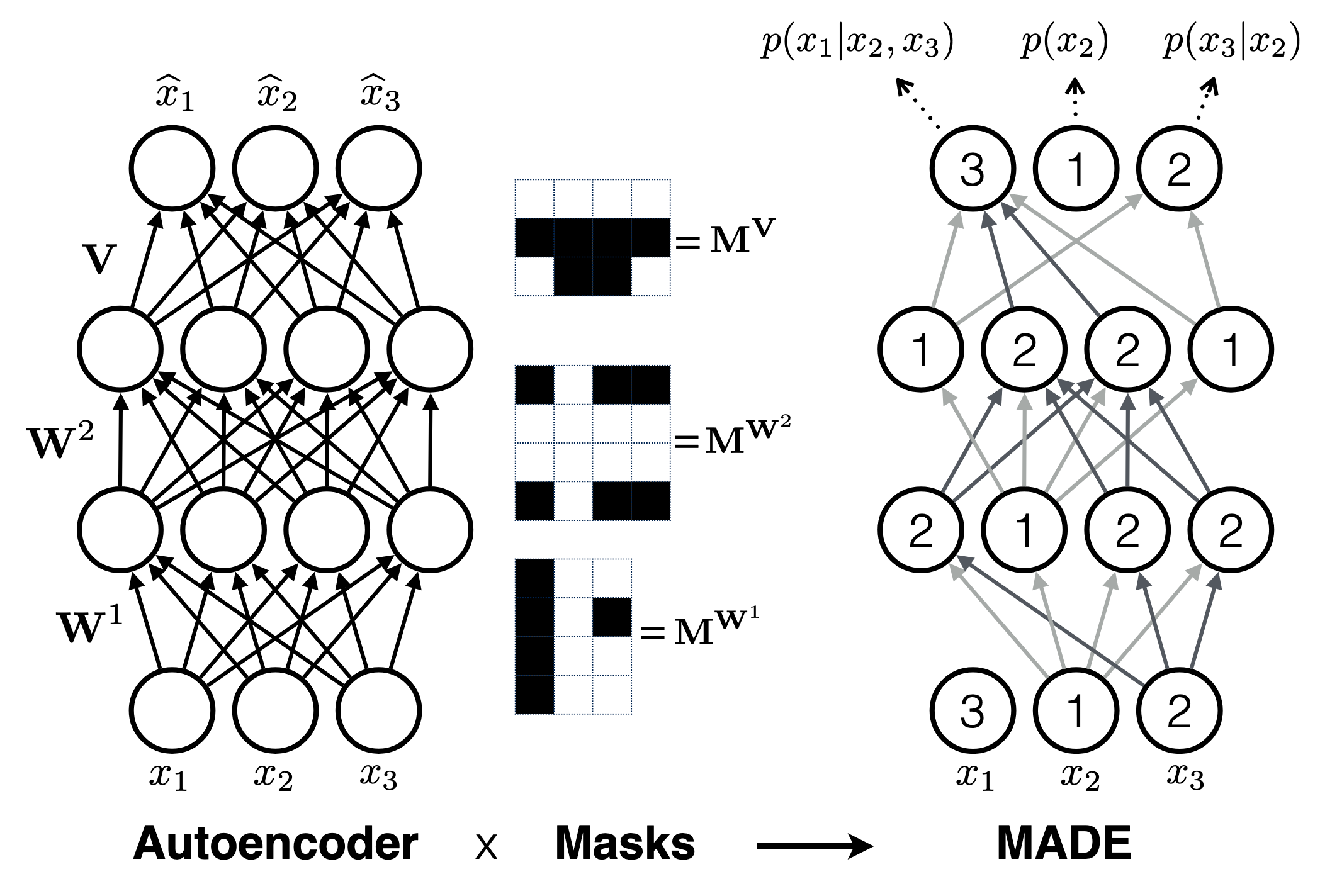}
    \caption{On the \textbf{left} there is the conventional autoencoder with three hidden layers. The networks are oriented from top to bottom and the input passes through three fully connected layers. Note that since the output attempts at reconstructing the inputs, and it depends entirely on the inputs, the standard Autoencoder is not suitable for data generation. On the \textbf{Right} the networks has the same structure as the standard Autoencoder, but some connections have been removed, such that each input unit is only predicted based on the previous ones, using multiplicative binary masks. In the example, the ordering of the inputs has been changed to 3,1,2. Masks are constructed following the discussion illustrated in Section \ref{sec:masks}.}
    \label{fig:made}
\end{figure}

Figure~\ref{fig:made} illustrates the conceptual transition from a conventional autoencoder to the Masked Autoencoder for Distribution Estimation (MADE). In a standard autoencoder (left), the network is fully connected, and each output unit depends on all input variables, making the model suitable for representation learning but not for autoregressive density estimation. In contrast, the MADE architecture (right) introduces binary masks that selectively remove connections between layers, enforcing an autoregressive property such that each output dimension is conditioned only on a subset of the inputs according to a predefined ordering. This masking mechanism allows the network to model the joint probability as a product of conditional distributions, thereby transforming a conventional feed-forward autoencoder into an efficient autoregressive model.

\subsection{Rational Quadratic Spline (RQS)} 
A spline is, in general, a piecewise-polynomial or a piecewise-rational function which is specified by K+1 points $\left \{(x_i^{(k)},y_i^{(k)})\right\}_{k=0}^K$ (K bins) called \emph{knots}, through which the spline passes. This notation is used to define the knots coordinates, defined for each dimension of the vectors $x_i$ and $y_i$. 
The requirement for the spline to be invertible is translated into the requirement for the spline to be monotone, meaning that $x_i < x_{i+1}$ and $y_i < y_{i+1}$. Usually, the spline is defined on compact intervals~\cite{splines}. 

An example of a spline, introduced by Ref.~\cite{rqs} is the Rational Quadratic Spline that introduces a more expensive transformation that remains invertible and computationally efficient. As the name suggests, the spline defines a different rational quadratic function for each bin defined by the knots. The knots are defined to be monotonically increasing between $\{(x_i^{(0)},y_i^{(0)}) = (-B,-B)\}$ to $\{(x_i^{(K)},y_i^{(K)}) = (B,B)\}$. We call the domain of the spline $\mathbb{B} = [-B,B]$. 
The spline is defined by following the method introduced in Ref.~\cite{rqs}. 
The method starts from defining: 
\begin{displaymath}
    h_i^{(k)} = x_i^{(k+1)} - x_i^{(k)}, \quad  \Delta_i^{(k)} = (y_i^{(k+1)} - y_i^{(k)})/h_i^{(k)}
\end{displaymath}
where $\Delta_i^{(k)}$ represents the variation of $y_i$ with respect to the variation of $x_i$ within the $k$-th bin. For this reason, being the spline monotonically increasing by construction, it is always non-negative. The goal will be to build a bijector $\mathbf{g}(x_i)$, mapping the $\mathbb{B}$ interval to itself, such that $\mathbf{g}(x_i^{(k)}) = y_i^{(k)}$ and derivatives $d_i^{(k)} = dy_i^{(k)} / dx_i^{(k)}$ satisfying the conditions: 

\begin{displaymath}
\begin{cases}
d_i^{(k)} = d_i^{(k+1)} = 0, & \text{if } \Delta_i^{(k)} = 0, \\[6pt]
d_i^{(k)} > 0,\; d_i^{(k+1)} > 0, & \text{if } \Delta_i^{(k)} > 0,
\end{cases}
\end{displaymath}
necessary and sufficient, in the case of a rational quadratic function, to ensure monotonicity~\cite{rqs,rqs2}. 
Notice that, in this specific instance, the derivatives at the boundaries are left unconstrained, assuming linear tails implementation outside the spline domain. Sometimes an identity map is used outside the spline domain and additional constraints must be introduced on the derivatives: $d_i^{(0)} = d_i^{(K)} = 1$. 

For $x_i \in [x_i^{(k)}, x_i^{(k+1)}]$, with $k = 0, \dots, K-1$, we define
\begin{displaymath}
    \theta_i = \frac{(x_i-x_i^{(k)})}{h_i^{(k)}}
\end{displaymath}
such that $\theta_i \in [0,1]$. We also define
\begin{equation} \label{eq:rational_quadr_fraction}
y_i = \frac{P_i^{(k)}(\theta_i)}{Q_i^{(k)}(\theta_i)},
\end{equation}
where the functions $P$ and $Q$ are defined by
\begin{displaymath}
\begin{aligned}
P_i^{(k)}(\theta_i) &=
\Delta_i^{(k)}\, y_i^{(k+1)}\, \theta_i^{2}
+ \Delta_i^{(k)}\, y_i^{(k)}\, (1-\theta_i)^{2} \\[4pt]
&\quad
+ \bigl( y_i^{(k)}\, d_i^{(k+1)} + y_i^{(k+1)}\, d_i^{(k)} \bigr)\, \theta_i (1-\theta_i), \\[8pt]
Q_i^{(k)}(\theta_i) &=
\Delta_i^{(k)}
+ \Bigl( d_i^{(k+1)} + d_i^{(k)} - 2\,\Delta_i^{(k)} \Bigr)\, \theta_i (1-\theta_i)\,.
\end{aligned}
\end{displaymath}
Equation \ref{eq:rational_quadr_fraction} can then be written in the simplified form
\begin{displaymath}
y_i = y_i^{(k)} +
\frac{ ( y_i^{(k+1)} - y_i^{(k)})(\Delta_i^{(k)}\, \theta_i^{2} + d_i^{(k)}\, \theta_i (1-\theta_i))}
     { \Delta_i^{(k)} + ( d_i^{(k+1)} + d_i^{(k)} - 2\,\Delta_i^{(k)} )\, \theta_i (1-\theta_i) } \,.
\end{displaymath}
The Jacobian $J_{\mathbf{g}}$ is then diagonal ($y_i$ only depends on $x_i$) and can be written as
\begin{displaymath}
\frac{\partial y_i}{\partial x_i}
=
\frac{
\bigl(\Delta_i^{(k)}\bigr)^{2}
\Bigl( d_i^{(k+1)} \theta_i^{2} + 2\,\Delta_i^{(k)} \theta_i (1-\theta_i) + d_i^{(k)} (1-\theta_i)^{2} \Bigr)
}{
\Bigl( \Delta_i^{(k)} + \bigl( d_i^{(k+1)} + d_i^{(k)} - 2\,\Delta_i^{(k)} \bigr)\, \theta_i (1-\theta_i) \Bigr)^{2}
}
\end{displaymath}
with $i = 1, \dots, D$. The inverse transformation can be obtained from \ref{eq:rational_quadr_fraction} by solving the quadratic equation with respect to $x_i$. 

\begin{figure}
    \centering
    \includegraphics[width=0.9\linewidth]{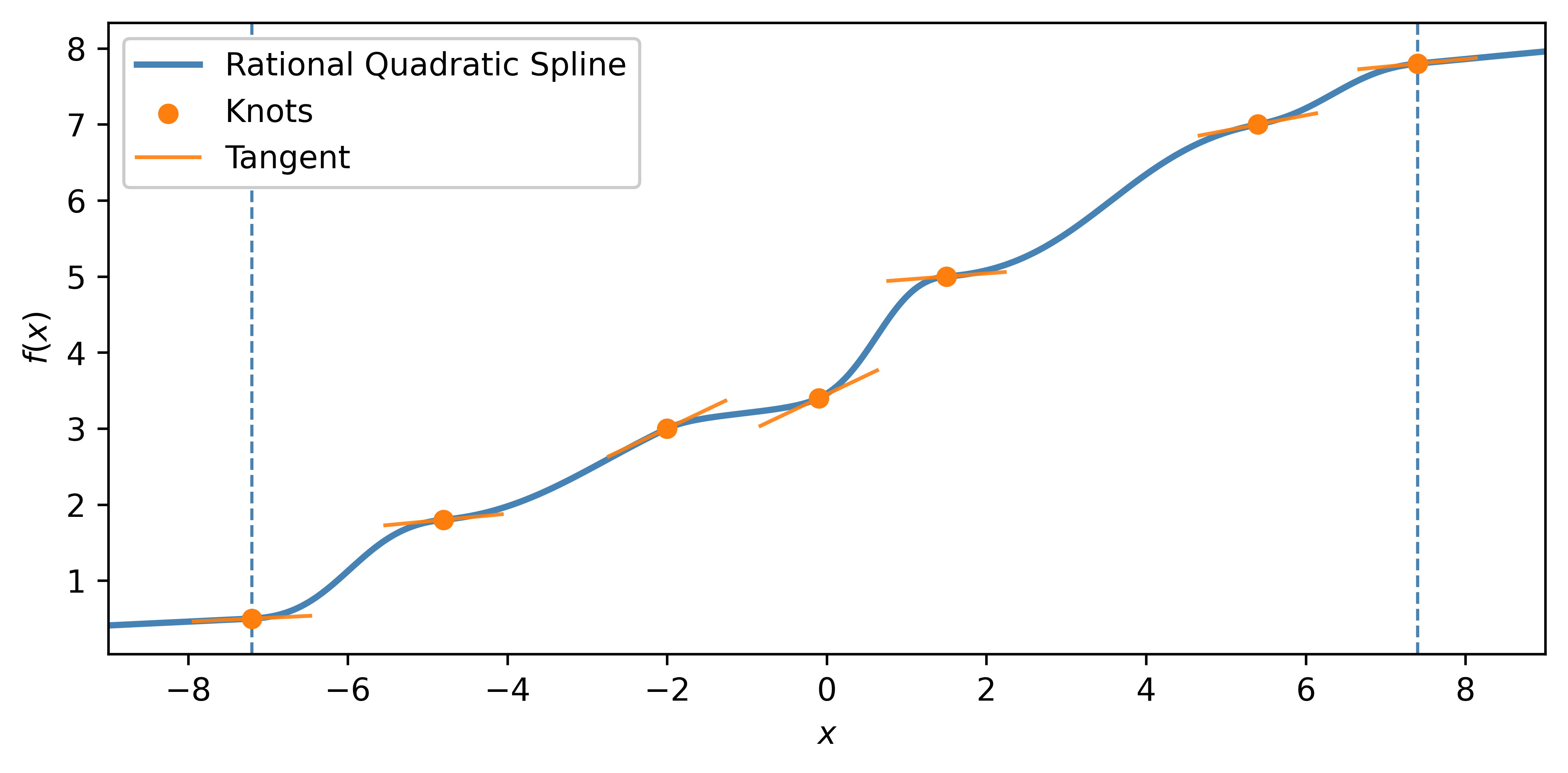}
    \caption{Example of a RQS transformation defined on the interval $\mathbb{B} = [-7,7]$, with 6 bins and 7 knots. }
    \label{fig:rqs}
\end{figure}

B and K are fixed variables (hyperparameters) to be chosen in the design phase. On the contrary, $\{ (x_i^{(k)}, y_i^{(k)}) \}_{k=0}^K$ and $\{ d_i^{(k)} \}_{k=0}^{K-1}$ are $2(K+1)$ plus $K-1$ parameters modeled by a NN, which determine the shape of the spline function. 
Figure~\ref{fig:rqs} shows an example of a RQS transformation defined on the interval $\mathbb{B} = [-7,7]$, with 6 bins and 7 knots.

\section{Evaluating Generative Models} \label{sec:eval_intro}
Before statistically grounded protocols, evaluation in physics used simple checks on a small set of hand-crafted observables. Typical studies compared one- or two-dimensional histograms and reported global scores built from physics features, such as the Fréchet and Kernel Physics Distances (FPD/KPD)~\cite{RKHS_kernel}. The CaloChallenge helped standardize this practice and made cross-model comparisons easier, sometimes adding classifier-based checks as a complement \cite{calochallenge,RKHS_kernel}. These tools were practical and interpretable, but they mixed heterogeneous metrics and had weak control of statistical uncertainty. Reported differences often depended on binning, sample size, or hyperparameters, and it was not clear whether they were significant or due to fluctuations.

Reference~\cite{ref_the_ref} addressed these issues by casting validation as a two-sample hypothesis test. The paper studies families of tests in a single framework: projection-based distances (e.g., sliced Wasserstein, mean-KS, sliced-KS) and multivariate baselines (e.g., unbiased MMD, Fréchet Gaussian Distance). It prescribes the construction of empirical null and alternative distributions for each test statistic and reports calibrated \(p\)-values and power at fixed sample cost. This gives a common language, clear uncertainty quantification, and reproducible comparisons across datasets and models. A key result is that well-designed one-dimensional projection tests can match the sensitivity of heavier multivariate metrics while being much cheaper and easier to parallelize.

\subsection*{Why new methods were necessary}
High-energy physics data are high-dimensional (from \(10^2\) up to \(10^4\)+ features per event). Naive multivariate distances are often too costly or unstable in this regime. Evaluation must therefore:
\begin{itemize}
    \item \textbf{Scale in dimension and sample size:} tests should remain efficient for large datasets and many features.
    \item \textbf{Provide calibrated error control:} report \(p\)-values at a chosen significance and power curves at fixed cost.
    \item \textbf{Be robust to finite samples:} avoid conclusions driven by binning choices or small fluctuations.
    \item \textbf{Capture correlations and tails:} assess structure beyond a few marginals and remain sensitive in rare regions.
    \item \textbf{Support reproducibility and comparability:} use a shared procedure, so results can be fairly compared.
\end{itemize}
The framework in \cite{ref_the_ref} meets these needs. It treats data and generated samples symmetrically, uses empirical nulls for calibration, and exposes the trade-off between sensitivity and compute. Physics-aware summaries such as FPD/KPD remain useful as diagnostics and headline indicators, but they are most reliable when embedded in a calibrated testing pipeline rather than used alone \cite{RKHS_kernel,calochallenge}.

\noindent
In this thesis, that calibrated framework is adopted. Null and alternative ensembles are built from truth and generated showers. Reported results include \(p\)-values, test power, and computational cost, with projection-based tests as the main high-dimensional tools and physics-aware summaries and classifier checks as complementary diagnostics.

\subsection{Two-sample hypothesis testing}
Given two random variables $x$ and $y$, defined on a space $\mathcal{X} \subseteq \mathbb{R}^d$, let us consider two samples $X = \{ x_i \}$ where $i = 1, \dots, n$ and $Y = \{ y_j \}$ where $j = 1, \dots, m$, with the assumption of them being independent and identically distributed according to the distributions $p$ and $q$, respectively. We can denote with $x_{i,I}$ ($y_{j,J}$) the scalar value of the $i\text{-th}$ ($j\text{-th}$) element of sample $X$ ($Y$) along the $I\text{-th}$ ($J\text{-th}$) dimension, with $I \text{($J$)}  = 1, \dots, d$. 

\begin{figure}
     \centering
     \includegraphics[width=0.5\linewidth]{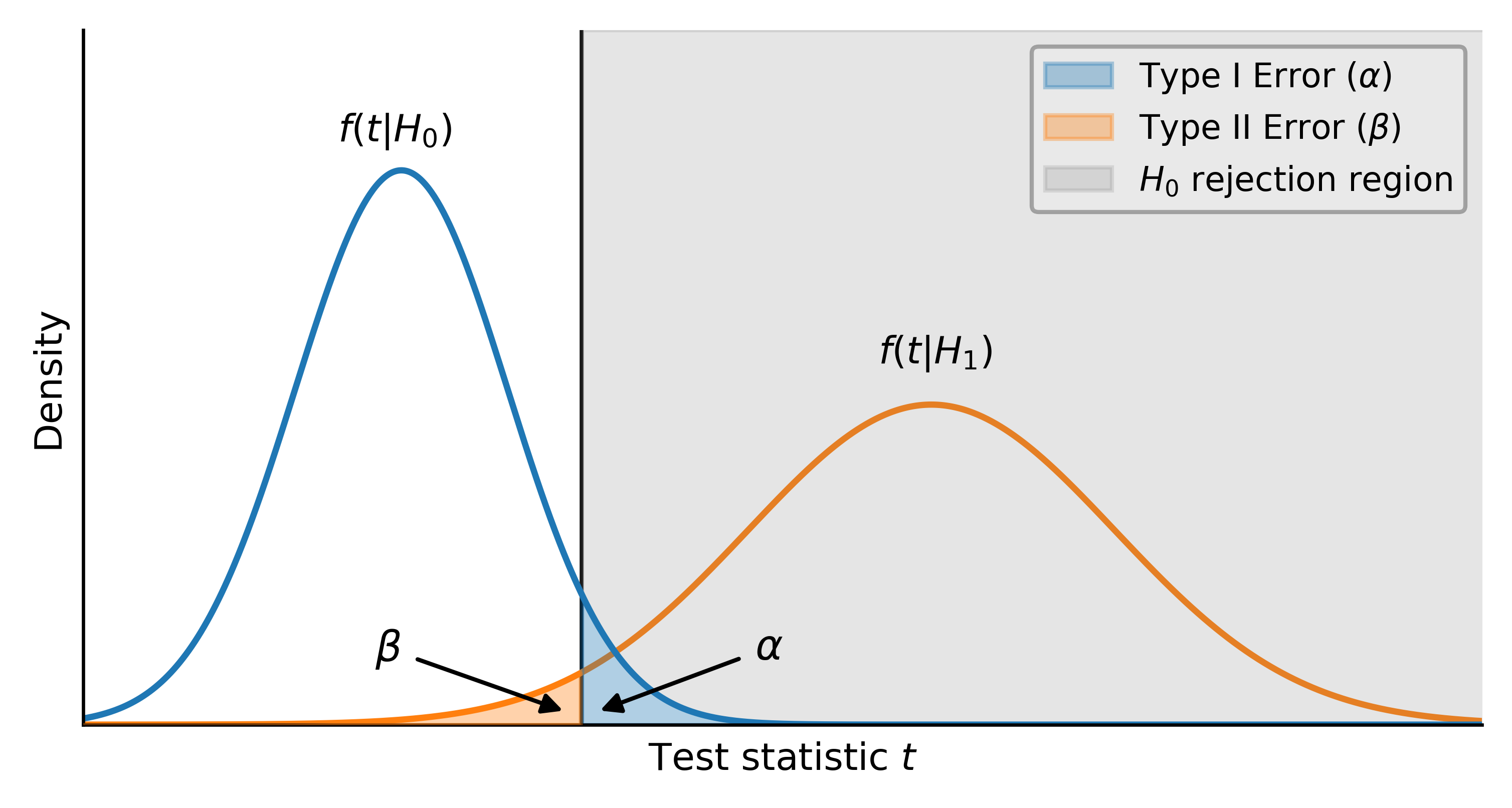}
    \caption{Illustration of Type~I and Type~II errors for a test statistic $t$. The blue curve is the sampling distribution $f(t\,|\,H_0)$ under the null hypothesis, and the orange curve is $f(t\,|\,H_1)$ under the alternative. A decision threshold $t^\star$ defines the shaded $H_0$ rejection region (to the right). The Type~I error rate $\alpha$ is the blue tail area beyond $t^\star$ (rejecting $H_0$ when $H_0$ is true). The Type~II error rate $\beta$ is the orange area to the left of $t^\star$ (failing to reject $H_0$ when $H_1$ is true).}

     \label{fig:errors}
 \end{figure}

Two-sample testing aims at determining whether the null hypothesis
\begin{equation}
    H_0 : p = q
\end{equation}
can be rejected based on finite data. The alternative hypothesis is the negation of the null:
\begin{equation}
    H_1 : p \neq q
\end{equation}
Given the hypotheses, we need to define the \emph{test statistic}, $t : (\mathcal{X})^n \times (\mathcal{X})^m \longrightarrow \mathbb{R}$, and calculate its value on the observed data: 
\begin{equation}
    t_{obs} = t(X,Y).
\end{equation}
Then, a binary test is defined by comparing the observed $t_{obs}$ to a threshold $t_{\alpha}$ defined by:
\begin{equation} \label{eq:significance}
    \alpha = P(t \geq t_{\alpha}\rvert H_0) = \int_{t_{\alpha}}^{\infty} f(t \rvert H_0) dt,
\end{equation}
 where $\alpha$ is defined as the \emph{significance} of the test and $f(t \rvert H_0)$ the distribution of the test statistic under the null hypothesis $H_0$. The significance represents a \emph{preselected} probability of type-I error, i.e., the \emph{rate of false positives} (rejecting the null hypothesis if true). The null hypothesis $H_0$ is rejected if $t_{obs} > t_{\alpha}$. In Fig~\ref{fig:errors} is reported a diagram showing the Type I and Type II errors.

 Another quantity can then be calculated, the \emph{p-value}, defined as the probability of observing a test statistic as extreme as $t_{obs}$ under $H_0$: 
 \begin{equation}
     p_{obs} = P(t\geq t_{obs} \rvert H_0).
 \end{equation}

 Given $\alpha$, the Neyman-Pearson construction provides a method to compare the performance of different tests. This consists in introducing type-II error, or the \emph{rate of false negatives}, defined as the probability of accepting the null hypothesis $H_0$ if the alternative hypothesis $H_1$ is true: 
 \begin{equation}
     \beta = P(t < t_{obs} \rvert H_1) = \int^{t_{\alpha}}_{- \infty} f(t \rvert H_1) dt,
 \end{equation}
 where $f(t \rvert H_1)$ is the distribution of the test statistic under the alternative hypothesis. The \emph{power} of the test can then be defined as 
\begin{equation} \label{eq:power}
    \text{power} = P(t \geq t_{\alpha} \rvert H_1) = 1-\beta
\end{equation}
The best statistic is usually the one with the higher power given a significance level $\alpha$, i.e. the one with the smallest rate of false negatives given a fixed rate of false positives. 
To compute the quantities in Eqs.~\ref{eq:significance} and \ref{eq:power}, the distribution of the test statistic under both the null and alternative hypothesis, $f(t \rvert H_0)$ and $f(t \rvert H_1)$, must be known or estimated. In some cases, analytical approximation for finite sample size or asymptotic behavior can be used, for the KS test and for likelihood ratio-based tests. In general, however, non-parametric testing often relies on empirical estimates. 

In this context a \emph{goodness-of-fit test} can be introduced, by designating one of the samples as the \emph{reference} one. We will be interested in testing the compatibility of a generated dataset with the true values obtained by Monte Carlo simulation, so we will fix as reference the Monte Carlo data and the generated samples as the alternative. 

\newdimen\W      \W=7cm      
\newdimen\Hgt    \Hgt=4cm     
\newdimen\gap    \gap=2mm     
\newdimen\HalfW  \HalfW=.5\W
\newdimen\HalfH  \HalfH=.5\Hgt
\definecolor{vibred}{rgb}{0.98, 0.40, 0.40}
\definecolor{vibblue}{rgb}{0.30, 0.60, 0.95}

\begin{figure}[h] 
\centering
\begin{tikzpicture}[
  font=\small,
  every node/.style={align=center}
]

\fill[vibblue, opacity=0.5]  (\HalfW, \HalfH) rectangle (\W, \Hgt);
\fill[vibred, opacity=0.5] (0, 0)          rectangle (\HalfW, \HalfH);

\draw[thick] (0,0) rectangle (\W,\Hgt);
\draw[thick] (\HalfW,0) -- (\HalfW,\Hgt); 
\draw[thick] (0,\HalfH) -- (\W,\HalfH);   

\node[anchor=south] at (.25\W, \Hgt+\gap) {\textbf{Accept $\mathbf{H_0}$}};
\node[anchor=south] at (.75\W, \Hgt+\gap) {\textbf{Reject $\mathbf{H_0}$}};

\node[anchor=east]  at (-\gap, .75\Hgt) {\textbf{$\mathbf{H_0}$ true}};
\node[anchor=east]  at (-\gap, .25\Hgt) {\textbf{$\mathbf{H_0}$ false}};

\node at (.25\W, .75\Hgt) {Correct decision\\(True negative)};
\node at (.75\W, .75\Hgt) {Type I error ($\alpha$) \\ (False positive)};
\node at (.25\W, .25\Hgt) {Type II error ($\beta$) \\ (False negative)};
\node at (.75\W, .25\Hgt) {Correct decision\\(True positive)};

\end{tikzpicture}
\caption{Possible outcomes of a statistical test. Colored cells highlight Type I and Type II errors.}
\label{tab:type_errors}
\end{figure}

\subsection{Test statistics}
With the exception of the Fréchet Gaussian Distance (FGD), all the test statistics are derived from or based on integral probability metrics (IPMs). IPMs are a broad class of probability distribution distances, defined as 
\begin{equation}
    \mathrm{d}_{\mathcal{F}}(p, q)
    = \sup_{f \in \mathcal{F}} 
    \left( 
        \mathbb{E}_{x \sim p}[f(x)] 
        - 
        \mathbb{E}_{y \sim q}[f(y)] 
    \right).
\end{equation}
where $\mathcal{F}$ is a class of real-valued, scalar functions defined over the set $\mathcal{X}$: $\forall f \in \mathcal{F}\, , \: f: \mathcal{X} \rightarrow \mathbb{R}$.
\noindent
We will now illustrate the relevant examples of test statistics introduced in \cite{ref_the_ref}.

\subsubsection{Sliced Wasserstein distance}
The sliced Wasserstein (SW) distance~\cite{SWD} involves averaging the 1D projections of the Wasserstein distance over ``all'' directions on the unit $d$-dimensional sphere. Being a computationally efficient variant of the W distance, it is a natural choice for two-sample testing.

We focus on the 1-Wasserstein distance, commonly referred to as \emph{earth mover's distance}. In one dimension, it is defined as
\begin{equation}
    W_{n,m} = \int_{\mathbb{R}} \lvert F_n(u) - G_m(u) \rvert du
\end{equation}
where $F_n$ and $G_m$ are eCDFs \footnote{A CDF of a random variable $X$, evaluated at $x$, is the probability of measuring $X$ less than or equal to $x$. In formulae: $F(x) = P (X \leq x)$
It is a non-decreasing function and, if the random variable is continuous, it is defined as $F_X(x)=\int_{-\infty}^{x}f_X(t)dt$, otherwise defined as $F_X(x) = \sum_{x_i\leq x}P(X = x_i) = \sum_{x_i\leq x}f_X(x_i)$, where $f_X$ is the PDF of the distribution of $X$. An empirical cumulative distribution function, is an approximation of the CDF built from an empirical measure of a sample. Given a sample $(X_1, \dots, X_n)$, it is defined by $F_n(x)= \frac{1}{n} \sum_{i=1}^n \mathbf{1}_{X_i \leq x}$, with $\mathbf{1}_{X_i \leq x} = 1$ if $X_i \leq x$ or $\mathbf{1}_{X_i \leq x} = 0$ if $X_i > x$.} (Empirical Cumulative Distribution Functions). This distance measures the minimal cost of transforming one distribution into another and depends linearly on the Euclidean distance between data points. It can be interpreted as an IPM by taking as $\mathcal{F}$ the space of 1-Lipschitz functions. In case of two 1D samples with an equal number of data points $m = n$, the W distance can be easily computed as
\begin{displaymath}
    W_n = \frac{1}{n} \sum\limits_{i=1}^n \lvert \underline{x}_i - \underline{x}'_i \rvert,
\end{displaymath}
where the underlined variables represent elements in the set obtained by permuting the original sample with a permutation $\mathcal{P}$ that sorts the points: 
\begin{equation}
    \{\underline{x}\} = \mathcal{P}(\{x\}\lvert \underline{x}_1 \leq \dots \leq \underline{x}_n )
\end{equation}

The Sliced variant takes the form:
\begin{equation} \label{eq:t_SW}
    t_{\mathrm{SW}} = \frac{1}{K} \sum\limits_{\theta \in \Omega_K} W^{\theta}_n = \frac{1}{K} \sum\limits_{\theta \in \Omega_K} \left( \frac{1}{n} \sum\limits_{i=1}^n \lvert \underline{x}_i^{\theta} - \underline{x}_i^{\prime \theta} \rvert \right)
\end{equation}
where $\Omega_K$ is a set of $K$ directions selected uniformly at random on the unit sphere $\Omega = \{ \theta \in \mathbb{R}^d \lvert \enspace \lVert \theta \rVert = 1 \}$ and $\{ \underline{x}_i^{\theta} \}_{i=1}^n = \{ \theta^T \underline{x}_i \}_{i=1}^n $ are the sorted data points projected on the direction $\theta$. It is important to note that the asymptotic behavior of the test statistic is not distribution-free, in fact in the limit $m,n \rightarrow \infty$ with $m/n \rightarrow c \neq 0, \infty$, the distribution of the test statistic in Eq.~\ref{eq:t_SW} under the null hypothesis will depend on the underlying data distribution.

\subsubsection{Kolmogorov-Smirnov (KS) inspired test statistics}
The KS test~\cite{ks2,KS1} is a widely used non-parametric method for both goodness-of-fit and two-sample testing. It measures the largest absolute difference between the two eCDFs of the samples. Defined as
\begin{equation} \label{eq:t_ks}
    t_{\mathrm{KS}} = \sqrt{\frac{nm}{n+m}} \sup_u \lvert F_n(u) - G_m(u) \rvert,
\end{equation}
it can be viewed as an IPM with $\mathcal{F}$ chosen to be the class of indicator functions $\mathbf{1}_{(-\infty,t]}$ for all $t \in \mathbb{R}$. In Eq.~\ref{eq:t_ks}, $F_n(u)$ and $G_m(u)$ are the eCDF of the two samples. The prefactor ensures that, under the null hypothesis (i.e. the two samples are drawn from the same distribution), and as $m,n \rightarrow \infty$ with $m/n \rightarrow c \neq 0, \infty$, the test statistic follows the Kolmogorov distribution, with the following CDF 
\begin{equation}
    F_K(x) = 1-2\sum\limits_{k=1}^{\infty}(-1)^{k-1}e^{-2k^2x^2},
\end{equation}
and PDF
\begin{equation}
    f_K(x)=\frac{d}{dx} F_K(x) = 8x\sum\limits_{k=1}^{\infty}(-1)^{k-1}k^2e^{-2k^2x^2}.
\end{equation}
Widely used for 1D data, the KS test has limited application in higher dimensions due to its computational cost. For this reason, two efficient multivariate extensions are introduced. 

\paragraph{Mean KS} The mean KS test extends the KS test to higher dimensions by averaging the KS statistic computed along each dimension of the data. The test statistic is defined by 
\begin{equation}
    t_{\overline{\mathrm{KS}}} = \displaystyle \frac{1}{d} \sum_{I=1}^d 
        \sqrt{\frac{n m}{n + m}} 
        \sup_{u} |F^I_n(u) - G^I_m(u)|, 
\end{equation}
where $F^I_n(u)$ and $G^I_m(u)$ are the eCDFs of the projected samples along the $I$-th dimension. This approach makes the KS distance computationally feasible in higher dimensions, with the downside that, since it is uniquely defined by the 1D marginals, it is not expected to be directly sensitive to correlations between dimensions. To improve the sensitivity to such correlation, another variant has been introduced in \cite{ref_the_ref}.

\paragraph{Sliced KS} Similarly to the Wasserstein distance case, the KS method is extended by projecting the original $d$-dimensional data onto 1D subspaces. The subspaces are chosen to be $K$ uniformly random directions sampled from the unit sphere. For each direction $\theta$, the KS test statistic is computed as 
\begin{equation}
    t_{\mathrm{KS}}^{\theta} = \sqrt{\frac{n m}{n + m}} 
        \sup_{u} |F^\theta_n(u) - G^\theta_m(u)|,
\end{equation}
where $F^\theta_n(u)$ and $G^\theta_m(u)$ are the eCDFs of the projected samples along the direction $\theta$. The Sliced KS statistic (SKS), is then computed as the average of the KS statistics across the $K$ random directions:
\begin{equation}
    t_{\mathrm{SKS}} = \displaystyle \frac{1}{K} \sum_{\theta \in \Omega_K}t_{\mathrm{KS}}^{\theta} = 
        \displaystyle \frac{1}{K} \sum_{\theta \in \Omega_K} 
        \sqrt{\frac{n m}{n + m}} 
        \sup_{u} |F^\theta_n(u) - G^\theta_m(u)|.
\end{equation}
This approach, due to the random directions of projections, might be sensitive to correlations between dimensions, but remains computationally feasible due to the 1D KS tests. 

\subsubsection{Maximum Mean Discrepancy (MMD)}
Introduced in \cite{MMD1, MMD2}, MMD is a statistical measure of the distance between two probability distributions. Its IPM formulation can be found by taking as $\mathcal{F}$ the unit ball in a reproducing kernel Hilbert space (RKHS) \cite{RKHS}. Following Ref. \cite{MMD2}, an unbiased empirical estimate of MMD is given by 
\begin{equation}
\begin{split}
    t_{\mathrm{MMD}} = & \displaystyle \frac{1}{n(n-1)} \sum_{i=1}^n \sum_{j\neq i} k(x_i, x_j) + \frac{1}{m(m-1)} \sum_{i=1}^m \sum_{j\neq i} k(y_i, y_j) \\
    &- \frac{2}{n m} \sum_{i=1}^n \sum_{j=1}^m k(x_i, y_j),
\end{split}
\end{equation}
where $k(x,x')$ is the kernel function defining the RKHS. In Ref. \cite{RKHS_kernel} a fourth-order polynomial kernel was used:
\begin{displaymath}
        k(x,x') = \left( \frac{1}{d} x^T x' + 1 \right)^4.
\end{displaymath}
This kernel is not \emph{characteristic}, meaning that $ k(x,x')$ is not a true metric on the space of probability measures. Specifically, for this kernel, the condition $p = q$ is sufficient for $t_{\mathrm{MMD}}(p,q) = 0$ but not necessary. In fact this kernel cannot distinguish between distributions that differ beyond their fourth moment. Since the polynomial kernel is not characteristic, we say that this instance of MMD is a \emph{pseudo-metric}. In contrast, characteristic kernels, such as Gaussian and Laplacian kernels, are capable of fully distinguishing between different distributions. However, they require tuning hyperparameters (like the kernel bandwidth) based on the data, not necessary for the polynomial kernel. 
From a computational point of view, the MMD test statistic between two datasets of size $n$ scales as $\mathcal{O}(n^2)$ due to the need to store the full kernel matrix $K$, where each element is given by $K_{i,j} = k(x_i,x'_j)$. This makes MMD computationally expensive, especially in large-scale scenarios or when the test needs to be evaluated multiple times. 

\subsubsection{Fréchet Gaussian Distance (FGD)}
The FGD\footnote{In \cite{RKHS_kernel} we take inspiration from the Fréchet Inception Distance (FID) (Ref. \cite{FDI}), which at the time (2023) was the standard metric for evaluation in computer vision (image generators, such as GANs or diffusion-based methods).} (introduced in Ref. \cite{RKHS_kernel}) is a pseudo-metric; specifically, it involves fitting a multivariate Gaussian distribution to the features of interest, and then calculating the Fréchet distance (or 2-Wasserstein distance). The fitted distributions are characterized by means $\mu_1, \mu_2 \in \mathbb{R}^d$ and covariance matrices $\Sigma_1, \Sigma_2 \in \mathbb{R}^{d \times d}$. The FGD between two samples of sizes $n$ and $m$ is given by: 
\begin{equation}
    \mathrm{FGD}_{n,m} = \displaystyle \sum_{I=1}^d (\mu^I_{1,n} - \mu^I_{2,m})^2 
        + \mathrm{tr}\!\left( \Sigma_{1,n} + \Sigma_{2,m} 
        - 2 \sqrt{\Sigma_{1,n}\Sigma_{2,m}} \right), 
\end{equation}
where $\mu^I_{1,n}$ and $\mu^I_{2,m}$ represent the $I$-th components of the sample means $\mu_{1,n}$ and $\mu_{2,m}$, respectively. 

As pointed out in Ref. \cite{FID_bias}, this distance happens to be biased when computed on finite samples. To mitigate this bias, an unbiased asymptotic extrapolation can be introduced, as proposed in Ref. \cite{Unbiased_FID}. This asymptotic value, denoted as 
\begin{equation}
    t_{\mathrm{FGD}} \vcentcolon = FGD_{\infty} = \lim_{n,m \to \infty} FGD_{m,n}
\end{equation}
is estimated by fitting a linear model to FGD values computed at different finite sample sizes. In the following, we refer to $\text{FGD}_{\infty}$ simply as FGD.

\subsubsection{Likelihood-ratio}
Detecting deviations from a reference model can be framed as a goodness-of-fit test between two competing statistical models. For simple hypotheses, the Neyman-Pearson lemma shows that the most powerful test in this scenario is the likelihood-ratio test \cite{NP_lemma}. The test statistic for the likelihood ratio will be found by following the approach of Ref.~\cite{ref_the_ref}. 

The likelihood function for the datasets $X$ and $Y$ under the null hypothesis (where both samples follow the reference distribution $p$) is written as
\begin{equation}
    \mathcal{L}_{H_0} = \displaystyle \prod_{x \in X}p(x) \prod_{y \in Y} p(y).
\end{equation}
Under the alternative hypothesis (where the sample $Y$ follows a different distribution $q$), the likelihood is: 
\begin{equation}
    \mathcal{L}_{H_1} = \displaystyle \prod_{x \in X} p(x) \prod_{y \in Y} q(y).
\end{equation}
The ratio of the likelihood under the null and alternative hypotheses is then given by 
\begin{equation}
    \Lambda = \frac{\mathcal{L}_{H_0}}{\mathcal{L}_{H_1}} = \prod_{y \in Y} \frac{p(y)}{q(y)}, 
\end{equation}
The test statistic for the LLR test is then defined as: 
\begin{equation}
    t_{\mathrm{LLR}} = -2 \log \Lambda.
\end{equation}
Notice how the test statistic for the likelihood ratio depends explicitly on both the distribution $p$ and $q$. Consequently, it can be used only if an analytic expression for the PDFs is available. 

\chapter{Calorimeter Physics and Detector Principles} \label{chap:calo}
In this chapter, we present an introduction to HEP experiments, specifically at LHC, emphasizing the central role played by calorimetry. A major part of the discussion has been inspired by Ref.~\cite{calorimetry1,calorimetry2}.

Calorimeters are one of the fundamental components of high-energy particle experiments. Originally, a \emph{calorimeter} is an instrument used to measure the heat produced in chemical reactions or physical transformations. For instance, the calories in food are measured by burning a sample inside what is called a \emph{bomb calorimeter}.
In this context, calorimetry (the science of calorimeters) is explored in its application to High Energy Physics (HEP), which is quite different from its applications in other sciences. 
In HEP, the purpose is to measure the energy of the particles resulting from a collision through the total absorption of the particle energy in a bulk of material, followed by the measurement of the deposited energy. The structure and working principles of calorimeters in HEP heavily depend on the specific particle for which they are designed. As will be discussed later in the text, the processes used to measure the energy of leptons and hadrons are quite different; thus, the calorimeter structure needs to address these differences. Before discussing the mechanism by which a calorimeter can measure the energy of incoming particles, we will discuss how an experiment at the LHC works. At the  LHC (Large Hadron Collider), very energetic protons are made to collide, and the resulting products are analyzed. The biggest detectors at the LHC have similar fundamental structures, and a significant part of the detector is occupied by calorimeters. Take, for example, the ATLAS detector~\cite{atlas2008overview, atlas-detector-tech}, which is the largest detector ever built at a collider; it is 46 meters long, 25 meters in diameter, and weighs $7000$ tons (similar to the weight of the Eiffel Tower). It is designed in concentric layers centered around the interaction point, with each layer specialized in different measurements. It can be divided into four major systems, each made up of different layers: 
\begin{itemize}
    \item Inner detector: begins a few centimeters from the proton beam to a radius of $1.2$ meters. Its basic function is to track charged particles by detecting their interactions with matter in discrete points. It has 3 sub-layers, each specialized in a specific task. The inner detector is surrounded by a solenoidal magnet, which is fundamental for measuring the momentum of charged particles by assessing their curvature in the magnetic field.
    \item Calorimeters, situated directly outside the solenoidal magnet, are responsible for measuring the energy of the incident particles. They are arranged in two sub-systems: the inner electromagnetic calorimeter and the outer hadron calorimeter. Calorimeters are usually built with a structure of \emph{volumetric pixels}, also called \emph{voxels}, allowing us to infer the incident particle direction and thus achieve better event reconstruction. 
  \item Muon spectrometer: the cross-section for the interaction of muons is too small for them to be absorbed. So muons need a dedicated system that, in this specific case, is extremely large; it starts at $4.25$ meters from the interaction point and extends to the full length of the radii (approximately $11$ meters). The large dimension of the muon section is necessary to precisely measure the momentum of muons by assessing the curvature in high magnetic fields. 
  \item Magnetic system: as already mentioned, there is a solenoid surrounding the inner detector that produces a $2$ Tesla magnetic field, allowing it to curve even very energetic particles. There is another superconducting toroidal magnet in the muon section that varies between $2$ and $8$ Tesla. A toroidal magnet is chosen because a solenoidal magnet of the requested size would have prohibitive costs. 
\end{itemize}

\begin{figure}
    \centering
    \includegraphics[width=0.8\linewidth]{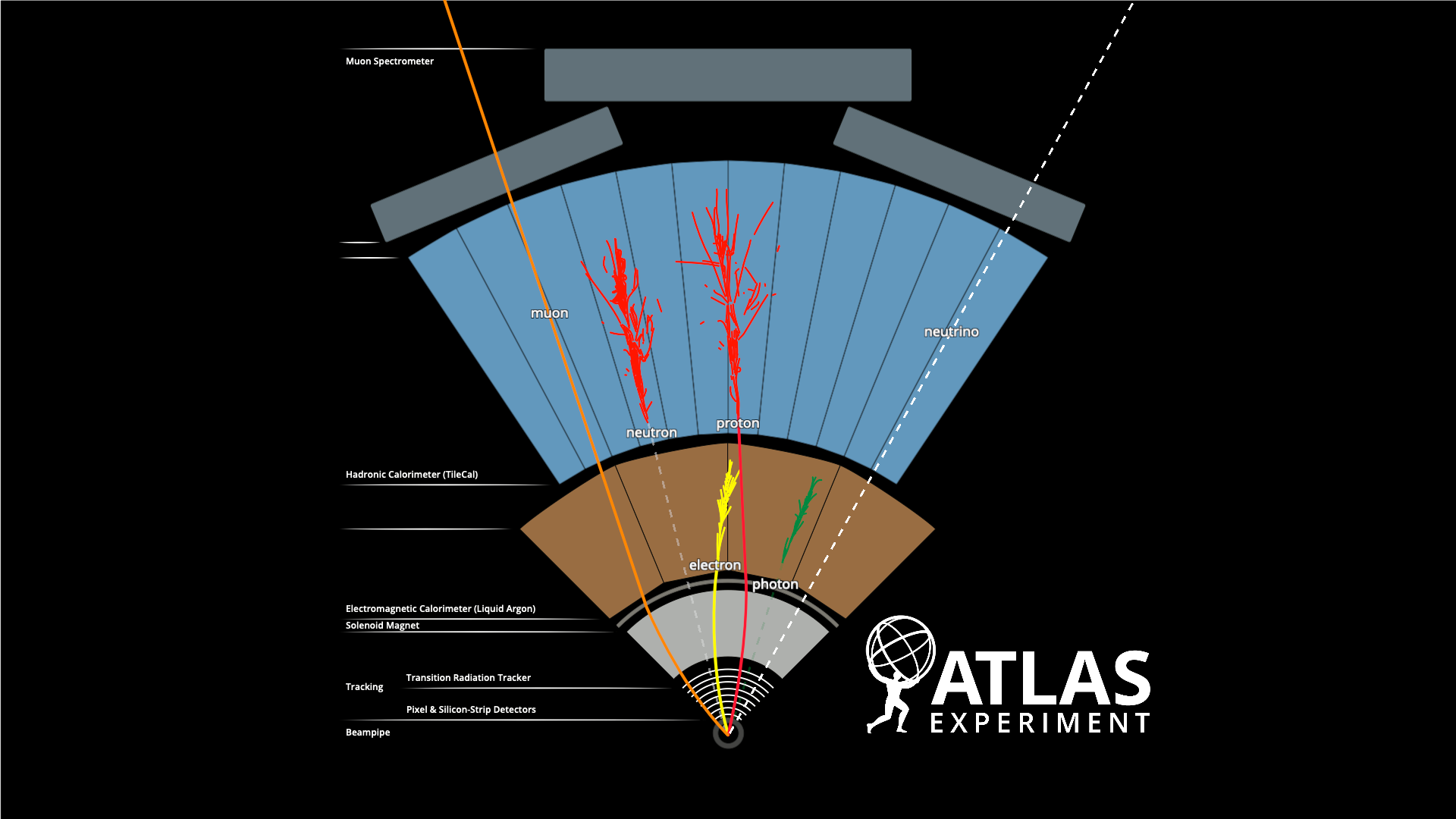}
    \caption{Schematic cross-section of the ATLAS detector (not to scale). From the beam pipe outward: silicon pixel and strip trackers and the Transition Radiation Tracker inside a solenoid; the liquid-argon electromagnetic calorimeter; the TileCal hadronic calorimeter; and the outer muon spectrometer. Example particle signatures are shown: an electron and a photon producing electromagnetic showers, hadrons showering in the hadronic calorimeter, a muon traversing to the muon system, and a neutrino leaving no visible trace (seen as missing transverse momentum). Adapted from ATLAS public outreach material.}
    \label{fig:atlas_detector}
\end{figure}

\section{Electromagnetic calorimetry}
The discussion about electromagnetic calorimetry must start with how electrons and photons can lose energy in matter. It is important to recall that the discussion is based on HEP applications, so the interactions described are prevalent at high energies. 
\subsection{Interaction with matter}
\subsubsection*{Electrons} lose energy in matter almost exclusively through \emph{bremsstrahlung} (from German \emph{bremsen} ``to brake'' and \emph{Strahlung} ``radiation''), i.e., radiation emitted from an accelerated charged particle. In a medium, an electron is deflected by other charged particles, typically by the electric fields of atomic nuclei. The radiative energy-loss rate can be written as:
\begin{displaymath}
  \left( -\frac{\text{d} E}{ \text{d} x} \right)_{\!rad} = 4\alpha N_A \frac{Z^2}{A} \, r_e^2 \, E \, \log\!\left(\frac{183}{Z^{\frac{1}{3}}}\right)
\end{displaymath}
conveniently rewritten as:
\begin{displaymath}
  \left( -\frac{\text{d} E}{ \text{d} x} \right)_{\!rad} = \frac{E}{X_0} \quad \text{with} \quad X_0 =  \frac{A}{4\alpha N_A Z^2 r_e^2 \, \log\!\left(\frac{183}{Z^{\frac{1}{3}}}\right)}
\end{displaymath}
where $X_0$ is the \textbf{Radiation Length}, defined as the length over which an electron reduces its energy by $1/e$ through radiation emission. 

\subsubsection*{Photons} On the other hand, at high energies, photons lose energy predominantly through \textbf{pair production}, i.e., a photon in the field of an atomic nucleus converts into an electron-positron pair. The mean free path of a photon before it converts is approximately: 
\begin{displaymath}
    \lambda_{pair} \approx \frac{9}{7}X_0
\end{displaymath}
where $X_0$ is again the radiation length.
\begin{figure}[h]
    \centering
    \begin{subfigure}{0.475\textwidth}
        \centering
        \includegraphics[width=\linewidth]{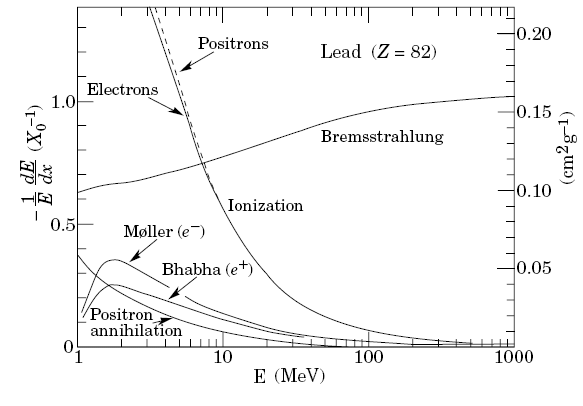} 
        \label{fig:first}
    \end{subfigure}
    \hfill
    \begin{subfigure}{0.425\textwidth}
        \centering
        \includegraphics[width=\linewidth]{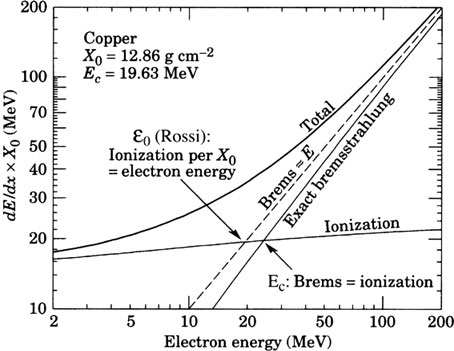} 
        \label{fig:second}
    \end{subfigure}

    \caption{Energy loss in lead (right) and copper (left). }
    \label{fig:side-by-side}
\end{figure}
\subsection{Electron-Photon cascades}
Based on the interactions with matter discussed above, an electron (or a photon) entering a medium can give rise to an electromagnetic shower~\cite{calorimetry2}: a high-energy electron entering a bulk of material emits a photon via bremsstrahlung. After roughly one radiation length, the photon converts into a pair; the electron and positron subsequently each emit a photon after about one radiation length, and the process continues. 
This is a simple explanation of how a single particle, either an electron or a photon, could create a shower of secondary particles. To analyze the shower development, we introduce a variable $t = \frac{x}{X_0}$ that is the distance normalized to radiation length. Let us assume that the energy is symmetrically shared among the particles at each step. The number of shower particles (electrons, positrons, or photons) at depth $t$ is
\begin{displaymath}
    N(t) = 2^t
\end{displaymath}
and the energy of each individual particle is given by
\begin{displaymath}
    E(t) = E_0 \cdot 2^{-t}
\end{displaymath}
The multiplication continues as long as $E_0/N > E_c$, after which the particles are absorbed through ionization for electrons and through Compton and photoelectric effects for photons. 
The position of the maximum of the shower is reached for
\begin{displaymath}
    E_c = E_0 \cdot e^{-t_{max}}
\end{displaymath}
and by solving for $t_{max}$
\begin{displaymath}
    t_{max} = \frac{\log(E_0/E_c)}{\log(2)}
\end{displaymath}
After the shower maximum, electrons and positrons are absorbed in a layer of about one radiation length, while photons  of the same energy can penetrate much longer distances. To absorb 95\% of the produced photons in a shower, one needs an additional $7-9 X_0$, which implies that the thickness of a calorimeter with high shower containment should be at least $14-16 X_0$.
Being a simple model to describe electromagnetic showers, it nevertheless captures the most important characteristics:
\begin{itemize}
    \item To absorb most of the incident energy, the calorimeter should be at least $14-16 X_0$
    \item The thickness of the calorimeter should increase logarithmically with the energy
  \item Leakage from the sides (\emph{lateral leakage}) and from the back (\emph{rear leakage}) is responsible for most of the missing energy.
\end{itemize}
At present, an accurate description of electromagnetic showers is obtained through Monte Carlo simulations.

Notice that the angles of emission for bremsstrahlung and pair production are very narrow. This implies that most of the lateral width of the shower is determined primarily by multiple scattering. It can be characterized by the \emph{Moli\`ere radius}
\begin{displaymath}
    R_M = \frac{21 \text{MeV}}{E_c} X_0 \text{\, \{g/$\text{cm}^2$\}}
\end{displaymath}

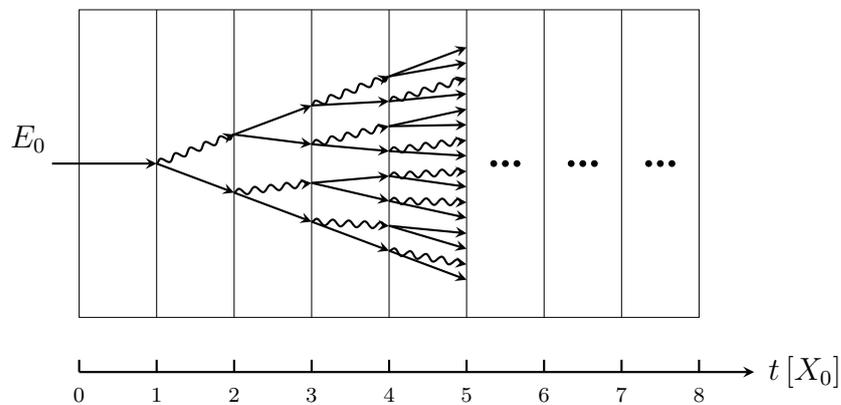
\begin{figure}[ht]
\centering
\begin{tikzpicture}[>=stealth, x=0.75\linewidth/12, y=0.75\linewidth/12]

\def\W{9}            
\def\H{4.5}             
\def\ninternal{7}     
\def\gens{4}          
\def\xoffset{1}       
\def\ycenter{0.5*\H}  
\def\spread{0.85}     
\def\seglen{6pt}      
\def\amp{1.2pt}       

\def\dotrad{0.045}    
\def\doty{\ycenter}   

\def\axisBelow{0.8}        
\def\ticklen{0.18}         
\def\axisRightExtra{0.8}   

\tikzset{
  electron/.style = {line width=0.8pt},
  photon/.style   = {decorate, decoration={snake, amplitude=\amp, segment length=\seglen}, line width=0.8pt},
}

\pgfmathsetmacro{\step}{\W/(\ninternal+1)}

\draw[line width=0.2pt] (0,0) rectangle (\W,\H);
\foreach \i in {1,...,\ninternal} {
  \draw[line width=0.2pt] ({\i*\step},0) -- ++(0,\H);
}

\pgfmathsetmacro{\xzero}{0.25*\step} 

\pgfmathsetmacro{\halfHmax}{0.5*\H*\spread}
\pgfmathsetmacro{\dhalf}{\halfHmax/(\gens+0.5)}

\foreach \g in {1,...,\gens} {
  \pgfmathtruncatemacro{\ng}{2^\g}
  \pgfmathsetmacro{\hh}{\dhalf*\g}
  \pgfmathsetmacro{\ymin}{\ycenter - \hh}
  \pgfmathsetmacro{\ymax}{\ycenter + \hh}
  \pgfmathsetmacro{\xg}{(\g+\xoffset)*\step}
  \foreach \j in {1,...,\ng} {
    \pgfmathsetmacro{\t}{(\ng==1) ? 0.5 : (\j-1)/(\ng-1)}
    \pgfmathsetmacro{\yj}{(1-\t)*\ymin + \t*\ymax}
    \coordinate (g\g-\j) at (\xg,\yj);
  }
}

\def\electronType{electron}
\def\photonType{photon}
\newcommand{\settype}[3]{\expandafter\gdef\csname ndtypeg#1j#2\endcsname{#3}}
\newcommand{\gettype}[2]{\csname ndtypeg#1j#2\endcsname}

\settype{0}{1}{electron}

\pgfmathsetmacro{\xfirst}{(\xoffset)*\step}

\draw[electron,->] ({\xzero-0.6*\step},\ycenter) -- (\xfirst,\ycenter) node[above left,xshift=-37pt,yshift=0pt] {$E_0$};

\foreach \g in {1,...,\gens} {
  \pgfmathtruncatemacro{\gprev}{\g-1}
  \pgfmathtruncatemacro{\ngprev}{2^(\g-1)}
  \foreach \j in {1,...,\ngprev} {
    \pgfmathtruncatemacro{\cA}{2*\j-1}
    \pgfmathtruncatemacro{\cB}{2*\j}

    \ifnum\g=1
      \coordinate (parent) at (\xfirst,\ycenter);
    \else
      \coordinate (parent) at (g\gprev-\j);
    \fi

    \expandafter\let\expandafter\ptype\csname ndtypeg\gprev j\j\endcsname

    \ifx\ptype\photonType
      \draw[electron,->] (parent) -- (g\g-\cA);
      \draw[electron,->] (parent) -- (g\g-\cB);
      \settype{\g}{\cA}{electron}
      \settype{\g}{\cB}{electron}
    \else
      \draw[electron,->] (parent) -- (g\g-\cA);
      \draw[photon,  ->]   (parent) -- (g\g-\cB);
      \settype{\g}{\cA}{electron}
      \settype{\g}{\cB}{photon}
    \fi
  }
}

\pgfmathtruncatemacro{\lastline}{\gens+\xoffset} 
\foreach \i in {\lastline,...,\ninternal} {
  \pgfmathsetmacro{\xL}{\i*\step}
  \pgfmathtruncatemacro{\iplus}{\i+1}
  \ifnum\i<\ninternal
    \pgfmathsetmacro{\xR}{\iplus*\step}
  \else
    \pgfmathsetmacro{\xR}{\W} 
  \fi
  \pgfmathsetmacro{\xA}{\xL + 0.35*(\xR-\xL)}
  \pgfmathsetmacro{\xB}{\xL + 0.50*(\xR-\xL)}
  \pgfmathsetmacro{\xC}{\xL + 0.65*(\xR-\xL)}
  \fill (\xA,\doty) circle[radius=\dotrad cm];
  \fill (\xB,\doty) circle[radius=\dotrad cm];
  \fill (\xC,\doty) circle[radius=\dotrad cm];
}

\pgfmathtruncatemacro{\nticks}{\ninternal+1} 

\draw[thick,->] (0,-\axisBelow) -- ({\W+\axisRightExtra}, -\axisBelow);

\foreach \k in {0,...,\nticks} {
  \pgfmathsetmacro{\xx}{\k*\step}
  \draw[thick] (\xx,-\axisBelow) -- ++(0,\ticklen); 
  \node[font=\scriptsize, below=2pt] at (\xx,-\axisBelow) {\k};
}

\node[anchor=west] at ({\W+\axisRightExtra+0.05}, -\axisBelow) {$t\,[X_0]$};

\end{tikzpicture}
\caption{A simplified diagram explaining the process of shower generation. Wavy lines represent photons while straight lines represent electrons. Diagram inspired by Ref.~\cite{calorimetry2}}
\label{fig:cascade-scheme}
\end{figure}

\subsection{Homogeneous calorimeters}
Homogeneous calorimeters are built from a material that combines absorber and detector properties. It means that the entire volume of the calorimeter is sensitive to the deposition of energy. These calorimeters are based on the measurement of scintillation light (scintillation crystals, liquid noble gases), ionization (liquid noble gases), and Cherenkov light (lead glass or heavy transparent crystals).

\begin{itemize}
  \item Crystal calorimeters are based on heavy scintillation crystals, i.e.\ materials that can convert absorbed energy into visible light and transfer this light to an optical receiver for measurements. Scintillation materials can be inorganic crystals, organic compounds, liquids, and gases. The general principle for a material to emit light is the excitation of atoms or bands by energetic particles, which emit photons to return to the ground state. The disadvantage of this type of calorimeter is the high cost of scintillation crystals and limitations in their production. 
  \item Ionization calorimeters are built as an array of ionization chambers immersed in liquid xenon or krypton. 
\end{itemize}

\subsection{Sampling calorimeters}
Sampling calorimeters, on the other hand, are far simpler and more economical. The key idea is to alternate thin layers of counters with layers of absorbers. The shower interacts with the absorber, and the products are measured by the counters. In this type of calorimeter, only a sample of energy is measured; hence the name \emph{sampling calorimeters}. The principal disadvantage of sampling calorimeters is that, in addition to the general leakage, the energy resolution is also affected by sampling fluctuations. The principal advantage is the significantly lower cost and simplicity. 

\section{Hadron calorimetry}
In principle, a hadron calorimeter works along the same lines as the electromagnetic ones; the principal difference is that, in most detector materials, the shower length is much larger. The longitudinal development is determined by the average interaction length defined by: 
\begin{displaymath}
  \lambda_I \approx 35\,A^{1/3}\,\mathrm{g/cm^2}
\end{displaymath}
For this reason, hadron calorimeters have to be much larger than the electromagnetic ones. 
\subsection{Hadronic showers}
Due to the \emph{strong interaction} in hadrons, in addition to the electromagnetic (EM) interaction, hadronic showers are much more complex than their EM counterparts. In fact, the interaction of hadrons with atomic nuclei produces many particles through various processes. 

There are two main components:
\begin{itemize}
  \item Hadronic component: populated by charged pions, kaons, protons, neutrons, etc. 
  \item EM component: primarily due to \emph{neutral pions}, which, on average, account for about one third of the total pions and decay into two photons, generating EM showers. Electrons, positrons, and photons can also appear in final states and will generate EM showers.
\end{itemize}
A large portion of the energy loss is \emph{invisible} to the calorimeter, including the energy used to release protons and neutrons from nuclei, as well as nuclear recoil. A significant fraction is also lost to evaporation neutrons. Given the complexity of the shower process, it does not possess a profile that can be parametrized. In fact, the first interaction of the incident particle will determine the EM fraction, which is not known a priori. 

Notice that in the hadronic process there is a large transfer of transverse momentum, making the hadronic shower much wider.

\section{Limitations}
In real calorimeters, the response drifts with time because the active scintillator tiles and wavelength-shifting (WLS) fibers accumulate \emph{total ionizing dose} (TID). As an example, we report the measured aging in the ATLAS Tile Calorimeter (TileCal). While the aging measured during Run~2 did not have a large impact on the measurements~\cite{calorimeter_deterioration}, here we consider the near-future HL-LHC scenario.  

\begin{figure}[t]
  \centering
  \includegraphics[width=0.8\linewidth]{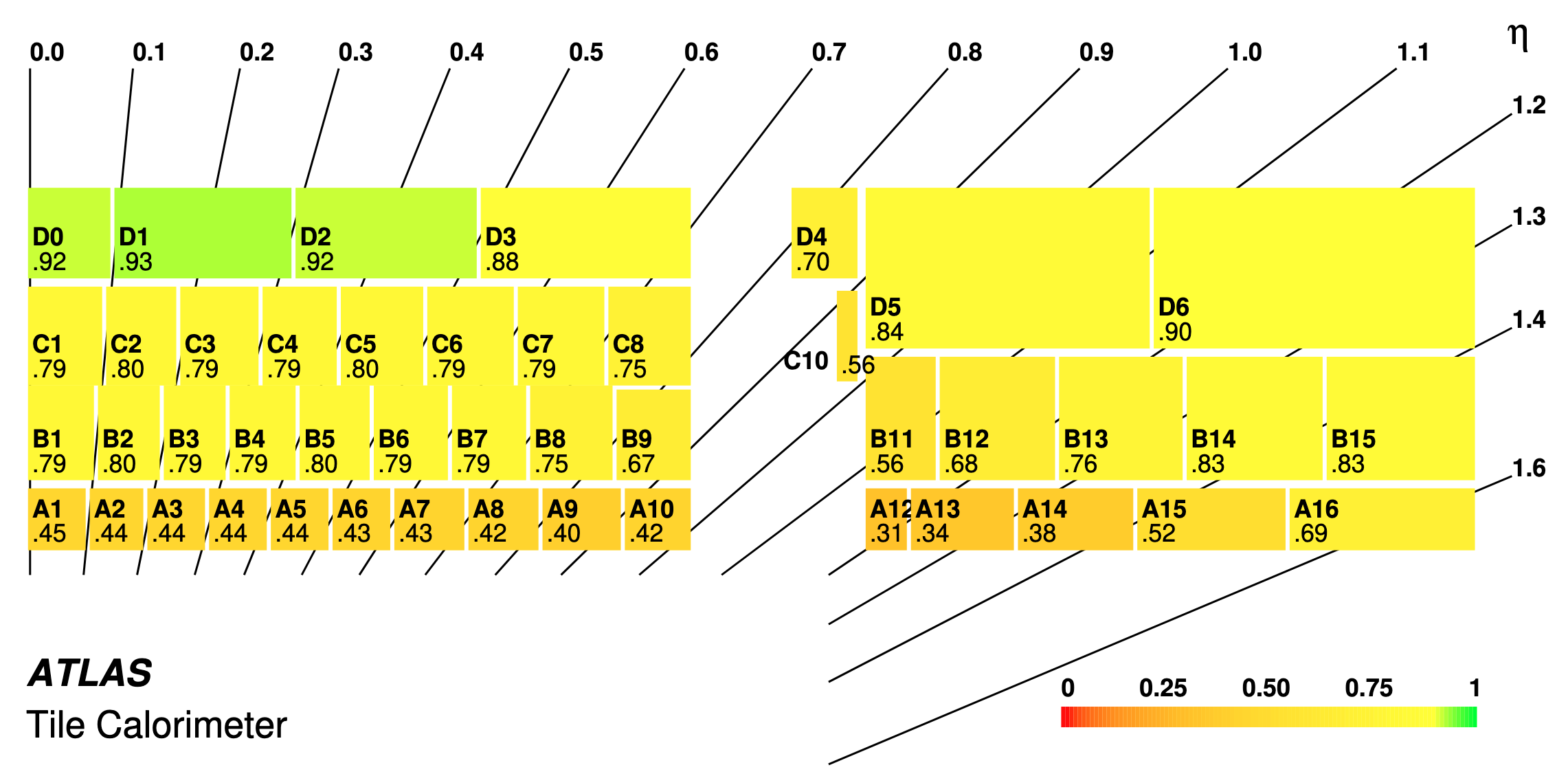}
  \caption{Expected residual light yield \(I/I_{0}\) at the end of HL-LHC (assumed \(4000~\mathrm{fb}^{-1}\) and a dose rate \(7\times\) Run~2) for \emph{each} TileCal barrel cell. 
  Layer letters encode the radial depth (\textbf{A} inner, \textbf{B/BC} middle, \textbf{D} outer); indices encode the \(\eta\)-bin. 
  Shaded ranges in \cite{calorimeter_deterioration} Table~2 summarize \emph{typical} cells per layer (A: \(48\text{–}60\%\) loss; B/BC: \(17\text{–}25\%\); D: \(7\text{–}16\%\)), while specific worst-case cells (e.g.\ A12/A13 with \(66\text{–}69\%\) loss) are called out explicitly. 
  Figure adapted from \cite[Figure 18]{calorimeter_deterioration}.}
  \label{fig:tile_hllhc_map}
\end{figure}

\begin{figure}[t]
  \centering
  \includegraphics[width=0.8\linewidth]{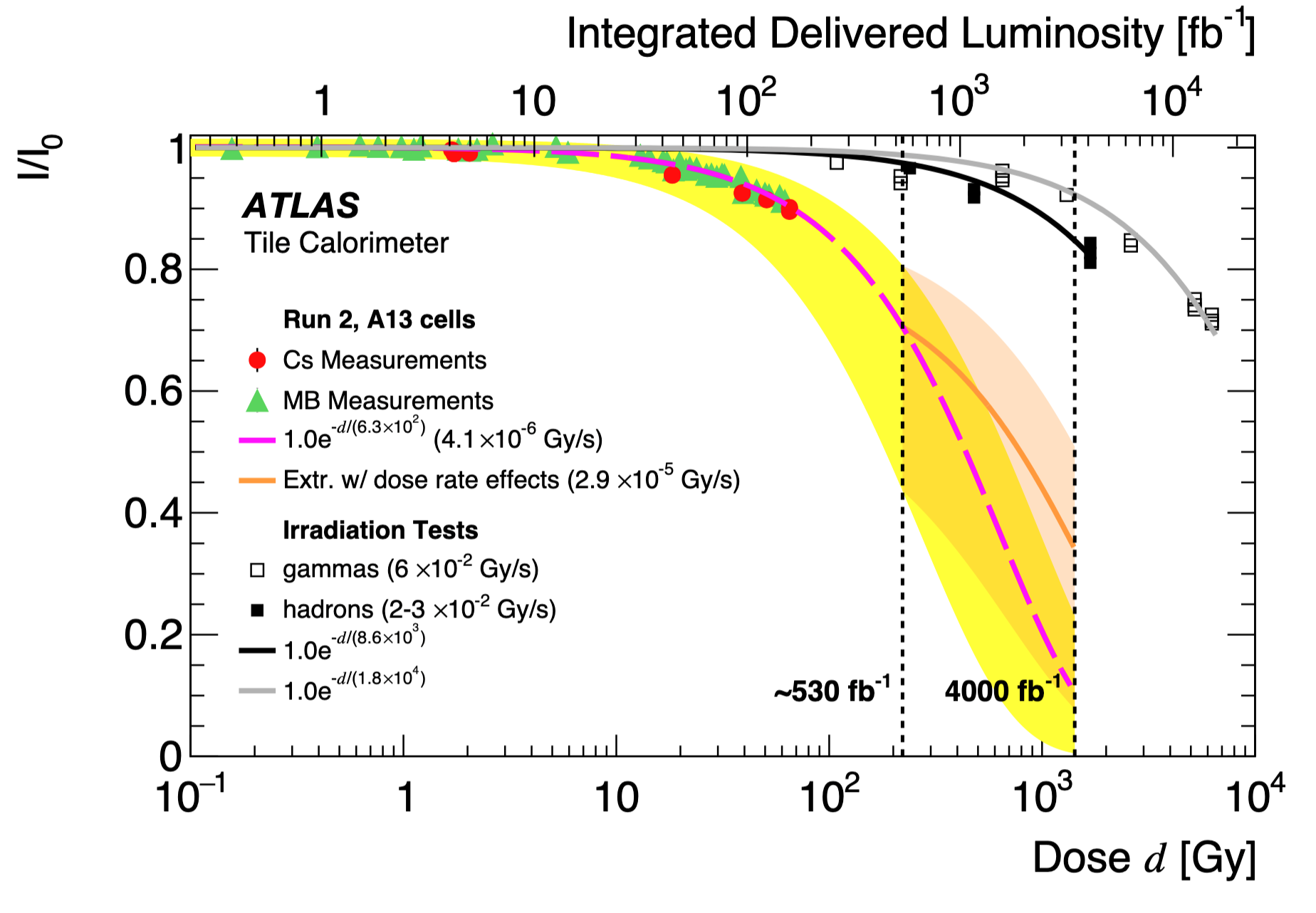}
  \caption{Average relative light yield \(I/I_{0}\) for TileCal A13 cells versus simulated dose \(d\) (bottom axis) and integrated luminosity \(L\) (top axis). Points show in-situ measurements: \({}^{137}\)Cs (dots) and minimum-bias currents (triangles). Vertical lines mark end of Run~3 (\(\sim 530~\mathrm{fb}^{-1}\)) and end of HL-LHC (\(4000~\mathrm{fb}^{-1}\)). The solid curve represents the extrapolation including dose-rate effects expected at HL-LHC, while the dashed line, represents the extrapolation neglecting dose-rate. Image taken from \cite{calorimeter_deterioration}.}
  \label{fig:tile_aging}
\end{figure}

In collider calorimeters, the response drifts with time because the active medium and the optical chain (plastic scintillator tiles, wavelength-shifting fibers, photodetectors) accumulate total ionizing dose, and parts of the front-end age; the practical signatures are a progressive loss of light yield, growing inter-cell non-uniformities, and occasionally, channels that become noisy or fail and must be masked. In the ATLAS Tile Calorimeter (TileCal) this behavior is conveniently monitored in terms of the relative light yield \(I/I_{0}\), and the detector granularity is organized in three radial layers: A (inner), B/BC (middle), D (outer), with cell names such as A13 denoting the layer and the \(\eta\) bin. The per-cell HL-LHC projection in Figure \ref{fig:tile_hllhc_map} makes this structure explicit while showing the spatial pattern of the expected attenuation \cite{calorimeter_deterioration}. During Run~2, the reconstructed energy scale was kept stable at the percent level by the laser, \({}^{137}\)Cs, and minimum-bias calibration systems, while the most irradiated A-layer cells exhibited an end-of-run light-yield loss of about \(3\text{–}10\%\) and the other barrel layers showed no measurable loss within the \(\sim 1\%\) sensitivity; the detector contributed \(99.65\%\) efficient high-quality data and fewer than \(1.1\%\) of cells were non-operational at the end of each data-taking year \cite{calorimeter_deterioration2}. Aging depends on both total dose and dose rate, therefore for a fixed integrated luminosity higher instantaneous luminosity increases the effective degradation, a behavior also observed in plastic scintillators of the CMS hadron endcap calorimeter \cite{calorimeter_deterioration_CMS,calorimeter_deterioration}. Extrapolating the Run~2 model to the HL-LHC scenario (\(L\simeq 4000~\mathrm{fb}^{-1}\), instantaneous luminosity about seven times Run~2), the typical A-layer is projected to lose roughly \(48\text{–}60\%\) of light response by the end of the program, with the most exposed A12/A13 cells reaching about \(66\text{–}69\%\) loss, while typical B/BC and D layers are expected within \(\sim 17\text{–}25\%\) and \(\sim 7\text{–}16\%\) loss, respectively, and cells with more than \(50\%\) expected loss carry a relative uncertainty of order \(50\%\) \cite{calorimeter_deterioration}. 
In terms of performance, response non-uniformities induced by aging contribute to the constant term in the usual calorimeter energy-resolution parametrization \(\sigma_E/E = a/\sqrt{E}\oplus b \oplus c/E\) \cite{PDG_Detectors}, and if not fully corrected they increase \(b\) and bias inter-calibration across cells, although in TileCal Run~2 measurements kept non-uniformity within specifications and simulations indicate a weak dependence of the muon response even for large light-yield reductions \cite{calorimeter_deterioration,calorimeter_deterioration2}, the HL-LHC projections in Figures \ref{fig:tile_hllhc_map} and \ref{fig:tile_aging} point to substantial attenuation in the inner layers together with a higher probability of disabled or noisy channels, which can degrade hadronic energy and position resolution, distort longitudinal and transverse shower shapes, and impair cluster splitting and jet or photon identification if not compensated, these considerations motivate generative super-resolution approaches that aim to reconstruct fine-grained energy patterns in regions with reduced or missing response and to regularize spatial non-uniformity, thereby stabilizing the effective constant term under HL-LHC irradiation.

\section{Calorimeter superresolution}
\label{sec:calo_superresolution}
We present our task formulation by starting with the notation: we denote by $\vec{E}_{\text{coarse}}$ the energy deposits measured on a coarse segmentation (larger cells or aggregated readout) and by $\vec{E}_{\text{fine}}$ the corresponding deposits on a finer segmentation (smaller cells resolving more structure). Then, the superresolution problem is the task of inferring the fine-grained shower given the coarse observation, which we formulate as learning the conditional distribution
\begin{equation}
p\left(\vec{E}_{\text{fine}} \mid \vec{E}_{\text{coarse}}\right).
\label{eq:sr_cond_pdf_general}
\end{equation}
This formulation is intentionally probabilistic: multiple fine-grained patterns can be compatible with the same coarse measurement, so a calibrated model should represent a distribution over plausible $\vec{E}_{\text{fine}}$ rather than a single deterministic map. We note that \eqref{eq:sr_cond_pdf_general} is one of the valid formulations but not the only one. In particular, when parts of the calorimeter are unavailable (dead or disabled channels) and only a subset of fine cells is recorded, we could pose the problem as \emph{filling in missing values}. Let $\vec{m}\in\{0,1\}^{N_{\text{fine}}}$ be a binary mask over the fine segmentation, with $m_j=1$ for observed cells and $m_j=0$ for missing ones; write $\vec{E}_{\text{fine}}=(\vec{E}_{\text{obs}},\vec{E}_{\text{miss}})$ accordingly, and learn the distribution
\begin{equation}
  p\!\left(\vec{E}_{\text{miss}} \,\middle|\, \vec{E}_{\text{obs}},\, \vec{E}_{\text{coarse}},\, \vec{m}\right).
  \label{eq:sr_inpainting}
\end{equation}
This way, the missing fine-cell energies are reconstructed consistently with the available fine measurements.
\chapter{Implementation and Experimental Setup} \label{chap:implementation}

\section{Dataset Description} \label{sec:dataset}
\subsection{The dataset}
The dataset is taken from the \emph{Fast Calorimeter Simulation Challenge} (\emph{CaloChallenge}). More specifically, this work is based on Dataset~2 of the \emph{CaloChallenge}, which is illustrated in detail below. The goal of the \emph{CaloChallenge} is to encourage the development of fast and high-fidelity calorimeter shower surrogate models. 

\begin{figure}[h]
    \centering
    \includegraphics[width=0.85\linewidth]{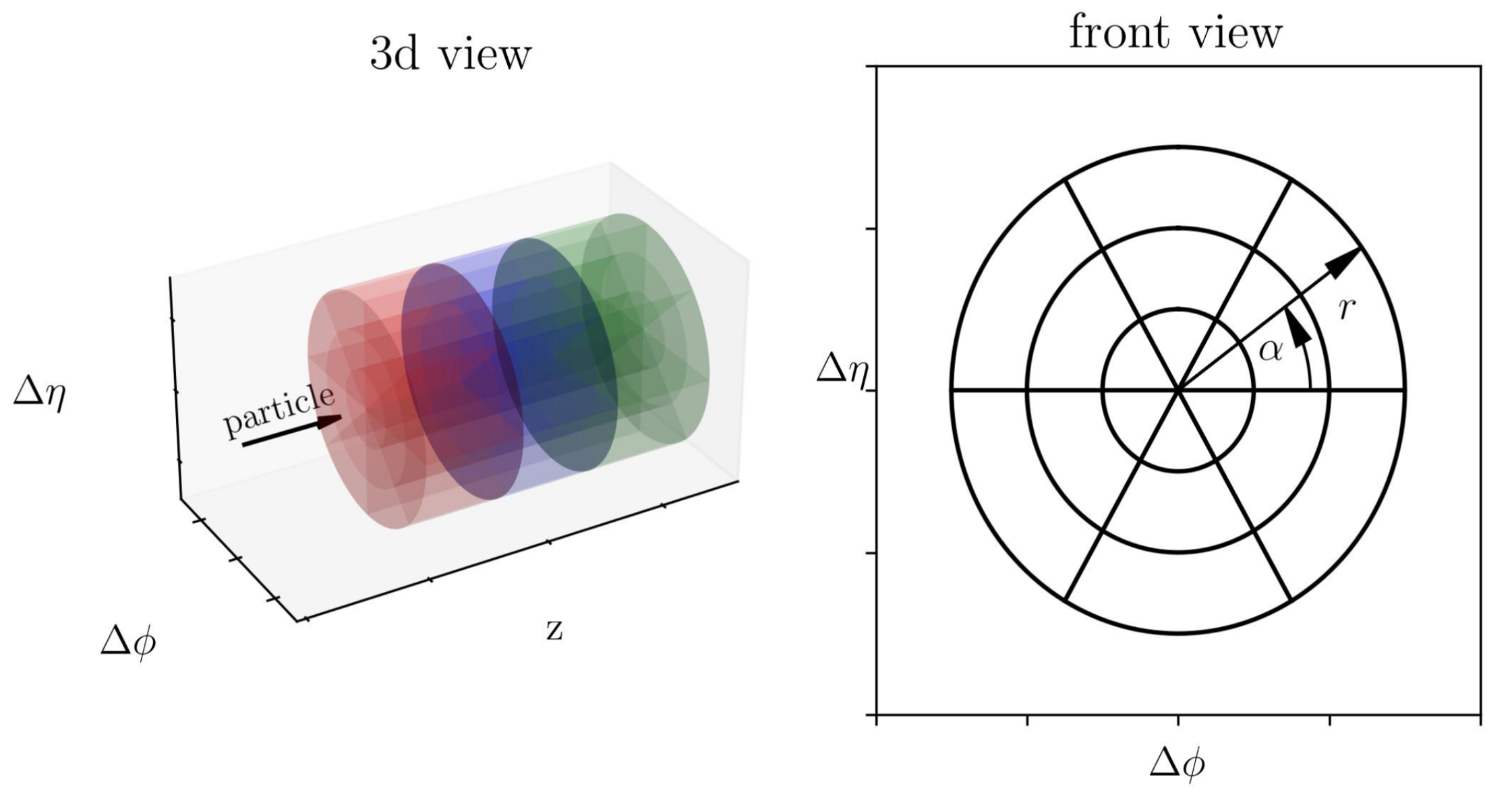}
    \caption{Schematic view of the voxelization of the detector. In the specific example the detector has 3 layers, 3 radial bins, and 6 angular bins.}
    \label{fig:calochallenge_detector}
\end{figure}

The detector geometry is made up of concentric cylinders with particles propagating along the \emph{z} direction. It is segmented into layers along the z axis. Each layer is sub-segmented into radial bins and angular bins. To clarify the detector geometry, Figure~\ref{fig:calochallenge_detector} shows the front view (right) illustrating the layer segmentation and a 3D view of the detector (left) showing the overall geometry. We refer to each individual segment of the detector as a \textbf{voxel} (volumetric pixel). 

\begin{figure}
    \centering
    \includegraphics[width=0.9\linewidth]{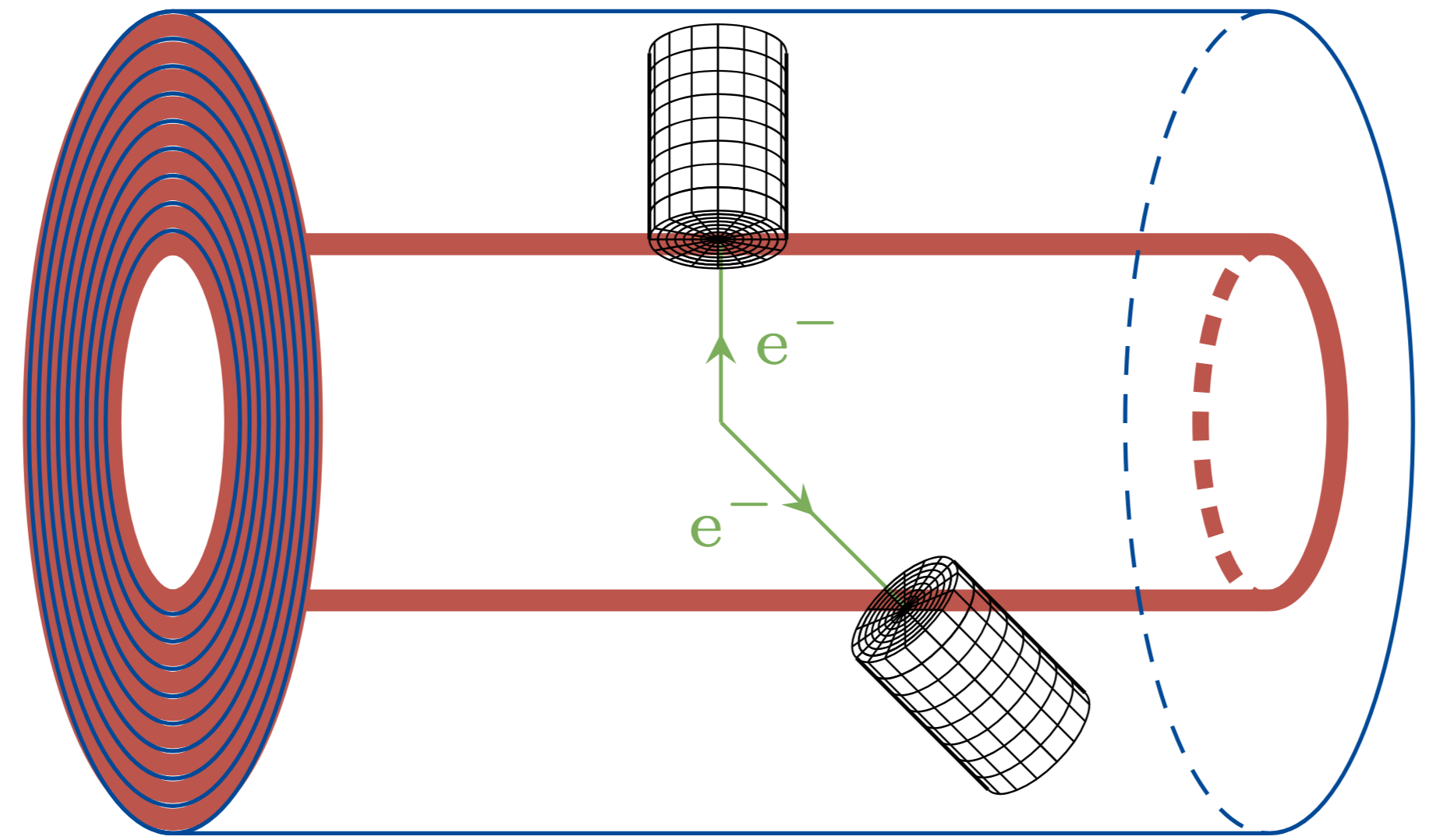}
    \caption{A schematic for the detector of the Par04 example. It consists of layers of absorber (red) and active material (blue). The energy deposit is recorded in a cylindrical readout (black).}
    \label{fig:Par04}
\end{figure}

Getting more in depth about the specific structure of \emph{Dataset~2}, it was simulated using the Par04 example of \textsc{Geant4}~\cite{geant4_par04_example}. The Par04 example implements an idealized calorimeter consisting of concentric cylinders of alternating absorber and active material (sampling calorimeter). The calorimeter has 90 layers, each composed of $1.4$\,mm of tungsten (W) as absorber and $0.3$\,mm of silicon (Si) as active material. A schematic of its layout is presented in Figure~\ref{fig:Par04}.
Particle showers are generated by electrons entering the detector perpendicular to the cylinder's axis, as shown by the top electron in Figure~\ref{fig:Par04}. Showers with axes at different incident angles are beyond the scope of the challenge.

As the Par04 example only writes out the energy deposited in the active material, it must be corrected for the deposit in the absorber. From the simulation, a simple scaling factor has been derived: $f = 1/0.033$, constant for all energies and cells of the detector. This means that, on average, only $3.3\%$ of the incident energy is registered in the detector. For \emph{Dataset~2} a threshold was fixed at a low value of $15.15$\,keV, below which cell energy is not registered. The same cutoff will be applied to the generated samples. 

The calorimeter geometry for \emph{Dataset~2} has 45 layers, 16 angular bins, and 9 radial bins, for a total of 6480 voxels. Therefore, the size of the readout cells is $\Delta r \times \Delta \phi \times \Delta z$, with $\Delta z = 3.4$\,mm, $\Delta \phi = 2\pi/16 \approx 0.393$\,rad, and, considering only the absorber's Moli\`ere radius, $\Delta r = 4.65\,\mathrm{mm}/9.327\,\mathrm{mm} = 0.5\,R_M$. Taking into account only the radiation length for the absorber ($X_0(\text{W}) = 3.504$\,mm), the size along the $z$-axis is approximately $\Delta z = 2 \cdot 1.4\,\mathrm{mm}/3.504\,\mathrm{mm} = 0.8\,X_0$.

Dataset~2 consists of two \texttt{.hdf5} files (built using Python's \texttt{h5py} package), each containing 100k showers (or events), with energies sampled from a log-uniform distribution ranging from 1\,GeV to 1\,TeV; one is used for training, the other for evaluation.   
Within each file there are two separate hdf5-datasets: \texttt{incident\_energies} has shape \texttt{(num\_events, 1)} and contains the energy of the incident particles in MeV; \texttt{showers} has shape \texttt{(num\_events, num\_voxels)} and stores the energy deposited in each voxel, flattened. The mapping of array index to voxel location follows the order (radial bins, angular bins, layer), so the first entries correspond to the radial bins of the first angular slice in the first layer. Then, the radial bins of the next angular slice of the first layer follow, and so on.

\section{Building the datasets}
In the previous section, the dataset was described in detail. In this section, we describe the preprocessing steps and the formulation of the problem to solve.
\subsection{The voxelization} \label{sec:voxellization}
\begin{figure}[h]
    \centering
    \includegraphics[width=0.7\linewidth]{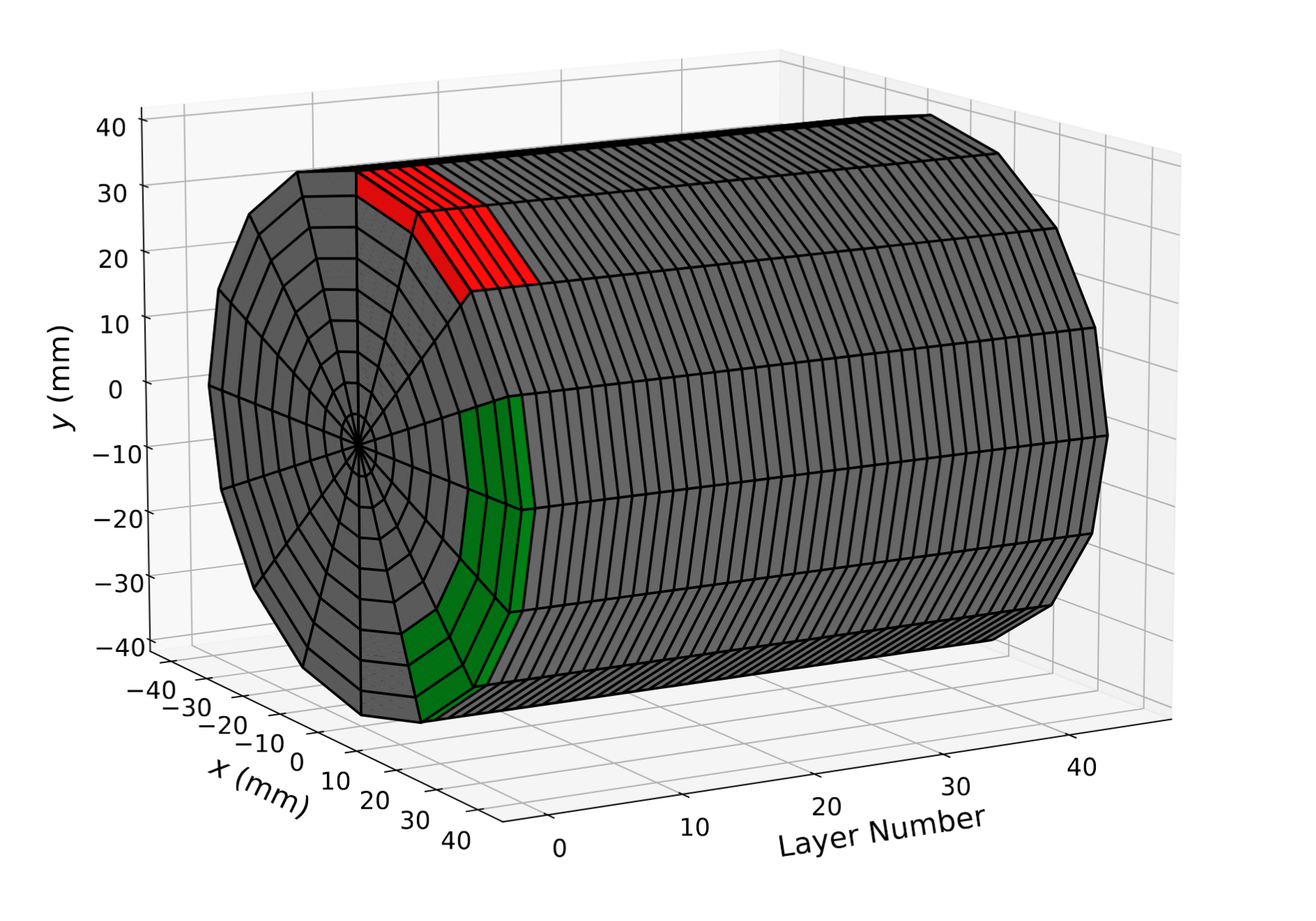}
    \caption{3D diagram from Ref.~\cite{supercalo} illustrating the calorimeter voxel geometry for Dataset 2 (fine representation) in Grey and the two choices of voxellization introduced in Section \ref{sec:voxellization}. In Red is shown the choice \texttt{A} and in Green the choice \texttt{B}.}
    \label{fig:voxellization}
\end{figure}

To replicate the results obtained in SuperCalo~\cite{supercalo}, we need to create the coarse voxels from the full-resolution showers. This is done by grouping neighboring voxels; however, there are many possible choices for grouping. Reference \cite{supercalo} considers two choices of coarse voxelization:
\begin{itemize}
    \item \textbf{Choice A:} 1 coarse voxel $= 1r \times 2\alpha \times 5z$. \\
    \\
    \noindent 
    The resulting coarse voxel geometry has nine layers along the $z$-axis, nine radial bins, and eight angular bins, for a total of 648 coarse voxels. This choice has the advantage of allowing the model to learn the inter-correlations among the fine voxels within each coarse voxel. The downside is that the correlation between fine voxels in neighboring coarse voxels along the radial and angular directions is lost.
    \item \textbf{Choice B:} 1 coarse voxel $= 3r \times 4\alpha \times 1z$. \\
    \\
    \noindent
    With this choice of coarse voxelization, the detector geometry has 45 layers, three radial bins, and four angular bins, for a total of 540 coarse voxels. This choice ensures that the total deposited energy in each fine layer is learned correctly. However, it does not guarantee the correct inter-layer correlations between fine voxels.
\end{itemize}
Figure~\ref{fig:voxellization} shows a 3D diagram of the calorimeter geometry with highlighted the voxellization choices. In red choice \texttt{A}, in green choice \texttt{B}.
The goal is now to recover the fine voxel energies $\hat{e}_{\text{fine},i}$ given the coarse voxel energy $E_{\text{coarse},i}$. We refer to \texttt{SuperCaloA} for choice \texttt{A} and \texttt{SuperCaloB} for choice \texttt{B}. Three-dimensional coarse graining has been explored in Ref.~\cite{supercalo}, but it has led to poorer-quality results.

In this thesis we focus on choice \texttt{A} and do not consider choice \texttt{B}.

\subsection{Conditional inputs} \label{sec:conditionals}
Since the energy distributions inside each coarse voxel are highly structured, depending on the incoming particle and the position in the calorimeter, additional conditional inputs are introduced. Here is a brief discussion of the input parameters: 
\begin{itemize}
    \item Deposited energy in the coarse voxel $i$, $E_{\text{coarse},i}$.
    \item Incident energy of the incoming particle, $E_{\text{inc}}$.
    \item Fine layer energies of layers spanned by coarse voxel $i$ (5 layers for \texttt{SuperCalo A} and 1 layer for \texttt{SuperCalo B})\footnote{This choice will be further discussed later in the text, due to the fact that this information is not obtainable from the coarse representation of the detector}.
    \item Deposited energy in neighboring coarse voxels in the radial, angular, and $z$ directions.
    \item One-hot encoded coarse layer number.
    \item One-hot encoded coarse radial bin.
\end{itemize}
\noindent\textbf{Scope note.} Following the prescription of Ref.~\cite{supercalo}, this thesis adopts an oracle-conditioned variant of super-resolution: the conditional vector includes the true fine per-layer energy totals for each coarse voxel (5 layers in \texttt{SuperCaloA}). These quantities are not available from the coarse representation and would be disallowed in a strict super-resolution setting. They are used here as a controlled first step to stabilize training and validate the evaluation pipeline. Results obtained under this choice should be interpreted as an \emph{upper-bound baseline} on what is achievable when accurate longitudinal aggregates are available (for example, when such sums are separately measured or provided by simulation). This configuration does \emph{not} represent missing-value or dead-channel scenarios in which the layer totals would be incomplete; therefore, it should not be interpreted as imputation of lost charge. Future work will repeat the analysis without these inputs (or with corrupted/noisy versions) to study the pure super-resolution case and quantify the gain provided by layer totals.

It is also important to note that every coarse voxel has a maximum of 6 neighboring voxels, and for coarse voxels with fewer than 6 neighbors, the missing energy values are set to zero. The number of values for each choice of coarse voxelization is reported in Table \ref{tab:conditional_numbers}. 
Finally, one-hot encoding is a method for converting categorical variables into binary format. It consists in associating a new column with each category value, where \texttt{1} means the category is present and \texttt{0} means it is not. As an example, take a calorimeter with 6 layers and take a voxel in the 2nd layer. To specify its position in the calorimeter we can use one-hot encoding and write \texttt{010000}.

\begin{table}[h]
    \centering
\begin{tabular}{ c | c | c | c | c | c | c | c}
 \thead{Choice} & \thead{Deposited \\ energy} & \thead{Incident \\ energy} & \thead{Fine layer \\ energies} & \thead{Neighbors \\ energy} & \thead{Coarse layer \\ number} & \thead{Coarse \\radial bin} & \thead{Total}\\ 
 \hline
 \hline
 A & \multirow{2}{*}{1} & \multirow{2}{*}{1} & 5 & \multirow{2}{*}{6} & 9 & 9 & 31\\ 
 B  & &  & 1 &  & 45 & 3 &  57 \\  
\end{tabular}
    
    \caption{Number of each component of the conditional input for each choice of coarse voxelization.}
    \label{tab:conditional_numbers}
\end{table}

\subsection{Preprocessing steps}
The incident energy of the incoming particle, $E_{\text{inc}}$, is preprocessed as
\begin{displaymath}
    E_{\text{inc}} \longrightarrow \log_{10} \frac{E_{\text{inc}}}{10^{4.5} \text{MeV}} \in [-1.5, 1.5]
\end{displaymath}
The coarse voxel energies are preprocessed according to a logit transform. The first step is defined as
\begin{displaymath}
    E_{\text{coarse},i} \longrightarrow x_i \vcentcolon = \log_{10} ((E_{\text{coarse},i} + \text{rand}[0,1 \text{ keV}])/E_{\text{coarse},\text{max}})
\end{displaymath}
and then the logit transform is applied
\begin{displaymath}
    y_i = \log \frac{u_i}{1- u_i} \text{,} \quad u_i \vcentcolon = \alpha + (1-2\alpha) x_i
\end{displaymath}
where $E_{\text{coarse},\text{max}}$ is the maximum coarse voxel energy in the whole dataset. The energies of neighboring coarse voxels are preprocessed in the same way but without any added noise. 
The fine layer energies are obtained by summing the fine voxel energies and preprocessed as
\begin{displaymath}
    E_{\text{layer},i} \longrightarrow \frac{E_{\text{layer},i}}{65 \text{ GeV}}
\end{displaymath}
One-hot encoded inputs are not preprocessed.

Lastly, the fine voxel energies $e_{\text{fine},ij}$ associated with the $i$th coarse voxel are preprocessed as
\begin{displaymath}
    e_{\text{fine},ij} \longrightarrow \hat{x}_i \vcentcolon = (e_{\text{fine},ij} + \text{rand}[0, 0.1 \text{ keV}])/ E_{\text{coarse} , i}
\end{displaymath}
and again a logit transform is applied
\begin{displaymath}
    \tilde{y}_i = \log \frac{\tilde{u_i}}{1- \tilde{u_i}} \text{,} \quad \tilde{u_i} \vcentcolon = \alpha + (1-2\alpha) x_i
\end{displaymath}
Note that the parameter $\alpha$ act as numerical regularizer and is fixed at $1\cdot10^{-6}$.
After the generation phase, the preprocessing is inverted for the outputs of the flow to recover the correct physical values. A lower cutoff at 15 keV is applied to the output values to adhere to the minimum fine voxel energy in the original data. This means that every sampled fine voxel energy value below 15 keV is set to zero. 

\begin{figure}
    \centering
    \includegraphics[width=0.8\linewidth]{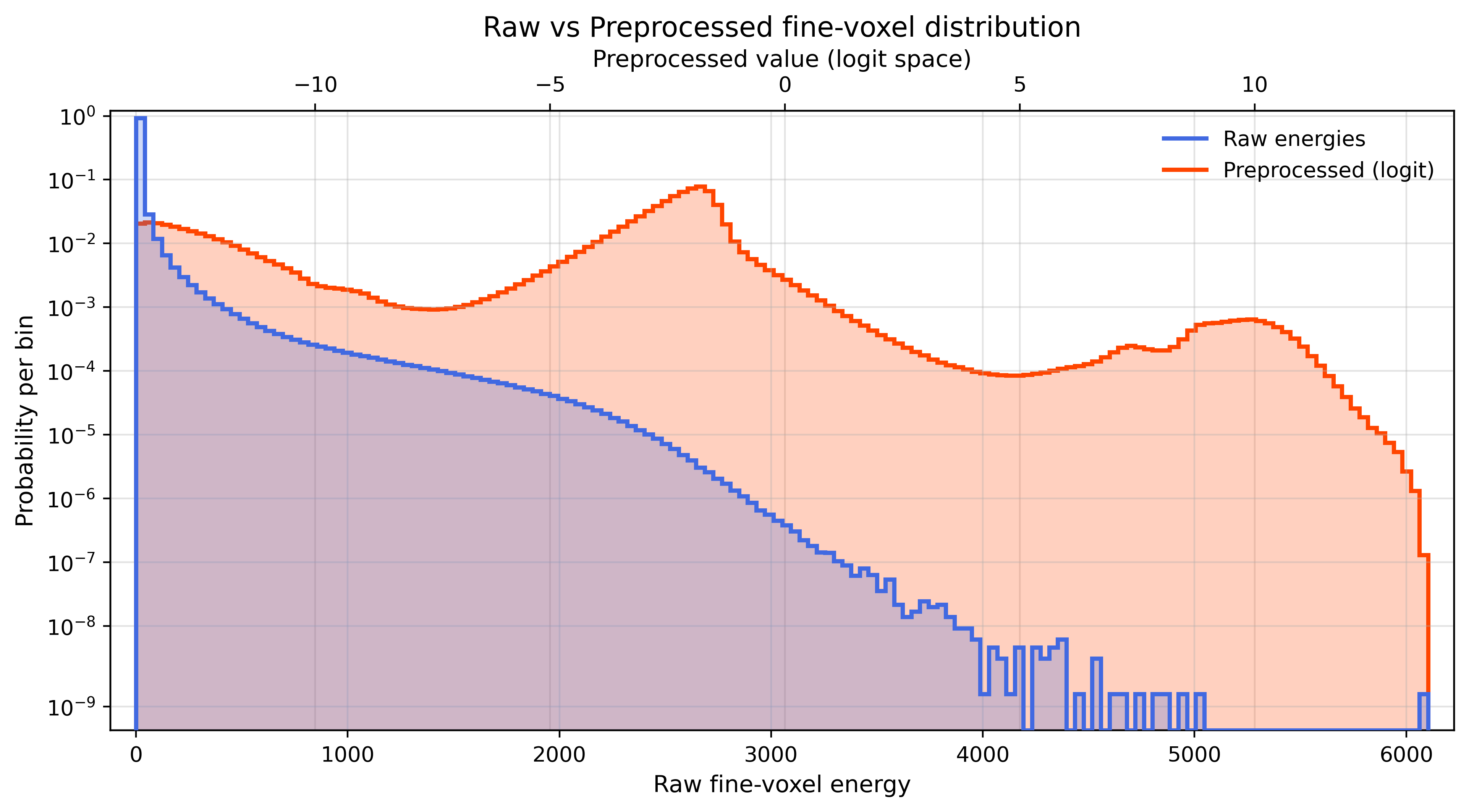}
    \caption{Effect of preprocessing on the raw distribution. In \textbf{Blue} the log-uniform distribution of the raw data and in \textbf{Red}, the distribution of the preprocessed data. Notice how the values of the preprocessed data are bound in the range $[-13.82,13.82]$ given by $\log(\alpha)$ and $\log(1/\alpha)$, with $\alpha = 1\cdot10^{-6}$.}
    \label{fig:preprocessing}
\end{figure}

\section{Architecture}\label{sec:model}
As introduced in the previous chapters, the task of super-resolution is achieved by learning the conditional PDF
\begin{equation} \label{eq:PDF_target}
    p(\vec{E}_{\text{fine}} \vert \vec{E}_{\text{coarse}})
\end{equation}
where it should now be clear that $\vec{E}_{\text{fine}}$ represents all the fine voxel energies and $\vec{E}_{\text{coarse}}$ is the coarse-grained representation of them. However, given the high dimensionality of the problem (6480 fine voxels to be generated), generating a full calorimeter shower in a single pass would have prohibitive computational costs. For this reason, the paradigm (\texttt{SuperCalo}) explored in this work attempts to overcome the high computational costs by exploring physically motivated approximations to the full density.

The motivated ansatz leveraged in Ref.~\cite{supercalo} is to assume that each coarse voxel is upsampled to its fine voxels with a universal super-resolution function that, as discussed in the previous section, may be conditioned on additional information:
\begin{displaymath}
    p(\vec{E}_{\text{fine}} \vert \vec{E}_{\text{coarse}}) = \prod^{N_\text{coarse}}_{i=1} p(\vec{e}_{\text{fine},i} \vert E_{\text{coarse},i}, \dots) 
\end{displaymath}
Notice that this ansatz assumes that, after upsampling, $\vec{e}_{\text{fine},i}$ are all independent of each other. The model is thus built to upsample a single coarse voxel to its fine voxel representation; however, thanks to parallel calculations on the GPU, it is possible to upsample multiple coarse voxels in one forward pass. The conditional inputs for each coarse voxel are built and later passed to the model in batches.

The model to approximate the PDF in Eq.~\ref{eq:PDF_target} is chosen to be a \emph{Masked Autoregressive Flow (MAF)} implementing a composition of \emph{Rational Quadratic Splines (RQS)} parametrized by \emph{MADE} (\emph{Masked Autoencoder for Distribution Estimation}) blocks. 

\begin{figure}[h!]
\centering
\begin{tikzpicture}[
  font=\small,
  node distance=7mm,
  >=Latex,
  box/.style={draw, rounded corners, fill=gray!20, minimum height=9mm, minimum width=40mm, align=center},
  perm/.style={draw, rounded corners, fill=Orange!30, minimum height=8mm, minimum width=32mm, align=center},
  base/.style={draw, rounded corners, fill=RoyalBlue!30, minimum height=9mm, minimum width=40mm, align=center},
  outb/.style={draw, rounded corners, fill=RoyalBlue!30, minimum height=9mm, minimum width=42mm, align=center},
  side/.style={draw, rounded corners, fill=OliveGreen!30, minimum height=8mm, minimum width=36mm, align=center},
  flow/.style={->, very thick},
  cond/.style={->, thick, dashed}
]

\node[base] (base) {Sample base\\$z$};

\node[box,  below=of base] (f1) {$B_1$:\; MADE + RQS};
\node[perm, below=of f1]   (p1) {$P_{\mathrm{rev}}$};

\node[box,  below=of p1]   (f2) {$B_2$:\; MADE + RQS};
\node[perm, below=of f2]   (p2) {$P_{\mathrm{rand}}$};

\node[box,  below=of p2]   (f3) {$B_3$:\; MADE + RQS};

\node[draw=none, below=of f3] (dots) {$\vdots$};

\node[perm, below=of dots]   (p7) {$P_{\mathrm{rev}}$};
\node[box,  below=of p7]     (f8) {$B_8$:\; MADE + RQS};

\node[outb, below=of f8] (out) {Output sample\\$x$};

\node[side, right=32mm of dots] (cond) {Conditioning $\vec{c}$};

\draw[flow] (base) -- (f1);
\draw[flow] (f1) -- (p1);
\draw[flow] (p1) -- (f2);
\draw[flow] (f2) -- (p2);
\draw[flow] (p2) -- (f3);
\draw[flow] (f3) -- (dots);
\draw[flow] (dots) -- (p7);
\draw[flow] (p7) -- (f8);
\draw[flow] (f8) -- (out);

\foreach \n in {f1,f2,f3,f8}{
  \draw[cond] (cond.west) -- ++(-8mm,0) |- (\n.east);
}

\end{tikzpicture}
\caption{Conditional MAF architecture with eight bijectors interleaved by alternating permutations.}
\label{fig:architecture}
\end{figure}
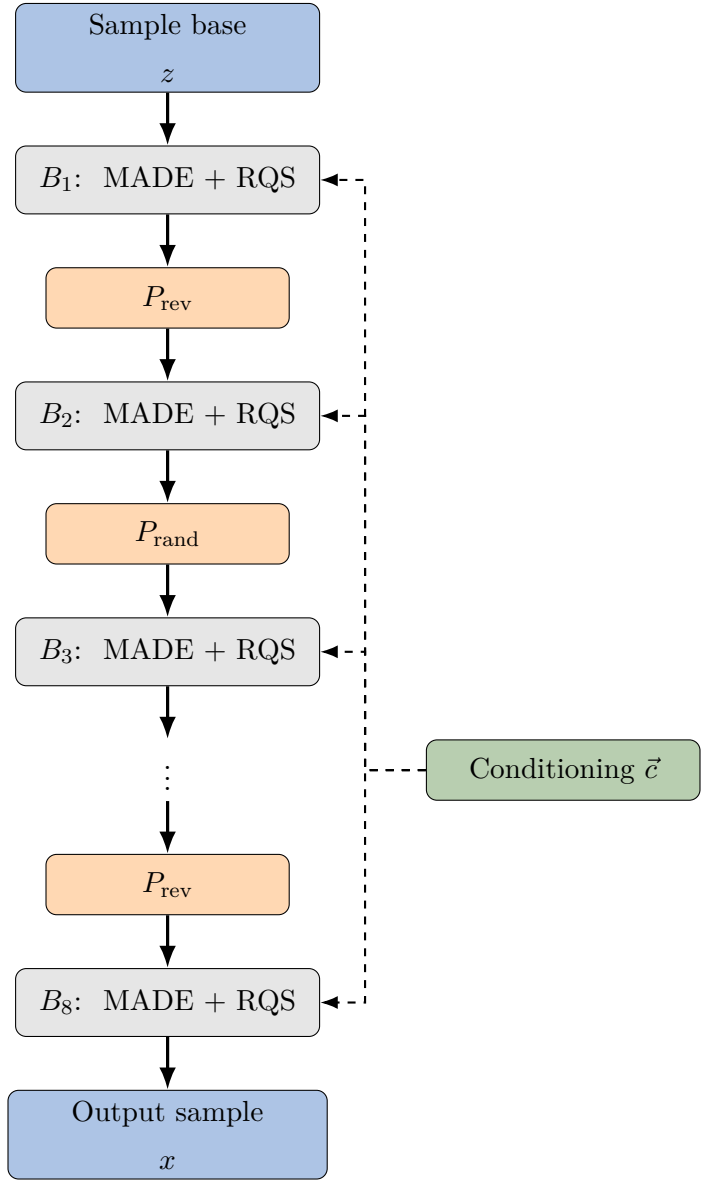

\paragraph{Splines.} 
Following the preprocessing results, splines are chosen to be defined in $[-14,14]$, while outside this domain linear tails are introduced with slopes matching the derivative of the RQS to maintain differentiability at the domain boundaries. For the specific application, 8 bins are used, for a total of 23 parameters: 
\begin{itemize}
    \item Coordinates for the knots account for 16 parameters ($9 \cdot 2 - 2 = 16$), since the first knot is fixed at $(x_0, y_0) = (-14,0)$.
    \item Derivatives at each inner knot account for 7 parameters. The derivatives at the boundaries are not fixed, since linear tails are used. 
\end{itemize}
\paragraph{Bijectors.}
Following Ref.~\cite{supercalo}, we employ a chain of eight autoregressive bijectors interleaved with permutations of the feature indices. The permutations alternate deterministically between a fixed reverse ordering and an independently seeded random permutation, yielding the sequence
\[
B_1 \;\to\; P_{\mathrm{rev}} \;\to\; B_2 \;\to\; P_{\mathrm{rand}} \;\to\; \dots \; \to B_7 \;\to\; P_{\mathrm{rev}} \;\to\; B_8,
\]
with no permutation after $B_8$. Each $B_i$ is a conditional MADE\,+\,RQS bijector; the intervening $P_{\mathrm{rev}}/P_{\mathrm{rand}}$ layers merely reshuffle coordinates (zero log-determinant) to promote mixing across dimensions while preserving training and likelihood evaluation mechanics. An illustrative diagram is reported in Figure \ref{fig:architecture}.

\paragraph{Base distribution.}
In a normalizing flow the data $\vec{x}$ are obtained by pushing forward latent variables $\vec{z}$ drawn from a simple \emph{base distribution} through an invertible map $f_\theta$ (here, the conditional MADE\,+\,RQS chain): $\vec{x}=f_\theta(\vec{z}\,;\,\vec{c})$. We take the base to be an isotropic Gaussian on $\mathbb{R}^{D_b}$,
\begin{equation}
  \vec{z} \sim \mathcal{N}\!\left(\mathbf{0},\,\mathbf{I}_{D_b}\right), \qquad D_b=\begin{cases}
  10 & \text{for choice \texttt{A}},\\
  12 & \text{for choice \texttt{B}},
  \end{cases}
  \label{eq:base_std_normal}
\end{equation}
implemented as a diagonal multivariate normal with zero mean and unit variance per component. The base is kept \emph{independent} of the conditioning variables $\vec{c}$; all dependence on $\vec{c}$ is introduced by the bijector $f_\theta(\cdot;\vec{c})$. This choice offers closed-form log-densities, full support on $\mathbb{R}^{D_b}$ (matching the RQS domain with linear tails), and efficient, numerically stable training. With this setup the conditional log-likelihood follows from the change of variables formula,
\begin{equation}
  \log p(\vec{x}\mid \vec{c}) \;=\; \log p_Z\!\big(f_\theta^{-1}(\vec{x}\,;\,\vec{c})\big)
  \;+\; \log \Bigl|\det \tfrac{\partial f_\theta^{-1}}{\partial \vec{x}}\Bigr|,
  \label{eq:flow_ll}
\end{equation}
where $p_Z$ is the standard normal density in~\eqref{eq:base_std_normal}. Because each $B_i$ is autoregressive, the Jacobian of $f_\theta^{-1}$ is triangular, so the determinant in~\eqref{eq:flow_ll} reduces to a sum over diagonal terms, preserving the usual $O(D_b)$ evaluation cost. In practice, we use double precision (\texttt{float64}) for the base and log-determinant computations to reduce numerical error during training. A detailed discussion about training instability is left to the following sections.

\begin{figure}
    \centering
    \includegraphics[width=0.6\linewidth]{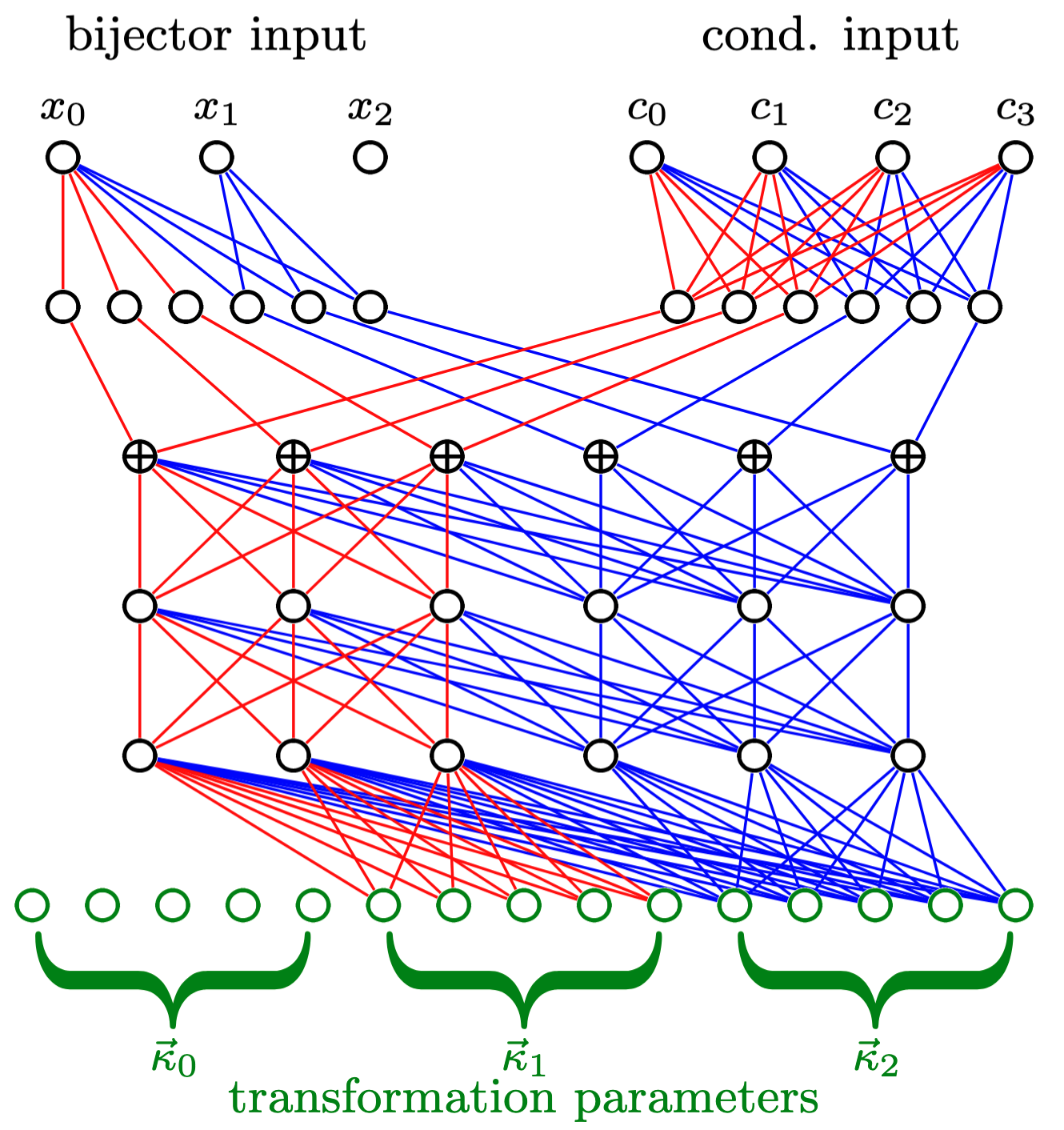}
    \caption{Schematic of a conditional MADE block used inside a MAF. Three variables $\vec{x}=(x_0,x_1,x_2)$ are transformed while four context variables $\vec{c}=(c_0,\dots,c_3)$ provide conditioning. Inputs are split into two parallel streams (one for $\vec{x}$, one for $\vec{c}$), each fully connected to a width-6 hidden layer; their pre-activations are then added element-wise (the “$+$” row), so conditioning enters without breaking the autoregressive masking. Two width-6 hidden layers are shown. The output layer produces three vectors $\vec{k}_i$ (one per transformed dimension $x_i$); each has 5 components implementing a two-bin Rational Quadratic Spline (one derivative and two knot coordinate pairs $(x_1,y_1)$, $(x_2,y_2)$). Colored edges illustrate autoregressive dependencies: in red, $\vec{k}_1$ depends only on $x_0$ and $\vec{c}$; in blue, $\vec{k}_2$ depends on $x_0,x_1$ and $\vec{c}$. Choice \texttt{B} is only reported for completeness, in this thesis we only explore choice \texttt{A}.}
    \label{fig:cond_input}
\end{figure}

In the previous section, we introduced the coarse voxelization choices and, as specified, choice \texttt{A} groups together 10 fine voxels, while choice \texttt{B} includes 12 fine voxels. This explains the number of output values reported in Table~\ref{tab:flow_configs}, since the RQS outputs 23 parameters for each dimension. In case of choice \texttt{A}, the parameters are $10 \cdot 23 = 230$, while in case of \texttt{B} the output parameters are $12\cdot23 = 276$. 
\begin{table}[h]
\centering
\small
\setlength{\tabcolsep}{5pt}
\renewcommand{\arraystretch}{1.0}

\begin{tabular}{@{}c|c|c|c|c|c|c@{}}
\multirow{2}{*}{Model} &
\multirow{2}{*}{\shortstack{Base distribution  \\ dimensions}} &
\multirow{2}{*}{\shortstack{Number of \\ MADE blocks}} &
\multicolumn{3}{c|}{MADE layer sizes} &
\multirow{2}{*}{\shortstack{RQS \\ bins}}\\
& & & Input & Hidden & Output & \\
\hline\hline
\textsc{A} & 10 & 8 & 41 & $2\times128$ & 230 & 8 \\
\textsc{B} & 12 & 8 & 69 & $2\times128$ & 276 & 8 \\
\end{tabular}

\caption{Configuration of the different flow models. The input layer dimension is defined to be the sum of the number of dimensions in the base distribution and the number of conditional inputs, discussed in Section \ref{sec:conditionals}. Note that $2 \times 128$ is a shorthand notation for 2 hidden layers of 128 nodes each.}
\label{tab:flow_configs}
\end{table}
In Figure \ref{fig:cond_input} is reported a diagram that is useful for explaining how conditional inputs are inserted in the model. It is shown as an example a MADE block that, as introduced in Chapter~\ref{generative_models}, is the fundamental building block of the MAF architecture. In the example, the MADE block has 3 variables ($x_i$) to be transformed and 4 additional variables ($c_i$) on which the transformation is conditioned. There is an input layer of 6 nodes and two hidden layers of 6 nodes each. The conditional inputs and the variables are fully connected to separate input layers, with the same dimensionality as the first hidden layer (6 in this case) and the logits (outputs of the layer) are summed term by term (the summing row in the diagram). This way the conditional inputs enter the MADE block without affecting the autoregressive property. 

The diagram in Figure \ref{fig:cond_input}, also gives the opportunity to better understand how the RQS transformation is implemented. In fact, the output layer presents three vectors, one for each dimension to transform ($x_i$) and each vector has 5 dimensions itself, necessary to implement a RQS with two bins: 1 derivative and 2 knots described by the coordinates $(x_1,y_1)$ and $(x_2,y_2)$.  
Furthermore, the red connections show that $\vec{k}_1$ only depend on $x_0$ and $\vec{c}$, while the blue connections show that $\vec{k}_2$ depends on $x_0, x_1$ and $\vec{c}$.

As introduced in Section \ref{sec:voxellization}, we explore only \texttt{SuperCaloA}, in fact, Ref.~\cite{supercalo} prioritized the inter-layer correlations between fine voxels.

\section{Training Strategy} \label{sec:training}
In this section, we introduce the hyperparameters used for the training process and describe the working environment.

\subsection{Hardware and Software Environment}
All experiments were executed on the INFN \texttt{TEOGPU} cluster, using a single NVIDIA L40S GPU equipped with 45\,GiB of memory.
The implementation is based on \texttt{TensorFlow~2.12} and \texttt{TensorFlow~Probability~0.20}, with numerical operations and data management handled primarily through \texttt{NumPy}, \texttt{h5py}, and \texttt{Matplotlib}.
Additional utilities for statistical evaluation and visualization rely on \texttt{SciPy}, \texttt{pandas}, and \texttt{seaborn}.
All the source code used in this work, including data handling scripts, model training routines, and evaluation utilities, is publicly available at
\href{https://github.com/andreacosso/Thesis}{\texttt{github.com/andreacosso/Thesis}}~\cite{acosso2025thesis}.

The training and evaluation pipelines were developed within a modular Python codebase designed for reproducible workflows and checkpointed model weights.
This framework extends the public implementation introduced in Ref.~\cite{Coccaro_2024}, originally built to compare normalizing-flow architectures for calorimeter shower modeling.
In the present work, the codebase was modified to incorporate conditional inputs and other details necessary for the architecture choice, such as the linear tails of the rational quadratic spline transformation.

\subsection{Loss Function}
The model is trained by minimizing the \emph{negative log-likelihood} (introduced in Section \ref{sec:loss}) on the flow output.
This corresponds to maximizing the likelihood of the data under the learned transformation and provides an exact, tractable density objective for normalizing flows. 
Following the prescription of Ref.~\cite{supercalo}, no regularization has been introduced.

\subsection{Optimizer and Learning Rate Schedule}
Training is performed using the \emph{Adam} optimizer (described in Section \ref{sec:adam}), with default exponential decay rates $(\beta_1,\beta_2) = (0.9, 0.999)$ and $\epsilon = 10^{-8}$.
Still following Ref.~\cite{supercalo}, we employed the \textit{OneCycle} learning rate (LR) policy, originally proposed in Ref.~\cite{onecycle} to achieve \emph{super-convergence} in deep networks.
The method is based on the intuition that training benefits from briefly exposing the model to high learning rates, encouraging rapid exploration of the loss landscape, followed by a smooth annealing phase that refines the parameters around wide, flat minima. To increase training stability at a high learning rate, an inverse cycle on $\beta_1$ momentum is implemented.

\begin{figure}[h]
\centering
\includegraphics[width=0.6\linewidth]{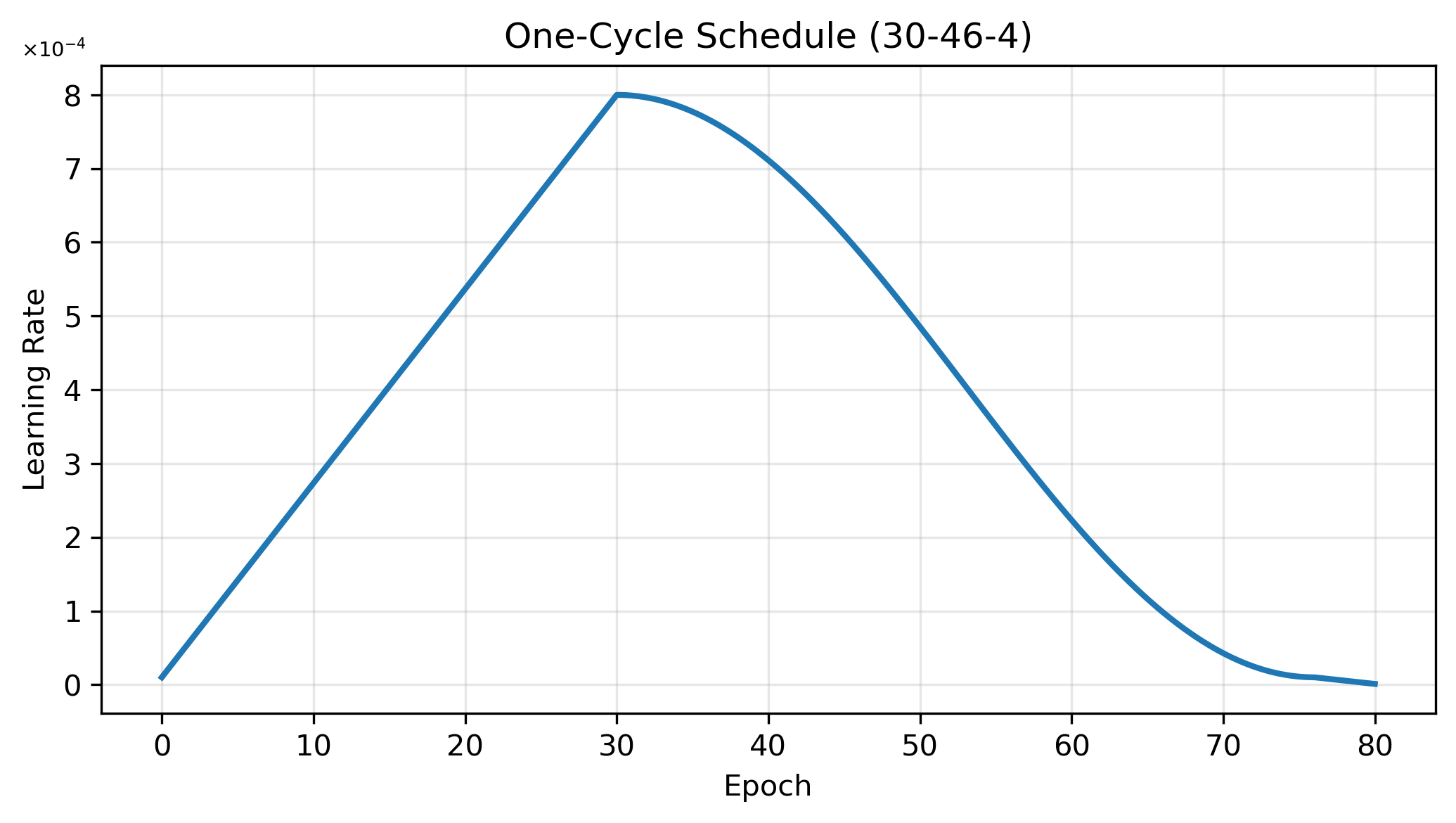}
\caption{Three-phase One-Cycle learning rate schedule used for model training. We chose 30 epochs of warm-up, 46 of cooldown, and 4 epochs of annihilation.}
\label{fig:onecycle_schedule}
\end{figure}

In our configuration, the LR increases linearly from a base value $\alpha_{\text{base}} = 1\times10^{-5}$ to a maximum value $\alpha_{\text{max}} = 8\times10^{-4}$ in 30 epochs, then decreases following a cosine profile back to $\alpha_{\text{base}}$ over 46 epochs, and finally undergoes a short linear annihilation phase of 4 epochs down to $\alpha_{\text{annihil.}} = 1\times10^{-6}$. Momentum starts from $\beta_1 = 0.95$ decreases in 30 epochs to $\beta_1 = 0.85$ and returns to $\beta_1 = 0.95$ in 50 epochs. 
This three-phase schedule (warm-up, cooldown, and annihilation) promotes faster convergence in the early stage and smoother stabilization at the end of training. 
In Figure \ref{fig:onecycle_schedule}, a plot of the LR cycle is reported. In Ref.~\cite{supercalo}, the authors were able to reach faster convergence by training for 40 epochs with 18 epochs of warm-up and 18 of cooldown, with 4 annihilation epochs, but our results following this procedure were worse than what we obtained with the parameters introduced.
No gradient clipping is applied during the training process.

Although many frameworks provide built-in One-Cycle schedulers, most notably \texttt{PyTorch}'s \texttt{OneCycleLR}, there is currently no standardized implementation in \texttt{TensorFlow}.
For this reason, a custom schedule was implemented for this work, faithfully reproducing the same learning rate and momentum dynamics.

Training is carried out with mini-batches of 10k events without early stopping.   
A summary of the training strategy is reported in Table~\ref{tab:training_config}.

\begin{table}[h]
    \centering
    \caption{Summary of training hyperparameters.}
    \begin{tabular}{l c}
        \toprule
        Parameter & Value \\
        \midrule
        Optimizer & Adam \\
        LR scheduler & OneCycle \\
        Initial learning rate & $1\times10^{-5}$ \\
        Max learning rate & $8\times10^{-4}$ \\
        $(\beta_1, \beta_2, \epsilon)$ & (0.9, 0.999, $10^{-8}$) \\
        Batch size & 10k \\
        Total epochs & 80 \\
        (warm-up, cooldown, annihilation) & (30, 46, 4)\\
        Regularization & None \\
        Gradient clipping & None \\
        Early stopping patience & None \\
        Framework & TensorFlow~2.12 + TFP~0.20 \\
        Hardware & NVIDIA L40S (45\,GiB) \\
        \bottomrule
    \end{tabular}
    \label{tab:training_config}
    \caption{Training parameters summary}
\end{table}

\section{Evaluation}\label{sec:eval_implement}
The evaluation procedure plays a central role in this work. In Ref.~\cite{supercalo}, the assessment of model performance was inspired by the procedure adopted within the \emph{CaloChallenge}~\cite{calochallenge}, although it does not fully reproduce the official evaluation pipeline. The proposed evaluation combines a classifier-based score with the Jensen–Shannon divergence, providing a practical but not yet statistically rigorous assessment of the generative quality.

Using a classifier to assess the performance of a generative model in HEP is not recommended. First, deep neural networks (DNNs) are often seen as ``black-box'' models~\cite{DNN_as_black_boxes}, which makes it difficult to understand which features of the generated data they consider different or similar. Second, their performance depends strongly on both the network architecture and the training dataset, and there is no clear way to define a single architecture that works well for all datasets and types of discrepancies. In addition, DNN training is stochastic, involving the optimization of a complex loss function with many local minima, and is usually slow. Because of this, it is sensitive to the initial conditions and hyperparameters that are not directly related to the problem, making the results hard to reproduce and the process less efficient.

The Jensen-Shannon divergence, belonging to the broader class of $f$-divergences, quantifies differences point-by-point in probability density and is invariant under reparameterization, but it doesn't account for the geometry of the underlying sample space. As discussed in Ref.~\cite{JSD_vs_ipm}, this limitation can lead to misleading conclusions when evaluating generative models.
For instance, two generated distributions that differ only by a small shift in a physically meaningful variable (such as jet mass or shower centroid) can be assigned the same $f$-divergence, despite one clearly providing a more realistic physical description. In contrast, metrics based on Integral Probability Metrics (IPMs), introduced in Section \ref{sec:eval_intro}, explicitly incorporate the metric structure of the space, allowing them to recognize when generated and real samples are close in a geometric or physical sense. 

As outlined in Section \ref{sec:eval_intro}, the evaluation pipeline adopted in this work is implemented following the statistical procedure introduced in Ref.~\cite{ref_the_ref}, for which the authors provide a Python module~\cite{GMetrics}. 

To summarize, the key idea is to take numerical samples, build the null and alternative hypotheses, and compare them using statistically robust, high-dimensional scalable two-sample tests.

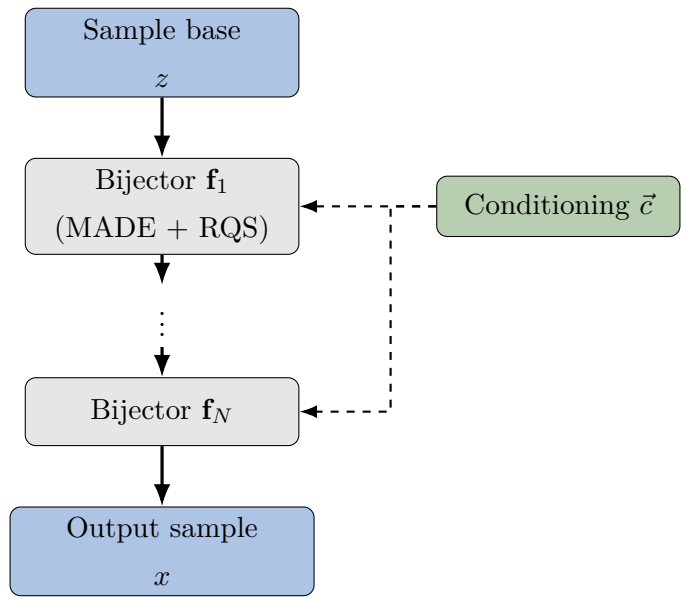
\begin{figure}[h]
\centering
\begin{tikzpicture}[
  font=\small,
  node distance=8mm,
  >=Latex,
  box/.style={draw, rounded corners, fill=gray!6, minimum height=9mm, minimum width=36mm, align=center},
  side/.style={draw, rounded corners, fill=OliveGreen!30, minimum height=8mm, minimum width=32mm, align=center},
  flow/.style={->, very thick},
  cond/.style={->, thick, dashed}
]

\node[box, fill=RoyalBlue!30] (base) {Sample base\\$z$};
\node[box, below=of base, fill=gray!20] (g1) {Bijector $\mathbf{f}_1$\\(MADE + RQS)};
\node (dots) [below=4mm of g1] {$\vdots$};
\node[box, below=4mm of dots, fill=gray!20] (gL) {Bijector $\mathbf{f}_N$};
\node[box, below=of gL, minimum width=40mm, fill=RoyalBlue!30] (out) {Output sample\\$x$};

\node[side, right=18mm of g1] (cond) {Conditioning $\vec{c}$};

\draw[flow] (base) -- (g1);
\draw[flow] (g1) -- (dots);
\draw[flow] (dots) -- (gL);
\draw[flow] (gL) -- (out);

\draw[cond] (cond.west) -- ++(-6mm,0) |- (g1.east);
\draw[cond] (cond.west) -- ++(-6mm,0) |- (gL.east);

\end{tikzpicture}
\caption{Conditional MAF (vertical layout): generative flow from a base draw $z$ through a sequence of bijectors $\mathbf{f}_1,\ldots,\mathbf{f}_N$ to obtain a sample $x$. Each bijector is parametrized by a MADE block (RQS transform) and receives the conditioning $\vec{c}$.}
\label{fig:maf_flow_simple}
\end{figure}

The model has been evaluated on a dataset of 100,000 full showers not presented to the model during the training process. The evaluation (sampling) process consisted of taking the dataset and building, for each fine shower, its coarse representation as presented in the previous chapter. Then one conditional input vector for each coarse voxel is constructed as described in the previous sections. The conditional inputs are then passed to the model to obtain the generated samples. The generative process is reported in the flowchart in Fig~\ref{fig:maf_flow_simple}. The model passes the samples from the base distribution through the chain of conditioned bijectors to obtain the sampled value $x$.

\subsubsection*{Metrics and testing strategy}
In what follows, we apply, for each feature introduced below, the family of test statistics reviewed in Section \ref{sec:eval_intro} (see the \emph{Test statistics} subsection). In practice, we combine complementary metrics so that sensitivity is not driven by a single notion of discrepancy:
\begin{itemize}
    \item \textbf{Sliced Wasserstein (SW)} captures geometry-aware shifts along random 1D projections, scalable to high dimensions via averaging over $K$ directions.
    \item \textbf{Kolmogorov–Smirnov variants} (mean-KS on marginals and sliced-KS on random projections) to detect localized CDF mismatches in 1D and along projected axes with low computational overhead.
    \item \textbf{Maximum Mean Discrepancy (MMD)} with a fixed polynomial kernel provides a kernel-based, moment-sensitive IPM that complements SW/SKS; used with the same kernel choice throughout for consistency.
    \item \textbf{Fréchet Gaussian Distance (FGD)} as a simple, low-variance Gaussian summary of mean/covariance mismatches in moderate dimensions.
\end{itemize}

The resulting evaluation is feature-wise and metric-wise: for each feature space we compute the observed test statistic on (truth, generated) samples, and we estimate the empirical distributions of the statistic under both $H_0$ and $H_1$. From these, we report the $p$-value taking as the observed statistic the \emph{median} of the alternative distribution. We also visualize the separation of null vs.\ alternative distributions. This multi-metric approach ensures robustness to different failure modes (global shifts, tail distortions, correlation changes), while remaining scalable through slicing/projection strategies and batched computation. 

Due to their poor scalability, we only evaluate the performance on FGD and MMD in low dimensionality scenarios.

\textit{Implementation notes.} The same random-seeded set of projection directions is used across runs for SW/SKS to reduce variance; kernel choices and projection counts $K$ are kept fixed across features to enable fair, cross-feature comparisons; batch size and the number of experimental replicates are chosen to balance statistical precision with runtime and are reported alongside the results.

\subsubsection*{Full dimensionality features}
Leveraging the high scalability of the evaluation procedure introduced in Ref.~\cite{ref_the_ref} with the dimension, we evaluate the performance of the model also on the full dimensionality of the shower and its coarse representation. Specifically, as introduced in the previous sections, the fine representation of the shower has $6480$ dimensions, while the coarse representation, for choice \texttt{A}, has a dimensionality of $648$. Due to the poor scalability of the metrics FGD and MMD, we chose to exclude them from the evaluation, and we only implemented the KS based statistics together with the SWD. 

This approach represents one of the methodological novelties introduced in this work. As highlighted in earlier sections, Previous studies in calorimeter shower generation typically relied on the evaluation of low-dimensional physics-inspired features (such as layer energies, centroids, or shower widths) to assess the performance of generative models. While such observables are physically interpretable, they inevitably compress the information contained in the original data and may conceal discrepancies detectable only in the full feature space. In contrast, the framework applied in this thesis enables a robust statistical comparison directly in the native, high-dimensional voxel space of the calorimeter, without resorting to handcrafted observables. This allows for a more faithful and unbiased assessment of the model’s ability to reproduce the underlying data distribution.

\subsubsection*{Physically inspired features}
We chose to evaluate the performance of the model also on low-dimensional, physically inspired features:

\begin{itemize}
    \item \textbf{Total incident energy}: Being 1D, it is fast and simple. We sum all the energy deposits in the fine voxels for both the true and sampled data and compare their distributions.
    \begin{equation}
        E_{inc} = \displaystyle \sum_{i=1}^n e_{\text{fine},i},
    \end{equation}
    where $n = 6480$ is the number of fine voxels in the calorimeter. 
    \item \textbf{Fine layer energies}: These are the distributions of the energies contained in each of the fine voxel layers. Given the structure of the calorimeter presented in Section \ref{sec:dataset}, this feature is 45D, one dimension for each layer. To build the distributions, we sum the energies in the radial and angular directions. 
    \begin{equation}
        E_{\text{layer}, j} = \displaystyle \sum_{i=1}^{\tilde{n}} e_{\text{fine},i}^{(j)}\,, 
    \end{equation}
    where $\tilde{n} = 144$ is the number of fine voxels in each layer (16 angular bins and 9 radial bins). We denoted with $e_{\text{fine},i}^{(j)}$ the $i$-th fine voxel contained in the $j$-th layer.
    \item \textbf{Energy centroid position}: we compare the distributions of the positions of the centroid in the radial, angular and longitudinal directions, defined by 
    \begin{equation}
        \bar{\phi} = \frac{\sum_i E_i \, \phi_i}{\sum_i E_i},
        \qquad
        \bar{r}   = \frac{\sum_i E_i \, r_i}{\sum_i E_i},
        \qquad
        \bar{z}   = \frac{\sum_i E_i \, z_i}{\sum_i E_i},
    \end{equation}
    where $\phi_i$, $r_i$ and $z_i$ are the positions of the fine voxel with energy $E_i$.
    \item  \textbf{Root Mean Square (RMS) along the longitudinal, lateral and angular axis}: energy-weighted RMS provides a compact and physically interpretable description of the shower’s spatial extent. It is therefore a natural choice for statistical comparison between real and generated showers. Note that RMS only captures the second moment of the distribution, leaving out asymmetries, tails, and multi-modal structures. RMS can be calculated as
    \begin{equation}
    \sigma_x = 
    \sqrt{
    \frac{\sum_i E_i \, (x_i - \bar{x})^2}
         {\sum_i E_i}
    },
    \qquad
    \bar{x} = 
    \frac{\sum_i E_i \, x_i}{\sum_i E_i}, 
    \end{equation}
    with
    \begin{displaymath}
        x \in \{z,\, r,\, \phi\}.
    \end{displaymath}
\end{itemize}
In the next section we introduce the hypothesis for the test and how the test statistics are computed according to the distributions.

\subsubsection*{Null hypothesis construction} \label{sec:null_hypo}
The evaluation procedure is based on the definition of a statistical null hypothesis, which states that the real and generated samples are drawn from the same underlying probability distribution,
\begin{equation}
    H_{0}: \; P_{\mathrm{real}} = P_{\mathrm{gen}}\,,
\end{equation}
while the alternative hypothesis assumes that the two distributions differ,
\begin{equation}
    H_{1}: \; P_{\mathrm{real}} \neq P_{\mathrm{gen}}\,.
\end{equation}
Under $H_{0}$, any discrepancy between the samples is attributed solely to random fluctuations due to finite statistics. To estimate the sampling distribution of each test statistic under this assumption, the combined dataset is repeatedly reshuffled, randomly exchanging elements between the real and generated sets. For each reshuffling, the test statistic is recomputed, resulting in an empirical distribution that approximates the true null distribution.

We adopt the following terminology. An \emph{experiment} denotes a single Monte--Carlo replicate in which we compute one realization of the test statistic from \code{batch_size} samples per distribution. We repeat experiments \code{n_iter} times. If the available dataset contains fewer than \code{n_iter} $\times$ \code{batch_size} samples, we form each batch by non-parametric bootstrap resampling (sampling with replacement) from the finite sample~\cite{bootstrap}. For sliced methods, Sliced Kolmogorov-Smirnov (SKS) and Sliced Wasserstein Distance (SWD), we set the number of random projections (\emph{slices}) to \code{nslices}. Projection directions are fixed for all experiments. 

An important thing to note is that larger \code{batch_size} makes the tests more sensitive due to the fact that sampling noise decreases as we use more points.
Indeed, both the null and the alternative sampling distributions include random sampling noise. If $T_n$ is the test statistic computed from batches of size $n$ per sample, you can think of it as \emph{signal + noise}. Under $H_0$ the signal is zero, so $T_n$ fluctuates around $0$; under $H_1$ there is a genuine difference between the two distributions (the signal is positive), so $T_n$ fluctuates around a nonzero value. In \emph{both} cases, the sampling noise shrinks at the canonical $1/\sqrt{n}$ rate as we increase \code{batch_size}, which tightens the null and the alternative histograms. Because the critical threshold is calibrated on the null, a tighter null lowers that threshold, while the alternative remains centered at a positive value due to the true difference; the overlap between the two distributions decreases and rejections become more likely. For slice-based methods (SKS, SWD) the same logic holds within each 1D projection: each slice uses $n$ points per sample, so its per-slice variability also falls like $1/\sqrt{n}$. Adding more \code{nslices} mainly averages out the randomness from the choice of directions, whereas increasing \code{batch_size} directly sharpens each slice’s estimate and thus improves detectability. As a toy example, imagine two coins: the real process has head probability $p=0.50$, while the model has $p'=0.52$, so the true gap is $\delta=p'-p=0.02$. If we estimate each probability from $n$ flips, the sampling error of the observed head rate is about $\sqrt{p(1-p)/n}$ (the binomial standard error). With $n=100$ per coin this is $\approx 0.05$ (5\%), which is larger than the signal $\delta$, so the two samples often look indistinguishable. With $n=10{,}000$ the error drops to $\approx 0.005$ (0.5\%), and the same fixed gap $\delta=0.02$ becomes large relative to the noise (about four standard errors), making it easy for any two-sample test to detect. This is precisely how a larger \code{batch_size} raises power: the discrepancy between the underlying distributions is fixed, while sampling variability shrinks like $1/\sqrt{n}$, improving the signal-to-noise ratio.

This data-driven construction allows one to evaluate how much the observed value of the statistic deviates from what would be expected if both samples originated from the same process. The resulting null distribution serves as a reference for all subsequent comparisons, including the computation of confidence levels, $p$-values, and the evaluation of statistical power. 

This approach ensures a statistically rigorous and reproducible estimation of the null and alternative distributions, providing a solid foundation for the subsequent comparison of real and generated calorimeter showers.

The Code introduce in Reference~\cite{ref_the_ref} is available at Reference~\cite{GMetrics}.

\section{Results}\label{sec:results}
In this section we present the results from the training and evaluation procedure thoroughly described in the sections above. First are illustrated the qualitative results of the training procedure by introducing qualitative plots. Secondly the results from the statistically robust evaluation procedure are presented. To present the results, we selected the model that achieved the lower validation loss across all the runs made to explore the hyperparameter space. The corresponding hyperparameters are presented in Section \ref{sec:training}.
\subsection{Training results}
\begin{figure}[h]
    \centering
    \includegraphics[width=0.8\linewidth]{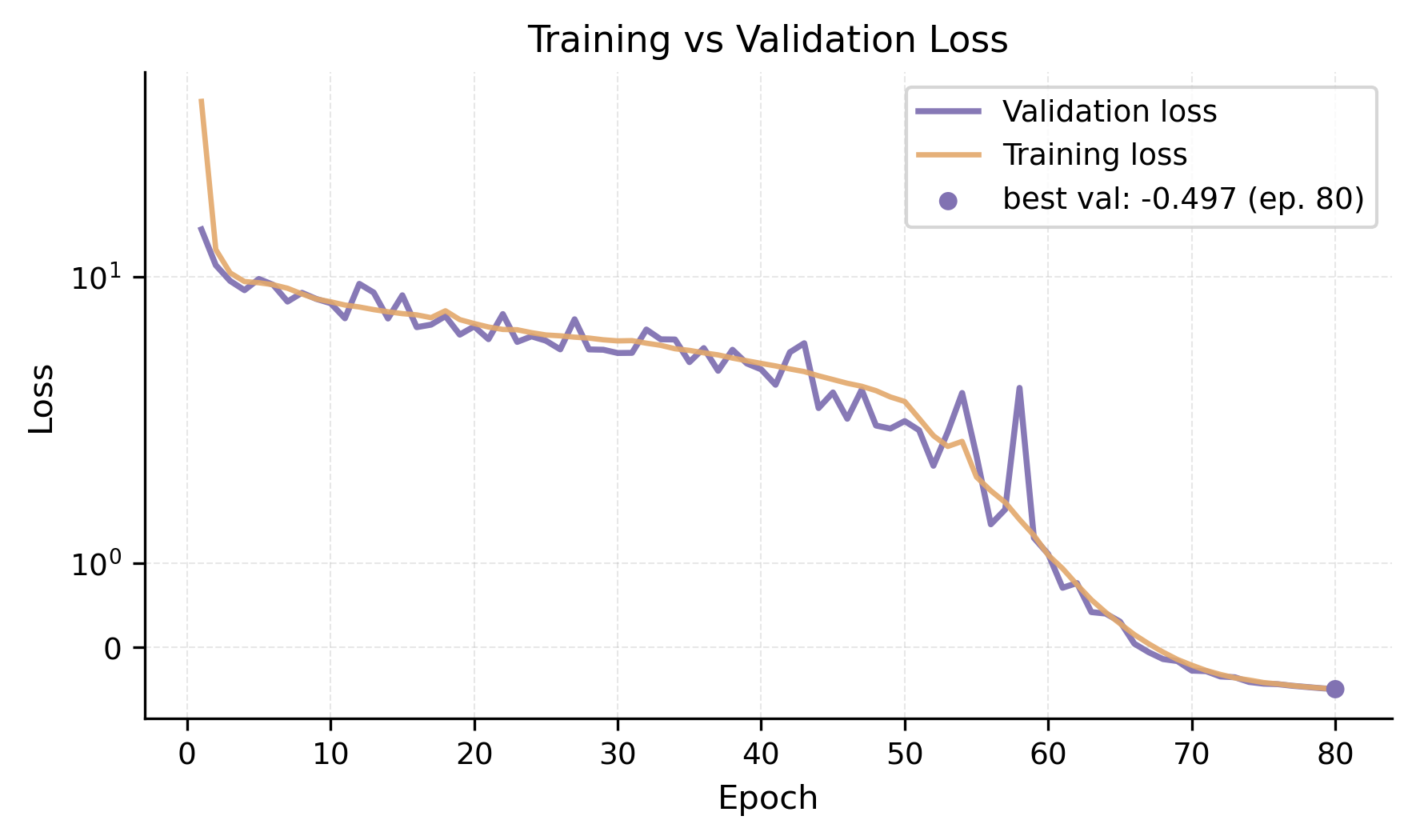}
    \caption{Plot of the validation and training losses over the course of training. Lower validation loss value was obtained at epoch 80.}
    \label{fig:loss}
\end{figure}

The model presented as detailed in Section \ref{sec:model} was trained for 80 epochs with details on the parameters illustrated in  Section \ref{sec:training}. The training and validation loss curve is reported in Figure \ref{fig:loss}. The lowest value of the validation loss was obtained in the last epoch (80). This suggests that the optimization task hasn't fully converge toward an optimal loss value and this leaves room for improvement. Another indication supporting this hypothesis is the very small generalization gap; in fact, the training loss reached its minimum at epoch 80 with $\mathcal{L}_{\text{min}}^{(\text{train})} = -0.497$, while the validation loss minimum (still at epoch 80), registered $\mathcal{L}_{\text{min}}^{(\text{val.})} = -0.496$. We tried increasing the cooldown phase and changing the base and max LR, without achieving better results. This indicates the need for a fine tuning of the hyperparameters.
The steep descent around epoch 55 in Figure \ref{fig:loss} can be explained by the \emph{super-convergence} effect of the OneCycle~\cite{onecycle} scheduler discussed in Section \ref{sec:training}. 

\begin{figure}[h]
    \centering
    \includegraphics[width=0.8\linewidth]{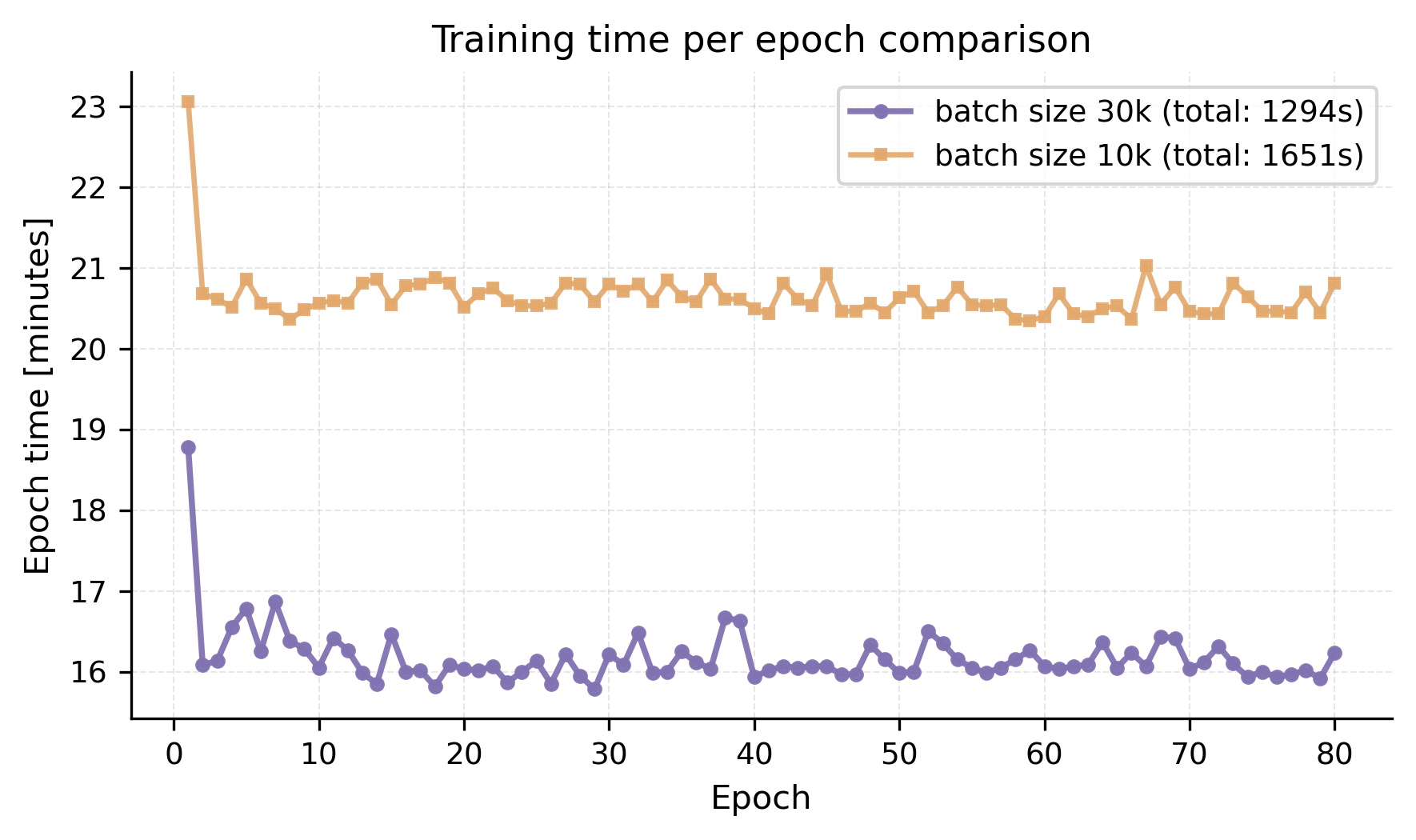}
    \caption{A comparison between runs with different batch size. There is a clear difference of training time if we compare a batch size of $10$k or $30$k. The absolute difference in total training time is about six hours.}
    \label{fig:epoch_time}
\end{figure}

We also notice that reducing the batch size yielded modest improvements probably due to increased noise in the gradient favoring exploration. We chose to keep a batch size of $10,000$ to balance exploration with training time. In fact, the model has spent approximately $27.5$ hours to complete the training process, while using a batch size of $30$k would have taken around $21.5$ hours. A comparison is illustrated in Figure \ref{fig:epoch_time}.

We have not been able to obtain similar results following the prescription of Ref.~\cite{supercalo}, that involves a OneCycle scheduler with 18 epochs of warm-up, 18 of cooldown and 4 epochs of annihilation, with a batch size of $60,000$ samples. They chose a base LR of $2\cdot10^{-5}$ and a max LR of $1\cdot10^{-3}$. A summary of the main differences in the training configuration is reported in Table \ref{tab:training_comparison}.

\begin{table}[h]
\centering
\small
\setlength{\tabcolsep}{10pt}
\renewcommand{\arraystretch}{1.2}
\caption{Comparison of training configurations between the present work and the reference implementation.}
\begin{tabular}{lcc}
\toprule
\textbf{Parameter} & \textbf{This work} & \textbf{Reference run (Ref.~\cite{supercalo})} \\
\midrule
Warm-up epochs & 30 & 18 \\
Cooldown epochs & 46 & 18 \\
Annihilation epochs & 4 & 4 \\
Batch size & 10\,000 & 60\,000 \\
Base learning rate ($\alpha_\text{base}$) & $1\times10^{-5}$ & $2\times10^{-5}$ \\
Maximum learning rate ($\alpha_\text{max}$) & $8\times10^{-4}$ & $1\times10^{-3}$ \\
Optimizer & Adam & Adam \\
Learning rate policy & One-Cycle & One-Cycle \\
Regularization & None & None \\
\bottomrule
\end{tabular}
\label{tab:training_comparison}
\end{table}
During training we encountered numerical instabilities that caused the model weights to diverge to \texttt{NaN}. This issue was resolved by increasing the numerical precision from \texttt{float32} to \texttt{float64}, at the cost of increased memory usage and longer training time.

Minor adjustments to the learning rate schedule or batch size could further refine performance, but the obtained model already provides a solid basis for the subsequent evaluation and comparison with reference results.

\subsection{Qualitative comparison of generated and reference showers}
\begin{figure}[h!]
    \centering
    \includegraphics[width=0.99\linewidth]{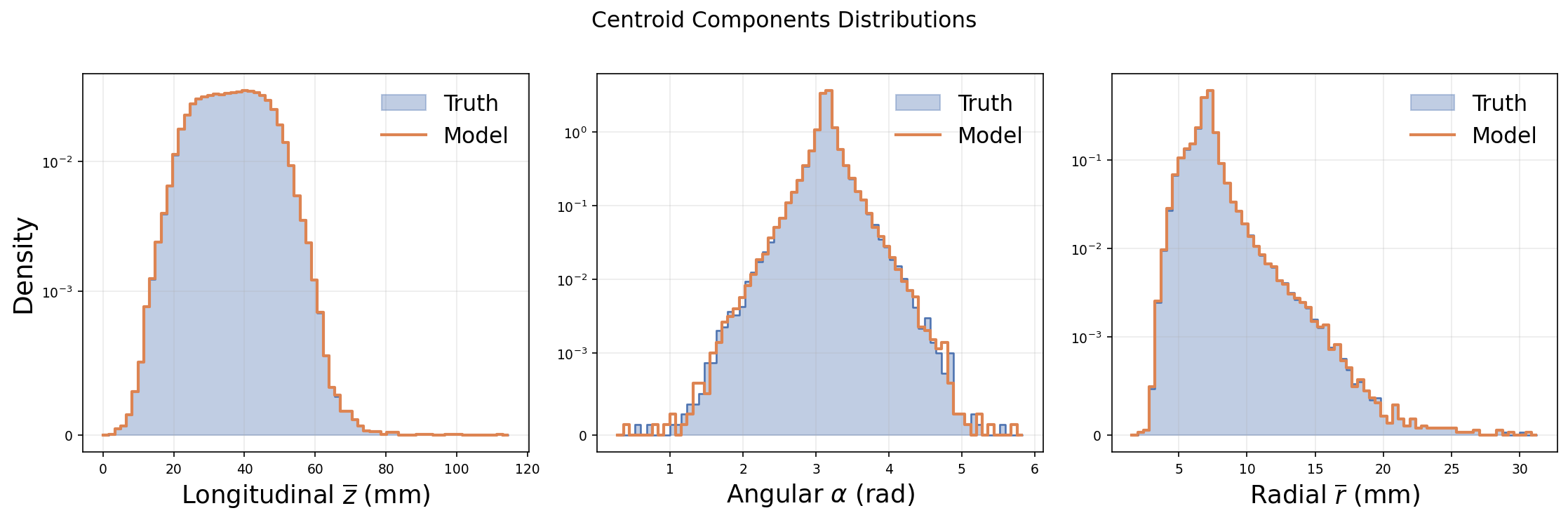}
    \caption{Centroids distributions along the longitudinal, angular and radial direction. Truth samples (\textsc{Geant4}) are filled in blue and generated samples are reported as orange line.}
    \label{fig:centroid}
\end{figure}

\begin{figure}[h!]
    \centering
    \includegraphics[width=0.99\linewidth]{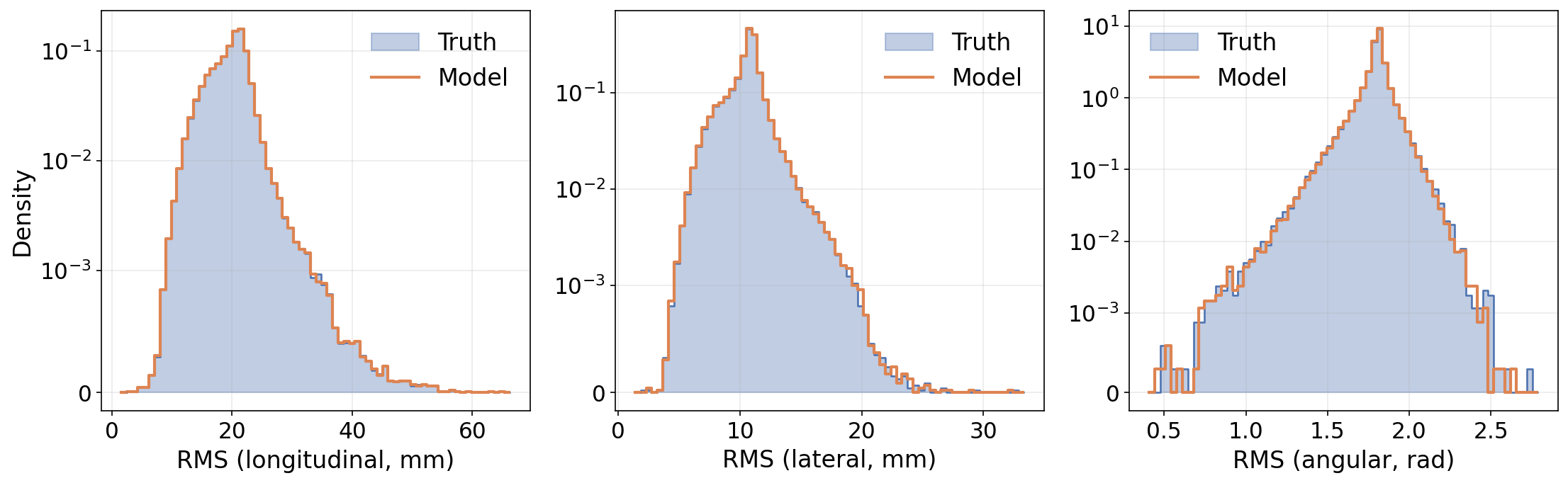}
    \caption{RMS distributions along the longitudinal, lateral and (radial), and angular direction. Reference distribution is filled in blue, while results from generated showers is reported in orange.}
    \label{fig:RMS}
\end{figure}

\begin{figure}[h!]
    \centering
    \includegraphics[width=0.99\linewidth]{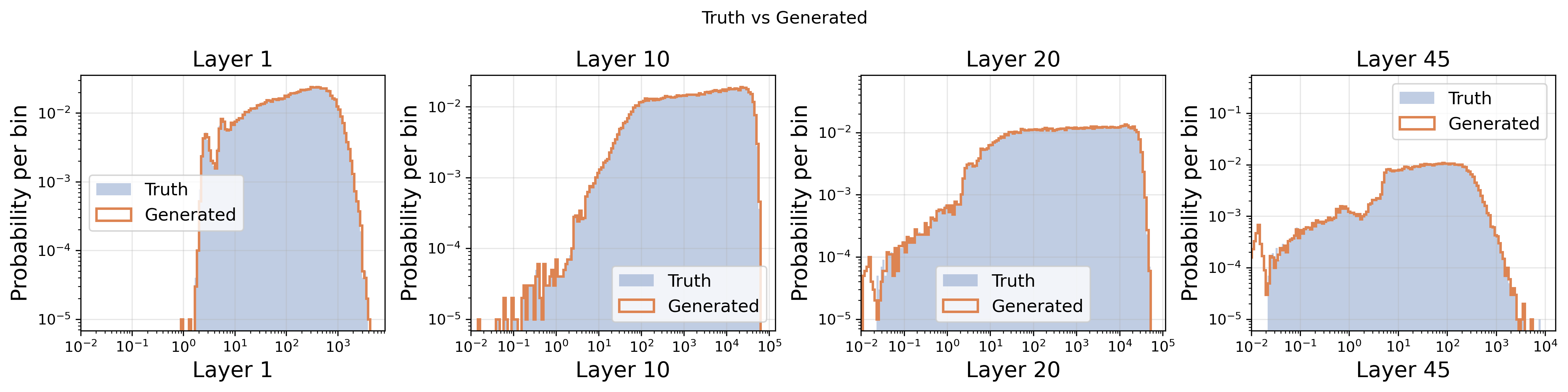}
    \caption{Distributions of the total energy deposited in layers 1, 10, 20 and 45. Can be obtained by summing the energy deposition in each fine voxel in radial and angular direction. The distribution of the \textsc{Geant4} is filled in blue while the orange line represent the generated samples.}
    \label{fig:layer_energies}
\end{figure}

To provide a qualitative visual assessment of the model performance before applying quantitative statistical tests, 
we compare a selection of distributions computed on the generated (sampled) and reference (true) showers. 
Figure~\ref{fig:centroid} shows the normalized histograms of the shower centroids; Figure \ref{fig:RMS} shows the RMS normalized distribution in the longitudinal, lateral, and angular directions. In Figure \ref{fig:layer_energies}, the total energy deposition in layers 1, 10, 20, and 45 is reported. In layers 20 and 45, some artifacts are visible near the small values of energy deposition. We have not been able to explain these artifacts, and they will require an in-depth analysis. 

Besides the artifacts in the layer distributions, we find a good overlap between the reference and generated data indicating that the model was able at capturing the main physical characteristics of the showers.

\subsection{Statistical evaluation}
In this section, the results of the statistical tests introduced in Section \ref{sec:eval_intro} and further discussed in Section \ref{sec:eval_implement} are presented. We begin the discussion by highlighting the fact that the two-sample tests evaluated have some hyperparameters that have only been briefly explored. The presented results are therefore only introductory, and an in-depth exploration is left to future work. The hyperparameters have been introduced in Section \ref{sec:null_hypo} and consist mainly in \code{batch_size} and \code{nslices}. 
\subsubsection{Results at full dimensionality}
Regarding high-dimensional features, which, as introduced previously, play a central role in this thesis, we tested the performance of the model with different hyperparameter values. A summary of the results is presented in Table \ref{tab:fine_pvalues_summary}.

\begin{table}[h!]
\centering
\begin{tabular}{ c  c  c  c  c  c }
 \thead{Feature} & \thead{Metric} & \thead{\texttt{batch\_size}} & \thead{\texttt{n\_slices}} & \thead{$p$-value} & \thead{Significance ($\sigma$)} \\
 \hline
 \hline
 \multirow{3}{*}{Coarse shower} 
   & KS  & \multirow{3}{*}{1000} & -    & 0.495   & $1.3\cdot10^{-2}$ \\
   & SKS &                       & 500  & 0.487   & $3.3\cdot10^{-2}$ \\
   & SWD &                       & 500  & 0.470   & $7.6\cdot10^{-2}$ \\
 \hline
  
 \multirow{3}{*}{Coarse shower} 
   & KS  & \multirow{3}{*}{5000} & -    & 0.503   & $8\cdot10^{-3}$ \\
   & SKS &                       & 500  & 0.505   & $1.3\cdot10^{-2}$ \\
   & SWD &                       & 500  & 0.489   & $2.8\cdot10^{-2}$ \\
  \hline
 \multirow{3}{*}{Fine shower} 
   & KS  & \multirow{3}{*}{1000} & -    & $7.99\cdot10^{-3}$ & 2.41 \\
   & SKS &                       & 1000 & 0.483   & $4.3\cdot10^{-2}$ \\
   & SWD &                       & 1000 & 0.473   & $6.8\cdot10^{-2}$ \\
 \hline
 \multirow{3}{*}{Fine shower} 
   & KS  & \multirow{3}{*}{1000} & -    & $7.99\cdot10^{-3}$ & 2.41 \\
   & SKS &                       & 7000 & 0.480   & $5.0\cdot10^{-2}$ \\
   & SWD &                       & 7000 & 0.469   & $7.8\cdot10^{-2}$ \\
 \hline
 \multirow{3}{*}{Fine shower} 
   & KS  & \multirow{3}{*}{5000} & -    & $9.99\cdot10^{-4}$ & 3.09 \\
   & SKS &                       & 1000 & 0.419   & $2.04\cdot10^{-1}$ \\
   & SWD &                       & 1000 & 0.464   & $0.90\cdot10^{-2}$ \\
 \hline
\end{tabular}
\caption{Summary of $p$-values obtained for different configurations and metrics. }
\label{tab:fine_pvalues_summary}
\end{table}

\begin{figure}[h!]
    \centering
    \includegraphics[width=0.95\linewidth]{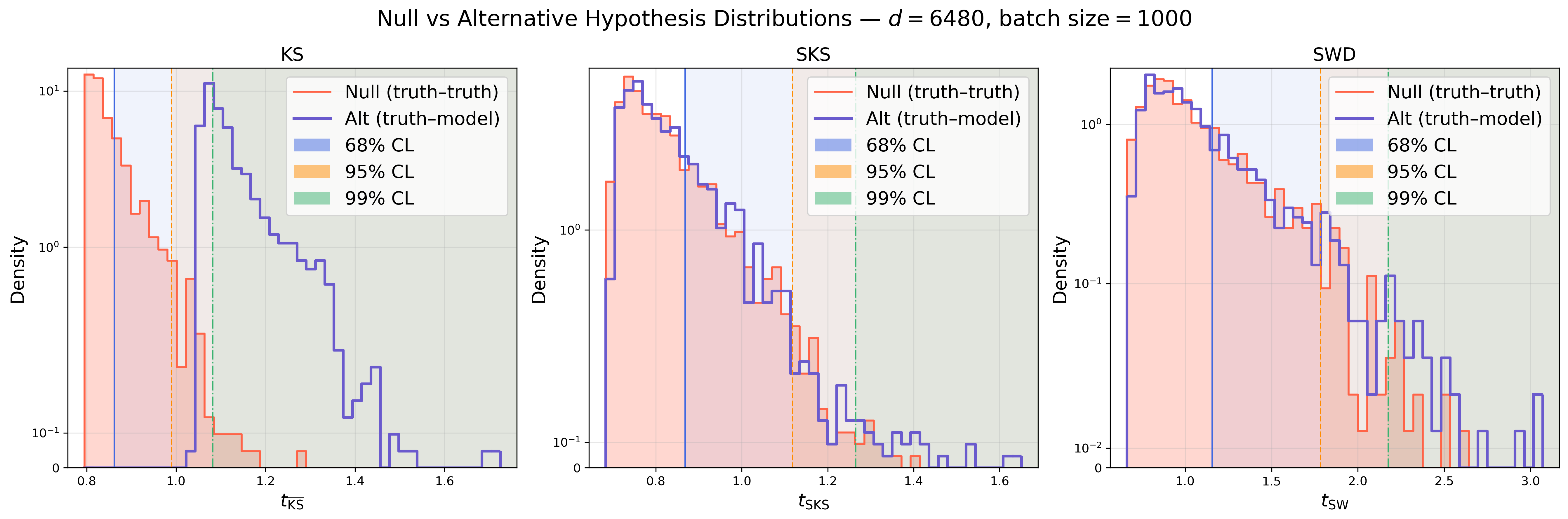}
    \caption{Null vs.\ alternative hypothesis distributions of three test statistics at full dimensionality ($d=6480$; \code{batch_size} $=1\cdot10^3$; \code{nslices} $=1\cdot10^3$). Each panel shows density-normalized histograms on a log $y$-axis: the null (truth–truth, red fill) and the alternative (truth–model, blue outline). From left to right: average Kolmogorov–Smirnov (KS), Sliced KS (SKS), and Sliced Wasserstein distance (SWD). Vertical dashed lines mark the 68\%, 95\%, and 99\% confidence levels estimated from the null; the shaded bands indicate the corresponding right-tailed rejection regions. A rightward shift of the blue curve relative to the red one signals increased power (greater separation from the null).}
    \label{fig:Full_1k_1k}
\end{figure}

\paragraph{Fine representation:} Figure~\ref{fig:Full_1k_1k}, compares the null (truth–truth) and alternative (truth–model) sampling distributions of the KS, SKS, and SWD test statistics for $d=6480$ and batch size $1\cdot10^3$. The separation is most pronounced for the average KS, with most of the bulk falling in the 99\%, while it is almost null for the SKS and SWD metrics. 
This indicates that $H_0$ is rejected with significance $2.41\sigma$ for the average KS, while the other two metrics accept $H_0$.

\begin{figure}[h!]
    \centering
    \includegraphics[width=0.95\linewidth]{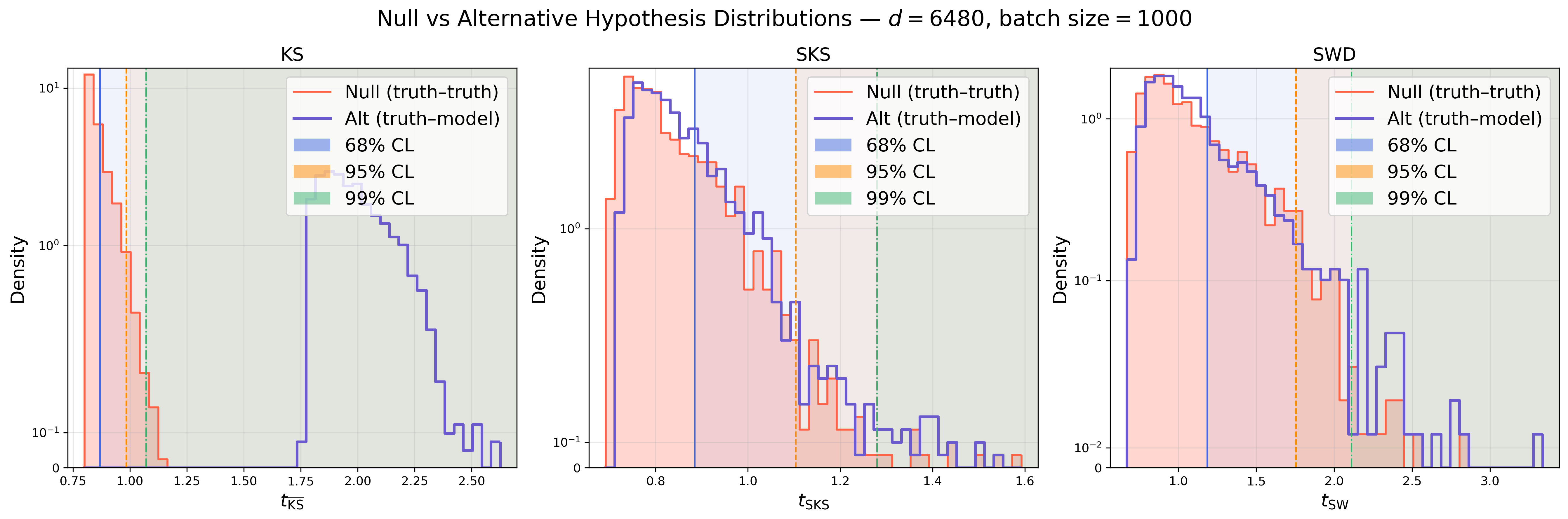}
    \caption{Same conventions as Figure \ref{fig:Full_1k_1k}. Here $d=6480$, \code{batch_size}~$=5\cdot10^3$ and \code{nslices}~$=1\cdot10^3$. Increased the batch size to probe the hyperparameter space.}
    \label{fig:Full_5k_1k}
\end{figure}

Figure~\ref{fig:Full_5k_1k}, shows another comparison between the null and alternative distributions of the same tests as in the previous comparison, for the same dimensionality ($6480$D) but with \code{batch_size}~$=5\cdot10^3$. As expected, the power of all the tests increases slightly (a small shift towards the right for the alternative distribution), even if the sliced test are still not able to detect differences. The power of the average KS test increases, rejecting the null hypothesis with significance $3.09\sigma$ (see Table \ref{tab:fine_pvalues_summary}).

\paragraph{Coarse representation.}
The same comparison implemented for the fine voxel representation is now discussed for the coarse representation. The coarse representation is obtained from the fine one by summing the fine voxel energies according to the process described in Section \ref{sec:voxellization}, resulting in $1\cdot10^5$ samples, each with 648 dimensions (recall Section \ref{sec:voxellization} where choice \texttt{A} was introduced), in which we group 10 fine voxels to build a single coarse voxel. We compare two different choices of \code{batch_size} and different choices of \code{nslices}.

\begin{figure}[h!]
    \centering
    \includegraphics[width=0.95\linewidth]{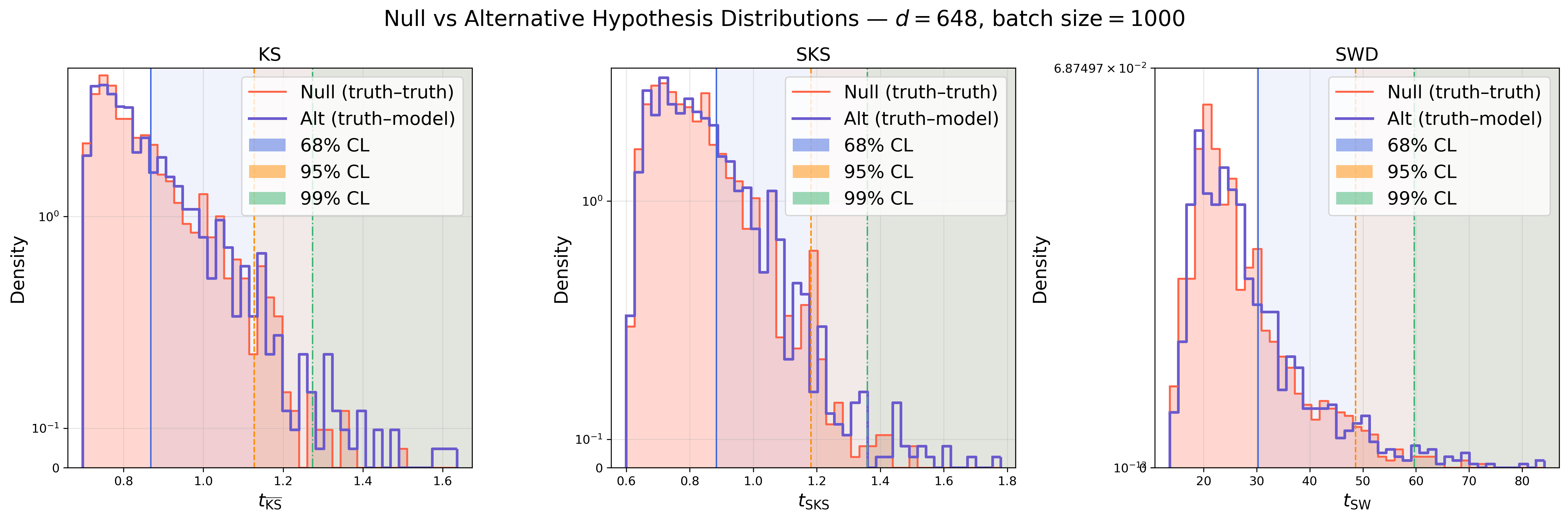}
    \caption{Same conventions as Figure \ref{fig:Full_1k_1k}. Here $d=648$, \code{batch_size}~$=1\cdot10^3$ and \code{nslices}~$=5\cdot10^2$. The metrics are not able to reject $H_0$.}
    \label{fig:Coarse_1k_500}
\end{figure}

\begin{figure}
    \centering
    \includegraphics[width=0.95\linewidth]{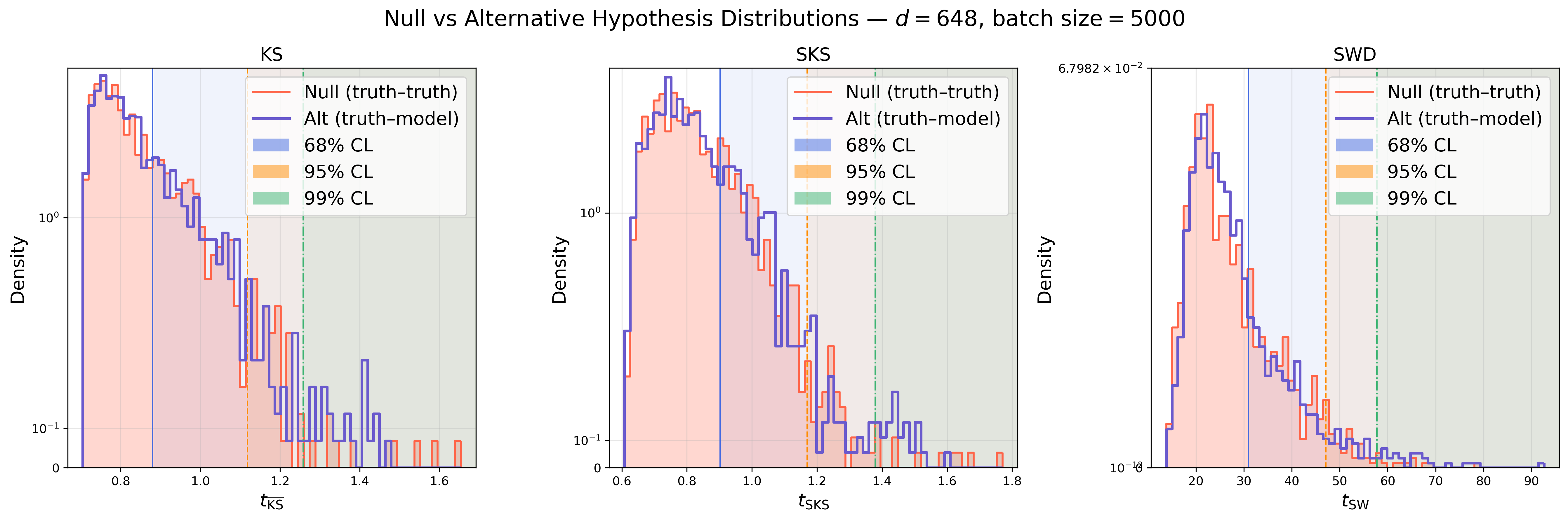}
    \caption{Same conventions as Figure \ref{fig:Full_1k_1k}. Here $d=648$, \code{batch_size}~$=5\cdot10^3$ and \code{nslices}~$=5\cdot10^2$. The metrics are not able to reject $H_0$.}
    \label{fig:Coarse_5k_500}
\end{figure}

Figure~\ref{fig:Coarse_1k_500}, shows the comparison between the test statistics distribution for the null and alternative hypotheses with \code{batch_size}~$=1\cdot10^3$ and \code{nslices}~$=5\cdot10^2$. The plot shows (from left to right: average KS, SKS, and SWD), that none of the test statistics evaluated were able to distinguish between the truth \textsc{Geant4} samples and the generated samples. 
Figure~\ref{fig:Coarse_5k_500} shows the same metrics evaluated on \code{batch_size} $=5\cdot10^3$. This plot shows, according to the previous results, that the metrics are not able to reject $H_0$.

\subparagraph{Learning the conservation of energy.} The difficulties of the two-sample tests described above in distinguishing the true data from the generated samples suggest that the model has learned the conservation of energy in each coarse voxel. Although no hard constraint forces the sums to match, maximum-likelihood training on data satisfying $\sum_{v\in b}E_v = C_b$ encourages the conditional model $p_\theta(\text{fine}\mid C,\ldots)$ to assign probability mass to fine patterns whose totals equal the provided $C$. Empirically, this yields $\mathbb{E}_\theta[\sum_{v\in b}\hat E_v\mid C_b]\approx C_b$ and small event-wise residuals, so the marginal distribution of re-aggregated coarse energies coincides with that of the conditioning inputs. Accordingly, coarse-level tests have little power here, and validation should focus on the fine-scale allocation and correlations within each block.

\begin{figure}[h!]
    \centering
    \includegraphics[width=0.95\linewidth]{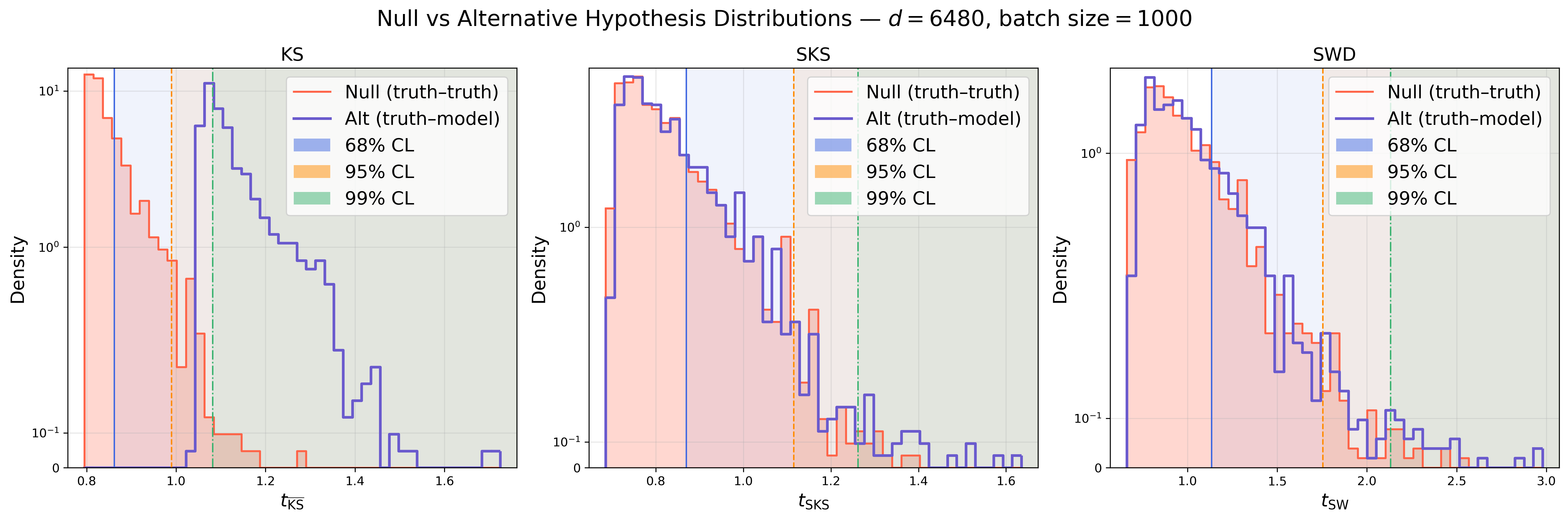}
    \caption{Same conventions as Figure \ref{fig:Full_1k_1k}. Here $d=6480$, \code{batch_size}~$=1\cdot10^3$ and \code{nslices} $=7\cdot10^3$ to investigate the role of the slicing in the discrimination performance of the tests. }
    \label{fig:Full_1k_7k}
\end{figure}

\paragraph{Under-performing of slice-based tests in this setting.}
Increasing the number of slices from $10^3$ to $7\times 10^3$ at fixed batch size does not change the conclusion (Figure \ref{fig:Full_1k_7k} versus Figure \ref{fig:Full_1k_1k}). This behavior is expected when the discrepancies between truth and model are \emph{sparse} across coordinates, i.e., confined to a comparatively small subset of the $D=6480$ marginals—rather than being global. Coordinate-wise tests such as the average KS examine each marginal directly and therefore “see” every affected voxel with the full sample size, which yields strong separation under $H_1$. In contrast, slice-based tests (SKS, SWD) first project the $D$-dimensional vectors onto random directions $\boldsymbol{u}\in\mathbb{S}^{D-1}$ and then apply a 1D test. Let $\boldsymbol{\delta}\in\mathbb{R}^D$ denote the (signed) discrepancy across coordinates, supported on $k\ll D$ entries. For a random slice, one has
\begin{displaymath}
\mathbb{E}_{\boldsymbol{u}}\!\left[(\boldsymbol{u}^\top \boldsymbol{\delta})^2\right]
= \frac{\|\boldsymbol{\delta}\|_2^2}{D}
\quad\Longrightarrow\quad
\lvert \boldsymbol{u}^\top \boldsymbol{\delta}\rvert \sim \|\boldsymbol{\delta}\|_2\,D^{-1/2}
\propto \sqrt{\tfrac{k}{D}},
\end{displaymath}

so the 1D signal in a typical slice is attenuated by a factor $\sqrt{k/D}$ due to \emph{dilution} over mostly well-modeled coordinates (a concentration-of-measure effect in high dimension). Averaging over more slices primarily reduces the Monte Carlo variance of this estimate ($\propto 1/\sqrt{\text{nslices}}$) but does not increase its median (or mean); hence moving from $10^3$ to $7\times 10^3$ slices leaves the null–alt separation essentially unchanged. By contrast, increasing the batch size $n$ boosts the power of all tests roughly as $\sqrt{n}$, which explains the modest rightward shift of the alternative curves when going from $n=10^3$ to $n=5\times 10^3$ (Figure \ref{fig:Full_5k_1k}). Summarizing, the strong performance of KS, together with the weak SKS/SWD separation at $d=6480$, is consistent with voxel-local (sparse) biases rather than broad geometry-level mismodeling.

\paragraph{Summary.} Regarding the evaluation of the full, fine-shower feature, the average KS was able to reject the Null hypothesis $H_0$ with a significance of $2.41\sigma$ for \code{batch_size}~$= 1\cdot10^3$ and $3.09\sigma$ for \code{batch_size}~$= 5\cdot10^3$; slice-based methods were not able to reject the hypothesis $H_0$, and we provided a possible explanation for the low power of these methods in such high-dimensional settings. In the evaluation of the coarse representation of the shower, we found that none of the tests, with the \code{batch_size} choices analyzed, were able to reject the null hypothesis, suggesting that a form of energy conservation was efficiently learned.

\subsubsection{Physically inspired observables}
In this section we present the results obtained from the evaluation of the physically-inspired features presented in Section \ref{sec:eval_implement} with the metrics introduce in Section \ref{sec:eval_intro}. Although the dimensionality of most of the features presented in this section is very low (with the exception of the layer wise energy deposits which is 45D), we chose to evaluate the performances mainly on the IPM-based statistics, i.e., the average KS, sliced KS and sliced WD. This choice is motivated by the large difference in execution time and by the fact that in Ref.~\cite{ref_the_ref}, comparable results have been found.

\begin{table}[h]
\tiny
\centering
\begin{tabular}{ c c c  c  c  c  c }
 \thead{Feature} & \thead{Dimensions} & \thead{Metric} & \thead{\texttt{batch\_size}} & \thead{\texttt{n\_slices}} & \thead{$p$-value} & \thead{Significance ($\sigma$)} \\
 \hline
\hline
 \multirow{5}{*}{Layer centroid} & \multirow{5}{*}{3}& KS  & \multirow{5}{*}{1000} & -    & 0.505 & 0.0125 \\
  & & SKS &   & 100.0 & 0.494 & 0.0150 \\
  & & SWD &   & 100.0 & 0.440 & 0.151 \\
  & & MMD &   & -     & 0.498 & 0.00501 \\
  & & FGD &   & -     & 0.476 & 0.0601 \\
 \hline
 \multirow{5}{*}{Layer centroid}& \multirow{5}{*}{3} & KS  & \multirow{5}{*}{5000} & -    & 0.462 & 0.0954 \\
  & & SKS &   & 100.0 & 0.534 & 0.0853 \\
  & & SWD &   & 100.0 & 0.486 & 0.0351 \\ 
  & & MMD &   & -     & - &  \\
  & & FGD &   & -     & - &  \\
  \hline
 \multirow{5}{*}{Layer centroid}& \multirow{5}{*}{3} & KS  & \multirow{5}{*}{10000} & -    & 0.385 & 0.292  \\
  & & SKS &   & 100.0 & 0.468 & 0.0803 \\
  & & SWD &   & 100.0 & 0.494 & 0.0150 \\ 
  & & MMD &   & -     & - & - \\
  & & FGD &   & -     & - & - \\
 \hline
 \multirow{5}{*}{Layer energy}& \multirow{5}{*}{45} & KS  & \multirow{5}{*}{1000} & -    & 0.521 & 0.0527 \\
  & & SKS &   & 100.0 & 0.507 & 0.0175 \\
  & & SWD &   & 100.0 & 0.482 & 0.0451 \\ 
  & & MMD &   & -     & - & - \\
  & & FGD &   & -     & - & - \\
 \hline
 \multirow{5}{*}{Layer energy}& \multirow{5}{*}{45} & KS  & \multirow{5}{*}{5000} & -    & 0.406 & 0.238 \\
  & & SKS &   & 100.0 & 0.476 & 0.0602 \\
  & & SWD &   & 100.0 & 0.474 & 0.0652 \\ 
  & & MMD &   & -     & - & - \\
  & & FGD &   & -     & - & - \\
 \hline
 \multirow{5}{*}{Layer energy}& \multirow{5}{*}{45} & KS  & \multirow{5}{*}{10000} & -    & 0.330 & 0.440 \\
  & & SKS &   & 100.0 & 0.496 & 0.0100 \\
  & & SWD &   & 100.0 & 0.418 & 0.207 \\ 
  & & MMD &   & -     & - & - \\
  & & FGD &   & -     & - & - \\
 \hline
 \multirow{5}{*}{RMS}& \multirow{5}{*}{3} & KS  & \multirow{5}{*}{1000} & -    & 0.460 & 0.100 \\
  & & SKS &   & 100.0 & 0.488 & 0.0300 \\
  & & SWD &   & 100.0 & 0.505 & 0.0125 \\ 
  & & MMD &   & -     & 0.496 & 0.0100 \\
  & & FGD &   & -     & 0.490 & 0.0250 \\
 \hline
 \multirow{5}{*}{RMS}& \multirow{5}{*}{3} & KS  & \multirow{5}{*}{5000} & -    & 0.408 & 0.233 \\
  & & SKS &   & 100.0 & 0.529 & 0.0200 \\
  & & SWD &   & 100.0 & 0.454 &   0.116  \\ 
  & & MMD &   & -     & - & - \\
  & & FGD &   & -     & - & - \\
 \hline
  \multirow{5}{*}{RMS}& \multirow{5}{*}{3} & KS  & \multirow{5}{*}{10000} & -    & 0.308 & 0.502 \\
  & & SKS &   & 100.0 & 0.434 & 0.166 \\
  & & SWD &   & 100.0 & 0.446 & 0.136 \\
  & & MMD &   & -     & - & - \\
  & & FGD &   & -     & - & - \\
 \hline
  \multirow{5}{*}{Total energy}& \multirow{5}{*}{1} & KS  & \multirow{5}{*}{5000} & -    & 0.474 & 0.0652 \\
  & & SKS &   & - & - & - \\
  & & SWD &   & - & - & - \\
  & & MMD &   & -     & 0.464 & 0.0904 \\
  & & FGD &   & -     & 0.454 & 0.116 \\
 \hline
\end{tabular}
\caption{Summary of $p$-values from all the evaluation of physically inspired features.}
\label{tab:low_dim_table}
\end{table}

Table~\ref{tab:low_dim_table} summarizes the hypothesis tests on the low-dimensional, physics-motivated features introduced in Section \ref{sec:eval_implement}. Overall, the $p$-values cluster around $0.3$–$0.55$ across features (layer centroids, layer energies, shower RMS, and incident energy) and metrics (average KS, SKS, SWD, with occasional MMD/FGD). Hence, for all tested configurations we do \emph{not} reject $H_0$ at conventional significance levels. In other words, none of the evaluated metrics, in these settings, was able to distinguish the sampled data from the truth.

Regarding the role of sample size, we observe the expected trend that increasing \code{batch_size} generally improves power (smaller $p$-values). This is most evident for KS on the layer-wise energy deposits ($45$D), where $p$ decreases from $0.521$ at \code{batch_size} $=10^3$ to $0.406$ at $5\cdot10^3$ and $0.330$ at $10^4$, and is also visible for KS on the shower RMS ($3$D), from $0.460$ to $0.408$ to $0.308$. SWD often follows a similar pattern (e.g., layer energy: $0.482 \to 0.474 \to 0.418$). By contrast, SKS can behave non-monotonically at fixed \code{nslices} (e.g., layer centroid: $0.494 \to 0.534 \to 0.468$), which is consistent with the additional Monte Carlo variability introduced by random projections and with the dilution effects already discussed for high-dimensional slices in Section \ref{sec:eval_intro}.

Finally, we note a tension between qualitative visual checks of some one-dimensional marginals (where small artifacts are visible) and the formal tests, which still return relatively large $p$-values. At present, we do not have a definitive explanation. Possible contributors include the small effect size of localized discrepancies (e.g., restricted to tails or to a subset of coordinates), averaging across coordinates (for KS) or across random directions (for SKS/SWD), and limited sensitivity of the chosen hyperparameters (\code{batch_size}, \code{nslices}) to such localized differences. A more systematic study, varying \code{batch_size} and \code{nslices} more widely, using targeted tests on the specific regions where artifacts appear, and exploring complementary metrics (e.g., tuned kernels for MMD/FGD), is left to future work.

\paragraph{Why many $p$-values exceed $0.5$.}
The frequent occurrence of $p>0.5$ in Table \ref{tab:low_dim_table} and Table \ref{tab:fine_pvalues_summary} can be explained by two effects of our setup and of the learned model.
First, we summarize the alternative by aggregating many replicates (e.g.\ taking the median across replicate values of the test statistic), while the empirical null is shown as single-replicate draws. This creates a variance asymmetry: the aggregation on the alternative side acts as a shrinker and removes occasional large excursions of sup- or distance-based statistics. A two-sample statistic can be sketched as
\[
T \;\approx\; S(F,G) \;+\; N,
\]
where $S(F,G)$ is a small structural shift between truth $F$ and model $G$, and $N$ is sampling noise. Aggregating many alternative replicates reduces the contribution of $N$, so the observed (aggregated) alternative can lie to the left of typical single null replicates even when $S(F,G)\neq 0$.

Second, the trained model shows signs of \emph{variance shrinkage} (milder tails and smoother fluctuations) compared to truth. In a truth--truth comparison, both samples explore the full tail variability, producing a wider envelope of empirical fluctuations. In a truth--model comparison, one side is tighter, so supremum-type distances (average KS) or sliced distances (SKS, SWD) can \emph{decrease} because random fluctuations are smaller, partly masking a small deterministic shift. Under our right-tailed convention (``larger statistic $\Rightarrow$ evidence against $H_0$''), this leads naturally to large $p$-values.

In summary, a large right-tailed $p$-value means we lack evidence for an \emph{inflation}-type discrepancy (increased distance). If the difference takes the form of \emph{contraction} (reduced spread or tail occupancy), a left-tailed or two-sided calibration may be more sensitive. A systematic study of these options is left to future work.
\chapter{Lessons Learned and What Comes Next} \label{chap:conclusion}
\section{Summary of Findings}

This thesis addresses calorimeter shower super-resolution by independently replicating the results of Ref.~\cite{supercalo}. Building upon the code framework from Ref.~\cite{Coccaro_2024}, we implemented a \emph{Conditional normalizing Flow}, specifically a \emph{Masked Autoregressive Flow (MAF)}~\cite{maf}. The architecture leverages the \emph{Masked Autoencoder for Distribution Estimation (MADE)}~\cite{made} to parametrize \emph{Rational Quadratic Spline (RQS)} transformations. Training was performed on Dataset~2 from the \emph{Fast Calorimeter Simulation Challenge (CaloChallenge)}, a community driven benchmark for advancing high fidelity fast calorimeter simulation. This dataset comprises two files, each containing $100$k full calorimeter shower events with $6480$ fine voxel energies per shower, one file for training and one for evaluation.

The task was formulated by aggregating neighboring fine voxels to construct a \emph{coarse representation}, from which the model generates the underlying fine voxel energies. Conditioning features include the incident particle energy, neighboring coarse voxel energies, one-hot encoded position in the calorimeter, the examined coarse voxel energy, and the sum of fine voxel energies within each coarse voxel. While this last feature deviates from a pure super-resolution setting, it was included to stabilize training and validate the pipeline as a proof-of-concept.

Training proceeded for 80 epochs using the \code{OneCycle}~\cite{onecycle} learning rate scheduler, requiring approximately $27.5$ hours on a single Nvidia L40S GPU. Initial training instability was resolved by increasing numerical precision from \texttt{float32} to \texttt{float64}. After hyperparameter tuning, we achieved stable, monotonic improvements throughout training.

Performance evaluation employed a statistically robust two sample testing methodology from Ref.~\cite{ref_the_ref}. Unlike many existing approaches for generative models, which lack rigorous statistical foundations, this framework provides reliable assessment. We applied this formalism to both very high dimensional features, the full shower (6480D) and its coarse representation (648D), as well as physically motivated lower dimensional features: energy centroid coordinates (3D), RMS along longitudinal, lateral, and angular directions (3D), total incident energy (1D), and layer wise energy deposits (45D).\\
\\
\noindent The preliminary statistical evaluation (Section \ref{sec:eval_intro}, Section \ref{sec:eval_implement}) indicates a clear pattern across dimensionalities and hyperparameters. At full dimensionality ($d=6480$), the average Kolmogorov--Smirnov (KS) test distinguishes the model from truth with increasing significance as the sample size grows: for \texttt{batch\_size} $=1\cdot10^3$ we observe $p=7.99\times10^{-3}$ ($2.41\sigma$; Figure \ref{fig:Full_1k_1k}), which strengthens to $p=9.99\times10^{-4}$ ($3.09\sigma$) for \texttt{batch\_size} $=5\cdot10^3$ (Figure \ref{fig:Full_5k_1k}); see Table \ref{tab:fine_pvalues_summary}. In contrast, slice-based tests (Sliced KS (SKS) and Sliced Wasserstein Distance (SWD)) do not show separation in this setting, remaining consistent with the Null hypothesis ($p\simeq 0.42$--$0.48$ across configurations), that means that the true and generated samples can not be distinguished. Increasing the number of slices from $10^3$ to $7\cdot10^3$ at fixed \texttt{batch\_size} leaves this conclusion unchanged (Figure \ref{fig:Full_1k_7k}), which is consistent with a discrepancy that is \emph{sparse} across coordinates at $d=6480$: coordinate-wise tests like the averaged KS retain sensitivity to voxel-local biases, while random projections dilute such localized effects and primarily benefit from larger sample size rather than more slices. At the coarse representation level ($d=648$), none of the tests reject $H_0$ across the explored hyperparameters (\texttt{batch\_size} $=1\cdot10^3$ and $5\cdot10^3$, \texttt{nslices} $=5\cdot10^2$; Figures ~\ref{fig:Coarse_1k_500}--\ref{fig:Coarse_5k_500}), which is compatible with the model having effectively learned coarse-block energy conservation from the conditioning: re-aggregated fine samples match the conditioning totals, leaving coarse-level marginals virtually indistinguishable from truth (Section \ref{sec:voxellization}). Overall, these results suggest (i) detectable but localized mismodeling at full dimensionality captured by the averaged KS, (ii) limited power of sliced tests under sparse discrepancies at fixed statistics, and (iii) strong agreement at the coarse level consistent with learned energy conservation. We emphasize that these findings are preliminary: only a modest region of the hyperparameter space (\texttt{batch\_size}, \texttt{nslices}) has been explored so far.

For physically inspired low-dimensional features (3D centroids, 3D RMS, 45D layer energies, 1D incident energy), $p$-values typically lie in the $0.3$--$0.55$ range and do not indicate mismodeling; we still see the expected trend that larger \texttt{batch\_size} improves power, e.g.\ KS on layer energies decreases from $0.521$ ($10^3$) to $0.406$ ($5\times10^3$) to $0.330$ ($10^4$), with a similar pattern for SWD ($0.482\to0.474\to0.418$), while SKS can be non-monotonic due to projection variability. Taken together, these findings point to localized discrepancies at the fine-voxel level (picked up by averaged KS at $d=6480$), strong agreement after coarse re-aggregation (consistent with learned energy conservation), and generally large right-tailed $p$-values in low dimensions, which we argued can arise from aggregation asymmetry and variance shrinkage in the model. A broader sweep over \texttt{batch\_size} and \texttt{nslices}, and targeted tests for specific marginal artifacts, are left to future work.

\section{Physics Implications}

As outlined in the Introduction, this approach has several implications for reconstruction quality, analysis sensitivity, and computing strategy. Learning a mapping from coarse to fine segmentation enables recovery of information that would otherwise require expensive high granularity simulation or hardware upgrades, while remaining compatible with existing reconstruction workflows.

Restoring fine-grained energy patterns reduces bias and variance in reconstructed energy while improving position estimates. Feature level metrics on centroids and longitudinal, lateral, and angular RMS (see Section \ref{sec:results}) indicate that salient shower shapes are retained, supporting improved calibration and more stable response across layers. This preservation of shower shape variables is particularly relevant for particle identification, as the method virtually increases segmentation, sharpening features that feed particle identification taggers and clustering algorithms.

Beyond reconstruction improvements, higher effective granularity helps mitigate pile up confusion at the HL-LHC by improving topological clustering, isolation definitions, and separation of nearby deposits. Learned upsampling based on coarse observables can assist reconstruction under high occupancy without detector modifications. The approach also offers a route to compensate for effective granularity loss from aging, masked cells, or noisy readouts. By inferring local energy patterns from available coarse measurements, it can potentially recover performance in degraded regions without hardware replacement, particularly attractive for long term operations where granular replacements are impractical.

From a computational perspective, using this method as a post processor for coarse, fast simulation can deliver fine-grained showers at a fraction of the Geant4 cost, reducing CPU, memory, and storage requirements. The generative nature also enables controlled variations for systematic studies (e.g., layer wise response or lateral spread), facilitating stress tests of analysis selections. Integration into reconstruction chains is straightforward: the method can be inserted after coarse clustering and calibration, before high level particle identification and object building, with minimal interface changes since it relies on observables already present in reconstruction. For online use, inference speed is the main constraint; autoregressive flows may require architectural modifications (e.g., IAF variants, distillation, batching, or hardware acceleration) before inclusion in trigger paths, whereas offline workflows can adopt the method sooner.

Conditional normalizing flows can thus enhance the fidelity of calorimeter information available to physics analyses, ease computing pressure by replacing portions of fine-grained simulation, and provide a practical tool to counteract pile up and aging effects. The approach represents a viable candidate for integration into end to end HEP workflows.

\section{Limitations}
This thesis's conclusions are bounded by assumptions in the data domain, modeling choices, and evaluation methodology. Training and testing were restricted to a single simulated geometry and dataset (Par04, CaloChallenge Dataset~2; see Section \ref{sec:dataset}), without explicitly modeling detector effects such as time dependent aging, coherent noise, out of time pile up, and masked channels. Generalization to other geometries, operating conditions, or real data therefore remains unverified.

For stabilization and pipeline validation, one conditioning feature included information unavailable in a strict super-resolution setting (the sum of fine voxel energies within each coarse voxel). This simplifies part of the task and should be viewed as a proof-of-concept choice to be removed in future work.

The adopted autoregressive architecture (MAF with MADE conditioner and RQS transforms) offers flexibility but incurs sequential sampling and higher inference latency. Numerical stability required \texttt{float64} precision (Section \ref{sec:training}), increasing memory and compute costs.

Performance may be sensitive to conditioning design, permutation strategy, spline hyperparameters, and preprocessing near zero energies. These factors were not exhaustively explored. Approximate energy conservation is encouraged through conditioning and evaluation rather than enforced by design, so small residual biases, particularly in rare tails, cannot be excluded.

The evaluation framework employs a statistically robust two sample approach with Truth–Truth baselines (Section \ref{sec:eval_intro}), controlling false discoveries but with finite statistical power. Depending on metric choice and hyperparameters (e.g., number of slices or batch size), very local mismatches may go undetected. Per event calibrated uncertainty estimates are not provided, and the response has not been anchored to control samples, limiting immediate use in precision analyses requiring uncertainty propagation.

Finally, computational budget constraints limited the hyperparameter search and training schedule. The reported behavior should be considered representative of the chosen configuration rather than globally optimal. Full bit wise reproducibility across platforms and drivers is not guaranteed.

\section{Outlook and Future Work}
High fidelity calorimeter shower super-resolution can provide valuable leverage to address challenges posed by the HL-LHC and future colliders such as the \emph{Future Circular Collider (FCC)} or the \emph{Muon Collider}. 
This thesis left many open questions and several promising directions for future research:
\begin{enumerate}
    \item Transition to a genuine super-resolution task by removing the conditioning on fine voxel energy sums, and addressing practical scenarios such as missing values from dead cells or noisy readouts.
    \item Replace MAF with an \emph{Inverse Autoregressive Flow (IAF)}, increasing sampling speed while maintaining fidelity.
    \item Explore the hyperparameter space systematically to achieve \emph{super-convergence} and reduce training time.
    \item Enhance computational efficiency through further optimization of the current \texttt{TensorFlow2} implementation.
    \item Conduct systematic performance assessment through structured exploration of the hyperparameter space.
    \item Update the architecture to state-of-the-art models such as \emph{Conditional Flow Matching} and \emph{Diffusion models}, which have demonstrated promising results in fast, high fidelity shower generation.
\end{enumerate}

\noindent The independent implementation, conditioning strategy, and statistically robust evaluation demonstrate that conditional normalizing flows can provide analysis grade super-resolution. With optimized inference and speed improvements, the approach represents a credible candidate for integration into modern HEP reconstruction chains.


\newpage
\thispagestyle{empty}
\mbox{}
\newpage

\cleardoublepage
\begin{acknowledgements}
  \addcontentsline{toc}{section}{Acknowledgements}
\thispagestyle{plain}  
I would like to thank my supervisors, \textbf{Dr.~Riccardo Torre}, \textbf{Dr.~Marco Letizia}, and my reviewer \textbf{Dr.~Andrea Coccaro}, for their guidance, patience, and many constructive discussions throughout this work. Their feedback and support were essential at every stage of the thesis.

I also thank the INFN Sezione di Genova and the University of Genoa for the opportunity to carry out this research. Computations were performed on the \textit{Teo} GPU cluster.
\end{acknowledgements}
\cleardoublepage

\newpage
\thispagestyle{empty}
\mbox{}
\newpage

\appendix

\chapter{More on Loss Functions} \label{app:loss}

This appendix provides a concise theoretical overview of additional loss functions widely used in machine learning and information theory. In particular, it focuses on the Kullback--Leibler (KL) divergence and the cross-entropy loss, both of which establish a formal connection between probabilistic modeling and optimization in statistical learning. These quantities play a fundamental role in understanding likelihood-based objectives such as the Negative Log-Likelihood (NLL).

\section*{Kullback--Leibler Divergence (KL)}

Given two probability density functions (PDFs) defined over the same variable $X$, denoted by $p$ and $q$, the KL divergence (or KL distance) measures how one distribution diverges from the other. In the context of unsupervised learning, it is defined as:
\begin{displaymath}
\begin{split}
    \mathcal{D}_{\text{KL}} (p \rVert q)
    &= \mathbb{E}_{X \sim p}\!\left[\log\!\left(\frac{p(X)}{q(X)}\right)\right]
    = \int p(\mathbf{x}) \log\!\left(\frac{p(\mathbf{x})}{q(\mathbf{x})}\right) d\mathbf{x}\\
    &= \int p(\mathbf{x}) \big(\log p(\mathbf{x}) - \log q(\mathbf{x})\big) d\mathbf{x}.
\end{split}
\end{displaymath}

An empirical estimation of this integral can be obtained by sampling $N$ points from the distribution $p(\mathbf{x})$ and computing the average of the logarithmic ratio between the PDFs:
\begin{displaymath}
    \mathcal{D}_{\text{KL}} (p \rVert q) \approx \frac{1}{N}\sum_{i = 1}^{N} \log\!\left(\frac{p(\mathbf{x}_i)}{q(\mathbf{x}_i)}\right).
\end{displaymath}

In the case of supervised tasks, each PDF becomes a conditional one, as discussed in the overview of the NLL loss. If the distribution of the output is assumed to be Gaussian, and we define $p(\mathbf{Y|X})$ as the true output PDF and $q(\mathbf{Y|X};\mathbf{\theta})$ as the model-predicted PDF, one can show that minimizing the KL divergence between the predicted and true PDFs again reduces to minimizing the MSE loss up to constant terms.

In information theory, the KL divergence belongs to a broader class of divergences used to measure the distance between probability distributions, known as \emph{f}-divergences.

\section*{Cross-Entropy}

The term \emph{cross-entropy} originates from the concept of \emph{differential entropy}, which generalizes the notion of Shannon entropy to continuous variables.

\subsection*{Differential Entropy}

Differential entropy is the continuous analogue of the \textbf{Shannon entropy}, which measures the average amount of information produced by a discrete random variable. It is defined as the expected value of the information content (also called \emph{self-information} or \emph{surprisal}) of the variable.  
For a variable $\mathbf{x}$ distributed according to the Probability Mass Function (PMF) $P(\mathbf{x})$, the information content is defined as:
\begin{displaymath}
    I(\mathbf{x}) = \log \frac{1}{P(\mathbf{x})} = - \log P(\mathbf{x}),
\end{displaymath}
and it quantifies the uncertainty associated with the occurrence of the event $\mathbf{x}$.

The Shannon entropy is then given by:
\begin{displaymath}
    H(\mathbf{X}) = \mathbb{E}_{\mathbf{X} \sim P}[I(\mathbf{X})]
    = \mathbb{E}_{\mathbf{X} \sim P}[- \log P(\mathbf{X})] 
    \approx - \sum_{\mathbf{x} \in \chi} P(\mathbf{x}) \log P(\mathbf{x}),
\end{displaymath}
where $\chi$ is the \textbf{sample space}, i.e.\ the set of all possible outcomes of the random variable $\mathbf{X}$. A higher entropy corresponds to a more unpredictable variable.

This concept can be extended to continuous variables. Given a continuous random variable $\mathbf{X}$ with PDF $p(\mathbf{X})$, the \textbf{differential entropy} is defined as:
\begin{displaymath}
    H(\mathbf{X}) = \mathbb{E}_{\mathbf{X} \sim p}[I(\mathbf{X})]
    = - \int p(\mathbf{X}) \log p(\mathbf{X}) \, d\mathbf{X}.
\end{displaymath}
In contrast to Shannon entropy, differential entropy can take negative values and is not invariant under a change of variables. Because it depends on the scale of the PDF, it is not directly comparable across different variables and cannot be interpreted as a direct measure of uncertainty. It quantifies the spread of a continuous distribution but not necessarily the amount of information it contains.

For a finite sample $\{\mathbf{x}_i\}$ of $N$ points drawn from $p(\mathbf{X})$, the differential entropy can be approximated as:
\begin{displaymath}
    H(\mathbf{X}) \approx - \frac{1}{N} \sum_{i = 1}^{N} \log p(\mathbf{x}_i).
\end{displaymath}
From this expression, we can see that, up to normalization factors, for a finite sample the differential entropy is approximated by the negative log-likelihood of the data under the true underlying distribution.

\subsection*{From KL Divergence to Cross-Entropy}

The cross-entropy loss can be obtained by rewriting the KL divergence as:
\begin{align*}
    \mathcal{D}_{\text{KL}} (p \rVert q) 
    &= \mathbb{E}_{X \sim p}[\log p(\mathbf{X}) - \log q(\mathbf{X})] \\
    &= -H(p) - \mathbb{E}_{X \sim p}[\log q(\mathbf{X})].
\end{align*}
We can identify the cross-entropy as the term
\[
    H(p,q) \vcentcolon= - \mathbb{E}_{X \sim p}[\log q(\mathbf{X})].
\]
Following the same reasoning as with differential entropy, if we have a finite dataset, the cross-entropy can be approximated as:
\begin{displaymath}
    H(p,q) = - \mathbb{E}_{X \sim p_{\text{data}}}[\log q(\mathbf{X}; \mathbf{\theta})]
    \approx - \frac{1}{N} \sum_{i = 1}^{N} \log q(\mathbf{x}_i; \mathbf{\theta}).
\end{displaymath}
Thus, the cross-entropy loss corresponds to the NLL (up to a constant factor) of the data given the model parameters. For supervised learning tasks, the only difference is that every PDF is replaced by its conditional counterpart.

Although the multiplicative and additive constants appearing across these different loss definitions are theoretically irrelevant to the optimization objective, they can significantly affect the numerical behavior of the algorithms used for training.

\chapter{Second order optimizer: The Newton-Raphson method} 
The Newton-Raphson method is one of the most appealing in its theoretical formulation; however, its application has some disadvantages.
The Newton-Raphson method (NR), under particular conditions, it is known to have \textbf{quadratic convergence}, making it a fast method (we will get to what this means later).  It is said to be a method of the \textbf{second order} since it leverages both first and second order derivatives, namely, the \textbf{gradient} and the \textbf{Hessian}, to iteratively find the extrema of the function. Given a multivariate function $f(\mathbf{\omega})$, with $\mathbf{\omega} \in \mathcal{R}^n$. The (n+1)-th step is expressed by the NR method as: 
\begin{equation} \label{eq: NR_method}
    \hat{\mathbf{\omega}}_{n+1} = \hat{\mathbf{\omega}}_{n} - [H(f(\hat{\mathbf{\omega}}_{n}))]^{-1} \nabla f(\hat{\mathbf{\omega}}_{n})
\end{equation}
where $\hat{\mathbf{\omega}}_{n}$ is the n-th step approximate solution, $\nabla f(\mathbf{\omega})$ is the gradient vector and $H(f(\mathbf{\omega}))$ the Hessian matrix, defined by: 
\begin{displaymath}
    (\nabla f(\hat{\mathbf{\omega}}_{n}))_i = \left. \frac{\partial f(\mathbf{\omega})}{\partial \omega_i} \right\rvert_{\mathbf{\omega} = \hat{\mathbf{\omega}}_{n}} \quad \text{and} \quad (H(f(\hat{\mathbf{\omega}}_{n})))_{ij} = \left. \frac{\partial^2 f(\mathbf{\omega})}{\partial \omega_i \partial \omega_j} \right\rvert_{\mathbf{\omega} = \hat{\mathbf{\omega}}_{n}}
\end{displaymath}
Equation \ref{eq: NR_method} leverages the Tailor expansion of the function around the n-th step, assuming the function is sufficiently differentiable, and the Hessian matrix is non-singular. The gradient vector $\nabla f(\hat{\mathbf{\omega}}_{n})$ indicates the direction of the steepest ascent of the function at $\hat{\mathbf{\omega}}_{n}$. So the algorithm will change the parameter by moving in the direction of the \textbf{steepest descent} by a quantity proportional to the inverse of the curvature (The Hessian matrix). The algorithm moves in the direction of the steepest descent with a large step if the curvature is small and a small step if the curvature is large. 
It can be shown that for functions that are smooth enough and for a sufficiently large $n$, the NR method allows you to: 
\begin{displaymath}
    \left\lVert \epsilon_{n+1} \right \rVert \leq C \left\lVert \epsilon_{n} \right \rVert^2,
\end{displaymath}
which is the definition of quadratic convergence. 
It is important to note that quadratic convergence is contingent upon the function being sufficiently smooth and the initial estimate being reasonably close to the true solution. Additionally, the requirements that the Hessian matrix be non-singular and positive (semi-)definite at every iteration to ensure convergence might not always be satisfied, particularly in high-dimensional optimization problems or when the function exhibits multiple local minima. Furthermore, for non-convex functions, NR method can quickly converge to saddle points or local minima, making it less effective. 
The practical challenges associated with the NR method that limit its use for machine learning models can be summarized in three points:
\begin{itemize}
    \item \emph{Non-Convex functions}: as already stressed, the NR method with non-convex functions quickly converges to a saddle point or local minima. 
    \item \emph{Computational cost}: the computational cost of calculating the Hessian matrix and its inverse at each step can be prohibitively expensive.
    \item \emph{Implementation challenges}: Ensuring the Hessian matrix remains positive definite and managing numerical stability can be difficult, the efficiency quickly diminishes in high-dimensional spaces.
\end{itemize}
In practice, for convex optimization problems, some Newton-inspired algorithms can mitigate the challenges associated with the standard Newton-Raphson method by approximating the Hessian matrix instead of calculating it at each step (Quasi-Newton methods) or by leveraging the curvature information to improve robustness and convergence.

\chapter{Use of AI-assisted tools}
Throughout this work, AI-based tools such as ChatGPT, Copilot, DeepSeek, and Gemini were used to support the research process. Their use was limited to technical and exploratory purposes, including clarifying aspects of code implementation, assisting in debugging and optimization, and verifying details of machine-learning architectures such as normalizing flows.

All conceptual and technical developments, analyses, and interpretations were entirely carried out by the author. Any AI-assisted material was critically reviewed, adapted, and integrated solely to enhance clarity or efficiency. The final text, results, tables, figures, and conclusions presented in this thesis reflect the author’s independent understanding and original work.

\printbibliography

\end{document}